\documentclass[11pt]{article}
\pdfoutput=1
\usepackage{amsmath,amssymb,epsfig,amsfonts}
\usepackage{graphicx}
\usepackage{subfig}
\usepackage[usenames, dvipsnames]{color}
\usepackage{cite}
\usepackage{multirow}
\usepackage{hyperref}
\usepackage{verbatim} 
\usepackage{enumitem}
\usepackage{rotating}
\usepackage{xcolor}
\usepackage{multirow}
\usepackage{bm}
\usepackage[normalem]{ulem}
\usepackage{cleveref}
\usepackage{slashed}
\usepackage{mathdots}

\usepackage{tikz}
\usepackage{diagbox}
\usetikzlibrary{positioning}
\usetikzlibrary{calc}
\usetikzlibrary{decorations.pathreplacing,calligraphy}

\graphicspath{ {figs/} }

\makeatletter

\newcommand{\be}{\begin{equation}}
\newcommand{\ee}{\end{equation}}
\newcommand{\ba}{\begin{aligned}}
\newcommand{\ea}{\end{aligned}}

\newcommand{\bea}{\begin{eqnarray}}
\newcommand{\eea}{\end{eqnarray}}

\newcommand{\R}{{\mathbb R}}
\newcommand{\Z}{{\mathbb Z}}

\def\Tr{\mathop{\mathrm{Tr}}\nolimits}

\def\half{{\frac{1}{2}}}

\def\p{\partial}

\def\unit{{1\kern-.65ex {\rm l}}}
\def\1{{1\kern-.65ex {\rm l}}}

\newcount\hour \newcount\minute
\hour=\time \divide \hour by 60
\minute=\time
\def\now{%
\ifnum \hour<13
  \ifnum \hour=0 \advance \hour by 12 \number\hour:\else \number\hour:\fi%
     \ifnum \minute<10 0\fi%
     \number\minute%
\ A.M.%
\else \advance \hour by -12 \number\hour:%
  \ifnum \minute<10 0\fi%
  \number\minute%
  \ P.M.%
\fi%
}

\makeatother

\def\mb{\mathbb}
\def\mbf{\mathbf}
\def\mc{\mathcal}
\def\bp{\begin{pmatrix}}
\def\ep{\end{pmatrix}}

\begin{document}

\baselineskip=18pt  
\numberwithin{equation}{section}  
\allowdisplaybreaks  


%
%


\thispagestyle{empty}

\vspace*{0.8cm} 
\begin{center}
{{\huge {Trifectas for $T_N$  in 5d}
}}

 \vspace*{1.5cm}
Julius Eckhard,  Sakura Sch\"afer-Nameki, Yi-Nan Wang\\

 \vspace*{1.0cm} 
{\it Mathematical Institute, University of Oxford, \\
Andrew-Wiles Building,  Woodstock Road, Oxford, OX2 6GG, UK}\\

\vspace*{0.8cm}
\end{center}
\vspace*{.5cm}

\noindent
The trinions $T_N$ are a class of 5d $\mathcal{N}=1$ superconformal field theories (SCFTs) realized as M-theory on $\mathbb{C}^3/\mathbb{Z}_N \times \mathbb{Z}_N$. 
We apply to $T_N$, as well as closely-related SCFTs that are obtained by mass deformations, a multitude of  
recently developed approaches to studying 5d SCFTs and their IR gauge theory descriptions. Thereby we provide a complete picture of the theories both on the Coulomb branch and Higgs branch, from various geometric points of view  -- toric and gluing of compact surfaces as well as combined fiber diagrams --  to magnetic quivers and Hasse diagrams.

\newpage

\tableofcontents




\section{Introduction}

5d gauge theories are not renormalizable. Nevertheless, overwhelming evidence exists that they can have strongly-coupled UV completions by 5d SCFTs. 
Starting with the work by Seiberg \cite{Seiberg:1996bd}, there have since been a multitude of approaches to study, and even potentially classify such 5d SCFTs. One approach is to characterize the SCFTs by M-theory on 
a Calabi-Yau threefold singularity \cite{Morrison:1996xf, Intriligator:1997pq}, where the space of resolutions of singularities has an interpretation in terms of the Coulomb branch of the IR-description of the SCFT. In this approach the Higgs branch, which corresponds to the deformations of the singularity, is somewhat obscure. The advantage of this approach is however, that it does give evidence for SCFTs that do not admit a weakly coupled gauge theory description. 
Alternatively, 5d SCFTs are realizable in terms of 5-brane webs \cite{Aharony:1997bh,DeWolfe:1999hj,Bergman:2013aca,Bergman:2014kza,Zafrir:2014ywa,Bergman:2015dpa,Ohmori:2015tka,Zafrir:2015ftn,Zafrir:2015rga,Zafrir:2015uaa,Hayashi:2018lyv,Hayashi:2018bkd}.
This approach is limited to theories that have a gauge theory description, but has the advantaged that both Coulomb and Higgs branch parameters are often manifest.

One of the key characteristics of 5d SCFTs is their flavor symmetry, which is generally enhanced compared to the IR flavor symmetry of any of its gauge theory descriptions. These features can be determined from both geometry and brane webs in a case-by-case analysis, but most systematically in the approach 
developed in \cite{Apruzzi:2019vpe,  Apruzzi:2019opn, Apruzzi:2019enx, Apruzzi:2019kgb} that encodes both the UV flavor symmetry, and the possible mass deformations in terms of a graph: each such  combined fiber diagram (CFD) is associated with a 5d SCFT. 
One goal of the present paper is to make this approach more accessible, and to put it into the context of other, perhaps more widely known approaches.

From a geometric point of view, one particularly well-studied subset {is the toric Calabi-Yau threefold singularity}~\cite{Xie:2017pfl,Closset:2018bjz}. Nonetheless, the class of models that can be studied using the toric geometry toolbox is relatively limited. Alternatively, one can start with the geometries underlying the 6d classification of SCFTs in F-theory \cite{Heckman:2013pva}. Although it remains an open problem, whether all 5d SCFTs arise from 6d by circle-reduction and mass deformations, this class of geometries reproduces all {of the known 5d SCFTs} -- and more. The elliptically fibered Calabi-Yau geometries that are used in F-theory, have a partial resolution, which characterize 5d SCFTs  and their Coulomb branch descriptions. This is the approach taken in \cite{DelZotto:2017pti,Apruzzi:2018nre, Apruzzi:2019vpe,  Apruzzi:2019opn, Apruzzi:2019enx, Apruzzi:2019kgb, Bhardwaj:2019fzv, Bhardwaj:2019xeg}. Alternatively one can directly classify the collection of surfaces that give rise to the Cartans of the 5d gauge theories, without constructing the full  non-compact CY geometry  \cite{Jefferson:2017ahm, Bhardwaj:2018yhy, Jefferson:2018irk, Bhardwaj:2018vuu,Bhardwaj:2019ngx,Bhardwaj:2020gyu}.

Another recent development is the structure of the Higgs branch of 5d SCFTs. Its magnetic quiver (MQ) is defined to be a 3d $\mc{N}=4$ quiver gauge theory, whose Coulomb branch gives the Higgs branch of the 5d SCFT \cite{Ferlito:2017xdq}. More recently, the toolkit of Hasse diagrams was developed from the magnetic quiver \cite{Bourget:2019aer}, which encodes the foliation structure of the Higgs branch as a hyper-K\"ahler singularity. In practice, the magnetic quiver and Hasse diagrams can be derived from the brane web description, for various classes of theories \cite{Cabrera:2018ann,Cabrera:2018jxt,Cabrera:2018uvz,Bourget:2019rtl, Bourget:2020gzi}.

The goal of this paper is to study all of these approaches in the context of a single class of theories -- the $T_N$ theories, which we believe will be useful in order to get an overview and understanding of the interconnection between various approaches. The $T_N$ theory was originally introduced in the context of 4d $\mc{N}=2$ class $\mc{S}$ theories~\cite{Gaiotto:2009we}, as $N$ M5-branes wrapping 2-sphere with three full punctures. It has central status in the classification of 4d $\mc{N}=2$ theories, as it can be used as building blocks to glue more complicated theories~\cite{Gaiotto:2009gz,Chacaltana:2010ks,Benini:2009mz}.

The theories $T_N$ in 5d were first introduced in \cite{Benini:2009gi}, where they were studied already in terms of toric geometry and brane webs. In the geometric picture, the $T_N$ SCFT is realized in terms of M-theory on the Calabi-Yau threefold singularity $\mathbb{C}^3/\mathbb{Z}_N\times \mathbb{Z}_N$. Its UV flavor symmetry is
\be
G_F = SU(N)^3 \,,
\ee
which is enhanced to $G_F= E_6$ in the $N=3$ case. The Coulomb branch dimension (rank) is
\be
r={1\over 2} (N-1)(N-2)\,.
\ee
{$T_2$ is a rank 0 theory and we will not further discuss it here. }
A weakly coupled description in terms of a linear quiver is known \cite{Hayashi:2014hfa}
\be \label{QuiverDescription}
{[N]- SU(N-1)_0  - SU(N-2)_0 - \cdots - SU(2) - [2 ]} \,,
\ee
and the partition functions of $T_N$ were studied in \cite{Bao:2013pwa,Hayashi:2013qwa,Hayashi:2014wfa,Hayashi:2015xla}.
{As a complementary approach, its holographic duals in $AdS_6$ }were determined in 
 \cite{Bergman:2018hin,Fluder:2018chf,Kaidi:2018zkx, Uhlemann:2019ypp}, based on the constructions in \cite{Apruzzi:2014qva, DHoker:2016ujz, DHoker:2017mds}.

In this paper, we will show how {various} approaches developed in the past two years are realized in this class of theories. In particular, {we will not only present the definition of CFDs and its derivation from the toric description}, but also the brane web approach can be used to determine the refined structure of the Higgs branch in terms of the Hasse diagram. Moreover, we provide new results on the BPS spectrum, the web of different SCFTs related by mass deformations and the Higgs branch structure of these theories. {Beyond $T_N$ theories, and its descendants obtained by mass deformations -- a {substantial} class of which have no weakly coupled description -- we also study their 5d and 6d ancestors, i.e. SCFTs, from which $T_N$ descends after decoupling hypermultiplet matter. }

The structure of this paper is as follows: in section 2, we present the definition of 5d $T_N$ theories in geometric language. We also review the geometric engineering rules of 5d SCFTs and gauge theories in M-theory, the toolkit of toric geometry and the definition of CFDs.  In section 3, we discuss the descendants and ancestors of the 5d $T_N$ theories related by mass deformations, which is encoded in the transitions of the CFDs. Furthermore, we will present new results on the low spin BPS spectrum of the $T_N$ theories in terms of $SU(N)^3$ representations. In section 4, we discuss the IR gauge theory descriptions of the 5d $T_N$ theories, from the perspectives of toric geometry, CFDs and box graphs. Finally, in section 5, we review the brane web descriptions, the magnetic quivers and present new results of the Hasse diagrams. We also show the magnetic quiver of some descendant theories of $T_N$.

\section{SCFTs, Geometry, and CFDs}

We begin with the realization of 5d gauge theories and SCFTs using M-theory on a singular Calabi-Yau geometry. 
The Calabi-Yau resolutions of these singularities have various characterizations, for $T_N$ in particular in terms of the toric resolutions. 
We connect the toric description, which is well-known, to the new approach using CFDs. 

\subsection{M-theory, Calabi-Yau Singularities and 5d SCFTs}

In this section we will give a brief introduction of the basics of 5d $\mathcal{N}=1$ superconformal field theories (SCFTs) and their descriptions in terms of M-theory on a non-compact Calabi-Yau threefold. 
Let us start with general properties of weakly coupled 5d $\mathcal{N}=1$ gauge theories.
In 5d the gauge coupling has negative mass dimensions, hence the Yang-Mills action is non-renormalizable. {However, the theory in the  UV can be a superconformal fixed point that flows to said gauge theory.} In other words, we can interpret a weakly coupled gauge theory in the IR as an effective description of an SCFT. In general, such an SCFT might have several inequivalent mass deformations, which can be interpreted as different ``UV dual" gauge theory descriptions.

We consider a 5d $\mc{N}=1$ gauge theory with the gauge group $\prod_{I=1}^N G_I$ and matter content $\oplus_f \bm{R}_f= \oplus_f (\bm{R}_1,\dots,\bm{R}_N)_f$, 
where the matter hypermultiplets transform in representations $\bm{R}_I$ of the simple factors $G_I$.
We only allow matter that is either charged under only one of the $G_I$ or in the bifundamental of $G_I\times G_J$.
The Coulomb branch of this theory can be described by the 1-loop IMS prepotential \cite{Intriligator:1997pq}. It is a function of the Coulomb parameters, i.e. the vacuum expectation values (vevs) of the real scalars in the vector multiplet. For a simple gauge group it is given by
\be \label{Prepotential}
\mathcal{F}_G(\phi_i)= \frac{1}{2g_0^2} h_{ij} \phi^i \phi^j + \frac{k}{6} d_{ijk} \phi^i \phi^j \phi^k +\frac{1}{6} \sum_{\alpha\in\Phi_+} \left|\alpha_i \phi^i \right|^3-\frac{1}{12} \sum_f \sum_{a=1}^{\dim \bm{R}_f} \left|\lambda^a_i  \phi^i -m_f \right|^3\,,
\ee
where $h_{ij}=\Tr_{\bm{F}}(T_i T_j)$, where $\bm{F}$ is the fundamental representation,  $d_{ijk}=\half\Tr_{\bm{F}}\left(T_i \left\{T_j,T_k\right\}\right)$, with the $T_i$ the Cartan generators of $G$, and the $\alpha_i$s are the positive roots $\Phi_+$ of $G$. Also, $g_0$ is the gauge coupling of $G$ and $k$ is the Chern-Simons level, which is only relevant for $G=SU(N>2)$. Finally, the $\lambda^a_i$ are the weights of the representations $\bm{R}_f$ and $m_f$ are the masses of matter hypermultiplets.
For a quiver theory, in the simplest case given by two simple gauge groups coupled by bifundamental matter $\bm{F}$, the prepotential is given by
\be
\mathcal{F}_{G_I - G_J}(\phi_i,\phi_j)= \mathcal{F}_{G_I}(\phi_i) + \mathcal{F}_{G_J}(\phi_j) - \frac{1}{12} \sum_{a=1}^{\dim \bm{F}_{G_I}} \sum_{b=1}^{\dim \bm{F}_{G_J}} \left|\lambda^a_i \phi^i + \lambda^b_j \phi^j -m_B\right|^3\,,
\ee
where each of the constituents has their individual gauge coupling and CS level as well as matter content and $m_B$ is the mass of the bifundamental. This naturally generalizes to longer quivers.

From the prepotential one can easily determine the 1-loop exact Lagrangian
\be
\ba
\mathcal{L}&=\tau_{ij} \left(F^i\wedge \star F^j + d\phi^i \wedge \star d\phi^j\right) +\frac{c_{ijk}}{24\pi^2} F^i \wedge F^j \wedge A^k+\dots\\
\tau_{ij}&= \frac{\p^2 \mathcal{F}}{\p\phi^i\p\phi^j}\,, \qquad c_{ijk}= \frac{\p^3 \mathcal{F}}{\p\phi^i\p\phi^j\p\phi^k}\,,
\ea
\ee
where the $A^i$ are the gauge fields with field strength $F^i$. For consistency, the effective gauge coupling $\tau_{ij}$ has to be positive definite and the effective Chern-Simons coefficients $c_{ijk}$ need to be integer-valued to ensure gauge invariance.
We see from \eqref{Prepotential} that the prepotential depends on the relative values of the masses and the Coulomb branch parameters. We can choose the Weyl wedge of $G$ such that $\left(\alpha_i \phi^i\right)\geq0$ for all roots, but the sign of $\left(\lambda^a_i  \phi^i -m_a \right)$ depends on the phase in the extended Coulomb branch, which is parametrized by $\phi^i$ and $m_a$.
This will be explained in more detail in section \ref{sec:BoxGraphs}. 

An important property of the gauge theories is their flavor symmetry. Let us consider the case where all the $m_f=0$, where the classical flavor symmetry $G_{\text{F,cl}}$ is maximal.
The classical flavor symmetry generated by $n$ fields in the representation $\bm{R}$ of $G_I$ is given by
\be
\bm{R}\text{ complex}\ \Rightarrow\ U(n)\,, \qquad
\bm{R}\text{ quaternionic}\ \Rightarrow\ SO(2n)\,, \qquad
\bm{R}\text{ real}\ \Rightarrow\ Sp(n)\,.
\ee
The other important matter type we will encounter is a single hypermultiplet in the bifundamental of $SU(N_1) \times SU(N_2)$ with classical flavor symmetry $U(1)$. Furthermore, each simple gauge group factor $G_I$ provides a topological symmetry $U(1)^I_T$ with current $J_T=\frac{1}{8\pi^2}\star \Tr(F\wedge F)$. In the strong coupling limit the flavor symmetry in the IR, $G_{\text{F, cl}}$,  can be enhanced by non-perturbative effects
\be
G_{\text{F,cl}} \times \prod_{I=1}^N U(1)^I_T \to G_{\text{F}}\,,
\ee
to the UV flavor symmetry with rank $\text{rk}(G_{\text{F}})=\text{rk}(G_{\text{F,cl}})+N$.

\begin{table}
\center
\begin{tabular}{c||c}
Gauge Theory & Geometry\\
\hline
\hline
SCFT & Singular $\text{CY}_3$ $Y$ \\
\hline
Extended Coulomb branch & Relative K\"ahler cone\\
\hline
Gauge bosons associated to Cartan subalgebra& Compact divisors $S_i$\\
\hline
Weakly coupled (IR) descriptions & Choice of ruling $f_i \hookrightarrow S_i \to \Sigma_i$\\
\hline
VEVs $\phi_i$ & Volume of $f_i$\\
\hline
Gauge group $G$ &  $C_{ij}^{\mathfrak{g}}=-S_i \cdot f_j$\\
\hline
W-Bosons & M2-branes on $f_i$\\
\hline
Gauge coupling $1/g^2$ & Volume of $S_i$\\
\hline
CS coefficients $c_{ijk}$ & $S_i \cdot S_j \cdot S_k$\\
\hline
Hypermultiplet matter& M2-branes on $(-1)$-curves in ruling\\
\hline
flavor masses $m_f$ & Volume of $(-1)$-curves\\
\hline
Change of gauge theory phases & Flops of $(-1)$-curves\\
\hline

\end{tabular}
\caption{Dictionary between 5d gauge theory data and the geometric description in M-theory on a CY$_3$.
\label{tab:Dictionary}}
\end{table}

A way to provide evidence for the strongly-coupled UV fixed points is the construction of such 5d gauge theories and their moduli spaces in  M-theory on a singular, non-compact $\text{CY}_3$ $Y$. In this paper, we require that the singularity has a crepant, i.e. Calabi-Yau, resolution $\widetilde{Y}$. 
The (relative) K\"ahler moduli space of $\widetilde{Y}$ corresponds to the extended Coulomb branch of the gauge theory. The origin of the extended Coulomb branch, that is associated to the SCFT, exactly corresponds to the singular limit of the CY.
$\widetilde{Y}$ is characterised by a set of compact divisors, i.e. complex surfaces, $S_i$, $i = 1, \cdots, r$, of finite volume, as well as non-compact divisors, which usually characterize the flavor symmetry. The compact divisors are Poincar\'{e} dual to harmonic $(1,1)$ forms, which can be used to expand the M-theory $C_3$ form and yield $r$ abelian gauge fields. If there is a non-abelian gauge theory description of the theory in question, the W-bosons arise from  M2-branes wrapping collapsed curves in a particular singular limit of the surfaces $S_i$. 
The key concept for this is a geometric ruling of the divisors by rational curves $f_i$, i.e.  $\mathbb{P}^1$s, over curves $\Sigma$ 
\be
f_i \hookrightarrow S_i \to \Sigma_i\,, \qquad f_i \cdot S_i=-2\,, \quad f_i \cdot f_i=0\,.
\ee
In summary, M-theory on this geometry yields two types of vector multiplets:
\begin{enumerate}
\item The $U(1)$ gauge bosons, from the expansion of $C_3$ in the $(1,1)$-forms dual to the $S_i$, such that $i=1,\dots,\text{rk}\ G$.
\item The W-bosons are obtained from M2-branes wrapping the fibers  $f_i$ of rulings, which are {rational curves with self-intersection 0 inside $S_i$.}
\end{enumerate}
The Cartan matrix of $G$ is given by the intersection number
\be
C_{ij}^{(\mathfrak{g})}=-S_i\cdot f_j\,.
\ee
We see that the W-bosons only become massless in the limit where the volume of the fibers $f_i$ goes to 0, and we obtain singularities of ADE-type over the curves $\Sigma$. This suggests that the volumes of the $f_i$, i.e. the K\"ahler parameters, are proportional to the vevs of the $\phi_i$ as the gauge group enhancement happens at the origin of the Coulomb branch.
Different rulings of the same resolution lead to ``dual" weakly coupled descriptions in terms of different gauge theories that however have the same UV completion.
Finally, the Chern-Simons coefficients of the gauge theory depends on the triple intersection number
\be
c_{ijk}^{\text{geo}}=S_i \cdot S_j \cdot S_k\,,
\ee
which encodes both the tree-level and one-loop contributions. 

The hypermultiplets that transform in representations of the gauge group correspond to rational curves with self-intersection $(-1)$ inside 
a surface $S_i$. Depending on their intersection with the $S_i$ we can determine the charges under the various $U(1)$s in the Cartan subalgebra, and thus the representations under which the matter transforms. The masses of these matter fields are given by the volumes of the $(-1)$-curves. 
As the volume of the $S_i$s is taken to zero the masses vanish, indicating the origin of the extended Coulomb branch. Different gauge theory phases are obtained by flops within the reducible compact surface $\cup_i S_i$. The corresponding UV fixed point is unchanged under such internal flops. 

Decoupling of hypermultiplet matter, and thus flowing to a different UV-fixed point is realized in terms of flops that take a $(-1)$-curve outside of $\cup S_i$. In the singular limit, its volume, and thus the mass of the associated hypermultiplet stays finite, and thus decouples from the SCFT sector. We will consider in the following both internal flops, as well as flops that correspond to decoupling/mass-deformations. The latter will give rise to the complete decoupling or RG-flow tree. The theories obtained after decoupling will be referred to as descendant theories.

\subsection{Toric Geometry}
\label{sec:toric}

Here we briefly review toric Calabi-Yau threefold singularities and the geometry of of their divisors. For a general introduction to toric geometry, see \cite{danilov1978geometry,fulton1993introduction,cox2011toric}. For the notations for toric Calabi-Yau threefold singularities, see for instance \cite{Xie:2017pfl}.

A \textit{toric threefold} $X_\Sigma$ is described by a \textit{toric fan} $\Sigma$, which is a set of cones in the 3d lattice $N=\mb{Z}^3$ with a common origin $(0,0,0)$. The 1d cones $\bm{v}_i=(v_{i,x},v_{i,y},v_{i,z})$ $(i=1,\dots,n)$ are 3d vectors with integral components,  also called  \textit{rays} of the toric fan. Geometrically, each ray corresponds to a \textit{toric divisor} (complex surface) $D_i$ of $X_\Sigma$. Similarly, the 2d cone $\bm{v}_i \bm{v}_j$ corresponds to the complete intersection curve $D_i\cdot D_j$, and the 3d cone $\bm{v}_i \bm{v}_j \bm{v}_k$ corresponds to the intersection point $D_i\cdot D_j\cdot D_k$. Here $v_i$, $v_j$ and $v_k$ are the boundary rays of these cones. It is required that the intersection of two cones is either empty or another cone in the toric fan. Two toric fans $\Sigma$ and $\Sigma'$ are equivalent if and only if the set of rays can be mapped to each other with an $SL(3,\mb{Z})$ rotation, while keeping the cone structure unchanged. The equivalence of two toric fans also induces the topological isomorphism between $X_\Sigma$ and $X_{\Sigma'}$.

There are three linear relations on the divisors $D_i$
\be
\label{linear-eq}
\sum_{i=1}^n v_{i,x}D_i=0\ ,\quad  \sum_{i=1}^n v_{i,y}D_i=0\ ,\quad  \sum_{i=1}^n v_{i,z}D_i=0\,.
\ee
The anticanonical divisor of $X_\Sigma$ is given by the sum of all the toric divisors
\be
-K_{X_\Sigma}=\sum_{i=1}^n D_i\,.
\ee
Hence if $X_\Sigma$ is Calabi-Yau, with $K_{X_\Sigma}=0$, all the rays $\bm{v}_i$ have to lie on the same plane in $\mb{Z}^3$. In this paper, after an $SL(3,\mb{Z})$ rotation, we take the form of all $\bm{v}_i$ to be
\be
\bm{v}_i=(v_{i,x},v_{i,y},1)\,.
\ee
A toric threefold is compact if and only if the cones in $\Sigma$ span the whole $\mb{Z}^3$. It is then easy to see that a toric Calabi-Yau threefold $X_\Sigma$ is always non-compact. The rays on the boundary of $\Sigma$ correspond to the non-compact divisors of $X_\Sigma$, which will be denoted by $D_\alpha$ in the remainder of the paper. 
 On the other hand, the rays in the interior of $\Sigma$ correspond to compact divisors of $X_\Sigma$, which will be denoted by $S_i$. In this notation, the curves $D_\alpha\cdot D_\beta$ are always non-compact, while the curves $S_i\cdot D_\alpha$ and $S_i\cdot S_j$ are always compact.

\begin{figure}
\begin{center}
\includegraphics[height=4cm]{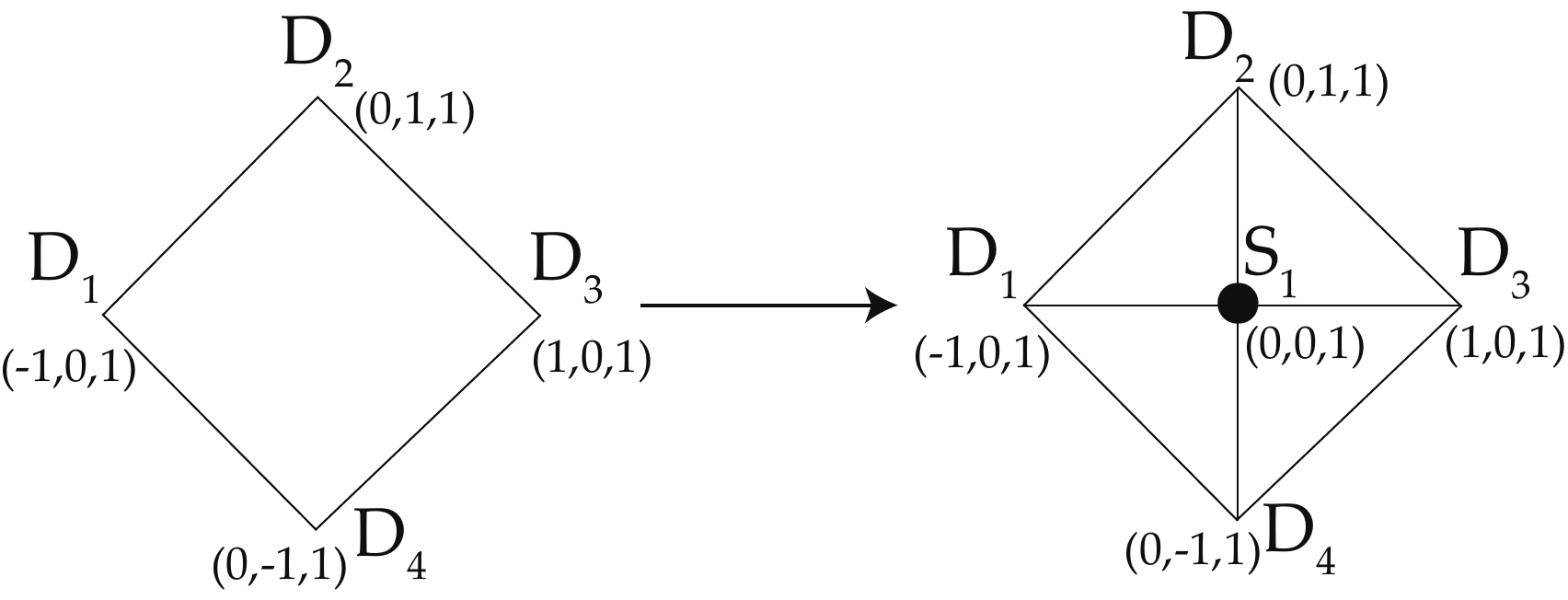}
\caption{An example of a toric Calabi-Yau threefold singularity and its crepant resolution. Each toric divisor is presented as a point on the $(x,y)$-plane, and we label their coordinates in $\mb{Z}^3$. Each 2d cone is presented as a line segment between the points and each 3d cone is given by a polygon. On the left hand side, there is a singular toric Calabi-Yau threefold with four non-compact divisors $D_1,\dots,D_4$. On the right hand side, after the crepant resolution, the toric fan is subdivided and each 3d cone (presented as a triangle) has unit volume. There is a new compact divisor $S_1$ in the middle.}\label{f:toric-resolution}
\end{center}
\end{figure}

A toric threefold is smooth if and only if each 3d cone has the form $\bm{v}_i \bm{v}_j \bm{v}_k$, and they all have  unit volume
\be
\bm{v}_i\cdot (\bm{v}_j\times \bm{v}_k)=\pm 1\,.
\ee
On the other hand, the toric threefold has a singularity if a 3d cone has more than three vertices, or its volume is greater than one. A crepant resolution of a toric Calabi-Yau threefold singularity exactly corresponds to a subdivision of the toric fan. After the resolution, all the 3d cones will have unit volume and $X_\Sigma$ will be smooth. We plot an example of a simple toric Calabi-Yau threefold singularity and its crepant resolution in figure~\ref{f:toric-resolution}.

\begin{figure}
\begin{center}

\includegraphics[width =5cm]{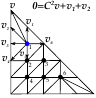}
\caption{The resolved toric Calabi-Yau threefold singularity corresponds to the $T_5$ theory. The toric fan of the compact toric divisor $S_1$ is constructed including all of its neighboring rays $\bm{v}$, $\bm{v}_i$ $(i=1,\dots,6)$.}
 \label{T5ToricIntersections}
\end{center}
\end{figure}

In this picture, the topology of compact toric divisor $S_1$ can be easily read off. $S_1$ itself is always a toric surface, and its toric fan $\Sigma(S_1)$ can be constructed as follows. Denote the 3d ray of $S_1$ in $\Sigma$ by $\bm{v}$, then each 2d cone $\bm{v}_i \bm{v}\in\Sigma$ gives rise to a ray
\be
\bm{v}^{(2)}_i=(v_{i,x}-v_x,v_{i,y}-v_y)
\ee
in $\Sigma(S_1)$. Similarly, each 3d cone $\bm{v}_i \bm{v}_j \bm{v}\in\Sigma$ gives rise to a 2d cone $\bm{v}^{(2)}_i \bm{v}^{(2)}_j$ in $\Sigma(S_1)$. From figure~\ref{f:toric-resolution}, it is easy to see that $S_1$ has the topology of Hirzebruch surface $\mb{F}_0=\mb{P}^1\times\mb{P}^1$. More generally, this rule can be applied to any compact toric divisor, see figure~\ref{T5ToricIntersections} for another example.

For a smooth toric threefold, $h^{1,1}$ and the self-triple intersection number of a compact divisor $S_i$ are given by 
\be
h^{1,1} (S_i) = \# (\text{neighboring vertices}) -2 \,,\qquad S_i^3= 10- h^{1,1}(S_i) \,.\label{self-tripleint}
\ee
The latter equation holds because $S_i^3=K_{S_i}\cdot_{S_i}K_{S_i}$, and $S_i$ is always a rational surface.

Then we consider the triple intersection number $D_{\bm{v}}^2\cdot S_i$, where $D_{\bm{v}}$ corresponds to the ray $\bm{v}\in\Sigma$, and it can be either compact or non-compact. $D_{\bm{v}}^2\cdot S_i$ can be computed by 
\be
D_{\bm{v}}^2\cdot S_i=C_{i,\bm{v}}\cdot_{S_i} C_{i,\bm{v}}\,,
\ee
where $C_{i,\bm{v}}=S_i\cdot D_{\bm{v}}$ is the complete intersection curve on $S_i$. The self-intersection number of $C_{i,\bm{v}}$ on $S_i$ can then be read off by a simple rule (see also~\cite{Morrison:2012js}): denote the 1d ray of $C_{i,\bm{v}}$ in $\Sigma(S_i)$ by $\bm{v}^{(2)}$, and its two neighbor rays by $\bm{v}_1^{(2)}$, $\bm{v}_2^{(2)}$. Then the self-intersection number $C_{i,j}^2 $ is given as a solution to 
\be
C_{i,\bm{v}}^2 \, \bm{v}^{(2)} + \bm{v}_1^{(2)} + \bm{v}_2^{(2)} =0\,.
\ee

\begin{figure}
\begin{center}

\includegraphics[height =6.5cm]{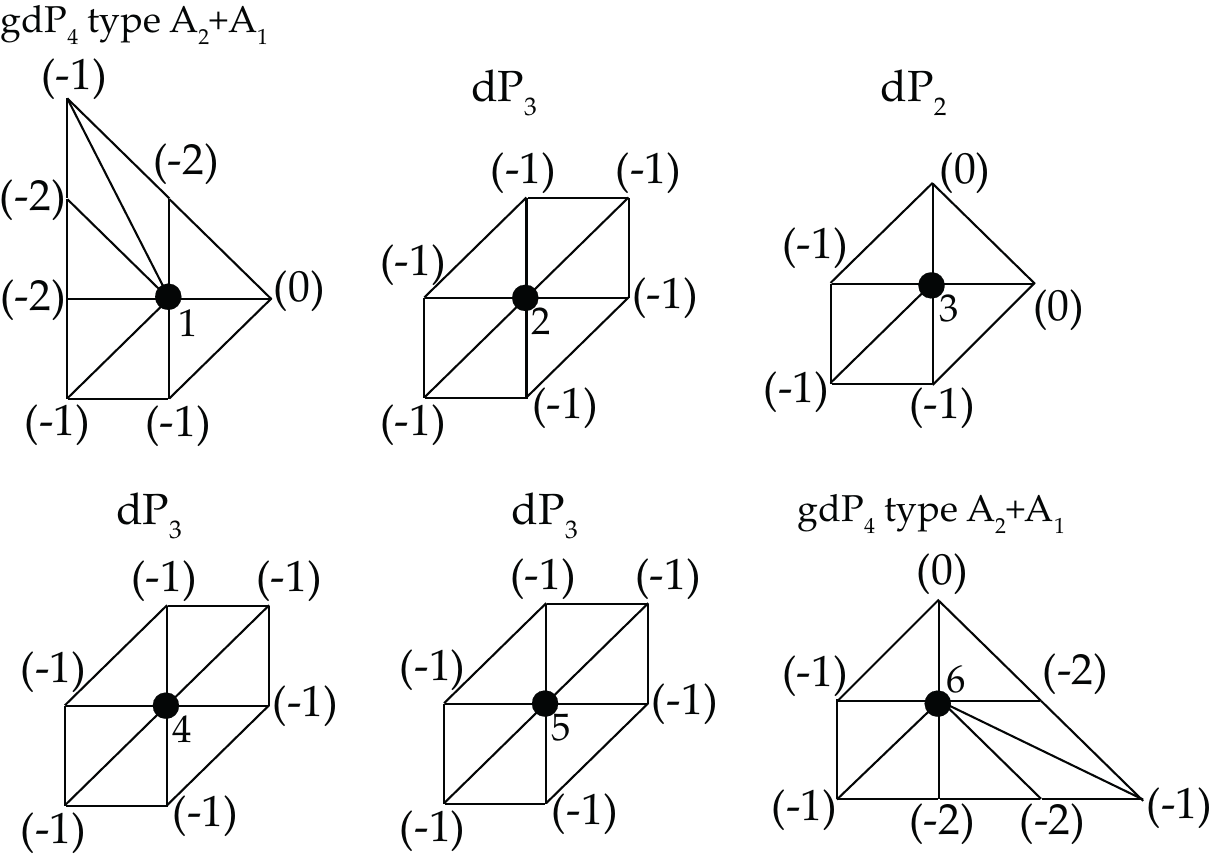}
\caption{The toric fan of each compact surface $S_i$. The subscript $i$ is labelled at the center of each toric fan, and the self-intersection number of each toric curve is written in the brackets. gdP$_n$ refers to the generalized del Pezzo surfaces.}
 \label{f:surface-topology}
\end{center}
\end{figure}

For the example in figure~\ref{T5ToricIntersections}, for the curves on $S_1$ (the blue node), we have $C_{1,\bm{v}}^2=-1$, $C_{\bm{v}_{4}}^2 = -1$, $C_{\bm{v}_{5}}^2 = -1$, $C_{1,\bm{v}_6}^2 = 0$ and $C_{1,\bm{v}_i}^2 = -2$ for the remaining nodes. From (\ref{self-tripleint}), we can compute $h^{1,1}(S_1)=5$ and $S_1^3=5$. Thus $S_1$ is a toric generalized del Pezzo surface $\text{gdP}_4$ with the following cycle of rational curves: $(-1,-2,0,-1,-1,-2,-2)$~\cite{Morrison:2012js,derenthal2014singular}. We present the topology of each surface $S_i$ in figure~\ref{f:surface-topology}. As one can see, the surface $S_3$ is a del Pezzo surface $dP_2$ and the surfaces $S_2$, $S_4$ and $S_5$ are all del Pezzo surfaces $dP_3$. The surface $S_6$ has the same topology as $S_1$. From figure~\ref{T5ToricIntersections} and figure~\ref{f:surface-topology}, we can also clearly see how the surfaces $S_i$ $(i=1,\dots,6)$ are glued together. For example, it is clear that $S_1$ and $S_2$ are glued along a ratioal curve with normal bundle $\mc{O}(-1)\oplus\mc{O}(-1)$. 

\begin{figure}
\subfloat{(a)}\includegraphics[height =5cm]{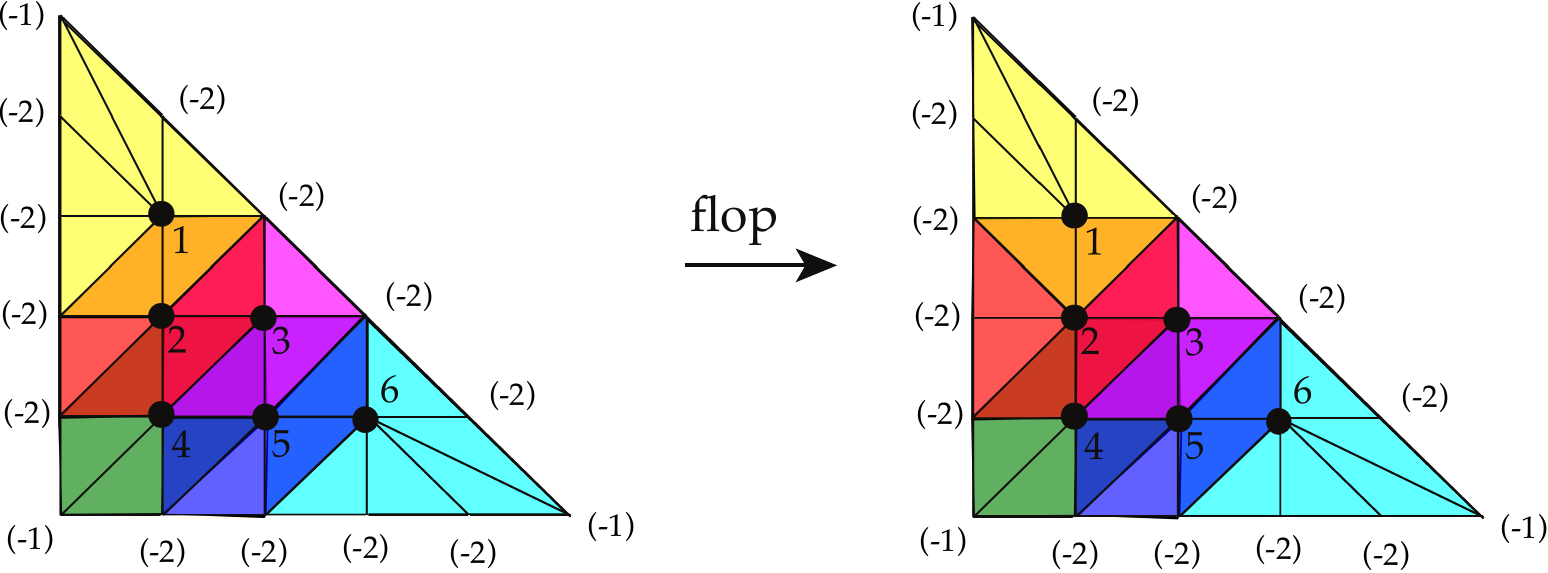}
\subfloat{(b)}\includegraphics[height =5cm]{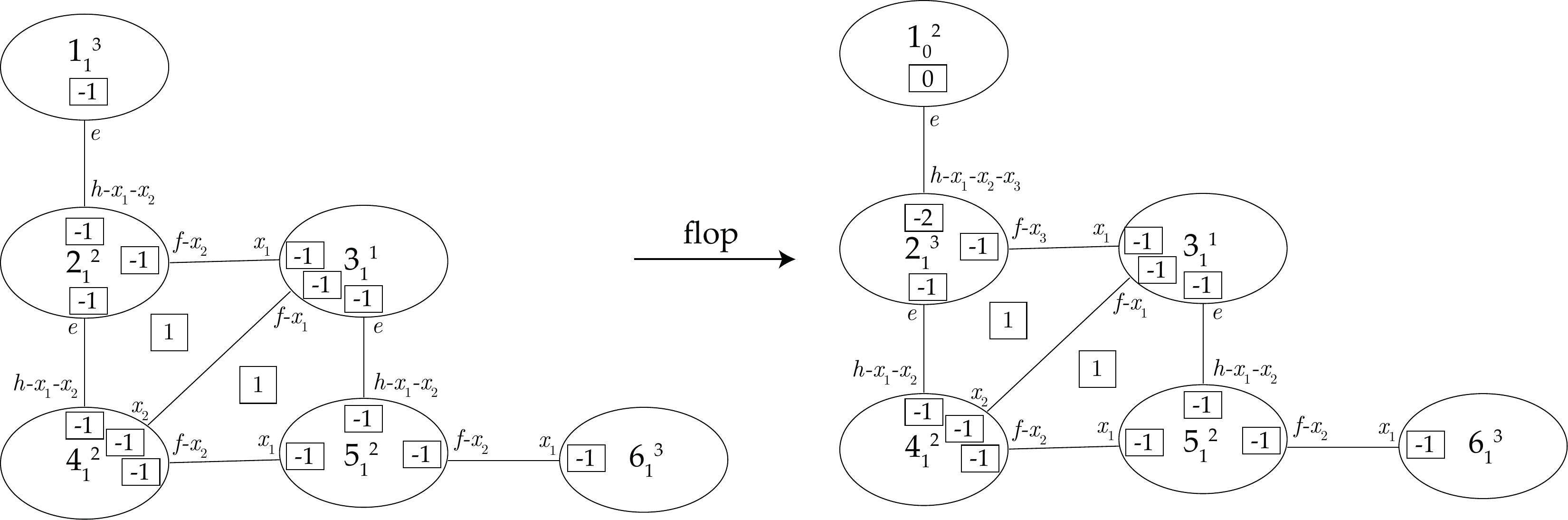}
\caption{(a) An example of toric flop of the resolved $T_5$ geometry among the compact divisors. 
 The triangulation in the picture changes, while the rays remain the same. We furthermore indicate how the geometry is glued from the rational surfaces such as in figure \ref{f:surface-topology}. The numbers $(-n)$ at the external vertices indicate the self-intersection numbers of curves that are intersections between compact and non-compact divisors and encode the $SU(5)^3$ flavor symmetry. 
  (b) Another presentation of the geometry, where the $i$th internal vertex $i_n^k$ is Bl$_k\mathbb{
  F}_n$. In this presentation the flavor symmetry is not manifest.}
 \label{f:toric-flop}
\end{figure}

Finally, for the triple intersection number among three distinct divisors $D_i\cdot D_j\cdot D_k$ on a smooth $X_\Sigma$ (no matter they are compact or not) is given by
\be
D_i\cdot D_j\cdot D_k=\left\{
\begin{array}{rl}1 & \bm{v}_i\bm{v}_j\bm{v}_k\in\Sigma\\
0 & \bm{v}_i\bm{v}_j\bm{v}_k\notin\Sigma\end{array}
\,\right. \,.
\ee
Here we list all the non-zero triple intersection numbers among the compact divisors $S_i$ in figure~\ref{T5ToricIntersections}:
\be
\ba
&S_1^3=5\ ,\ S_2^3=6\ ,\ S_3^3=7\ ,\ S_4^3=6\ ,\ S_5^3=6\ ,\ S_6^3=5\ ,\ S_1^2\cdot S_2=S_2^2\cdot S_1=-1\cr
&S_2^2\cdot S_3=S_3^2\cdot S_2=-1\ ,\ S_2^2\cdot S_4=S_4^2\cdot S_2=-1\ ,\ S_3^2\cdot S_4=S_4^2\cdot S_3=-1\cr
&S_3^2\cdot S_5=S_5^2\cdot S_3=-1\ ,\ S_4^2\cdot S_5=S_5^2\cdot S_4=-1\ ,\ S_5^2\cdot S_6=S_6^2\cdot S_5=-1\cr
&S_2\cdot S_3\cdot S_4=S_3\cdot S_4\cdot S_5=1\,.
\ea
\ee
A toric flop among compact surfaces changes the 2d and 3d cones in $\Sigma$, while leaving the rays unchanged, see figure~\ref{f:toric-flop} for an example. After this flop, the triple intersection numbers change as:
\be
(S_1^3\,,S_1^2\cdot S_2\,,S_2^2\cdot S_1\,,S_2^3)\rightarrow (S_1^3+1\,,S_1^2\cdot S_2-1\,,S_2^2\cdot S_1+1\,,S_2^3-1)\,.
\ee

As a point of reference, we also present the triple intersection numbers $S_i\cdot S_j\cdot S_k$ before and after the flop in the notations of ~\cite{Jefferson:2018irk,Bhardwaj:2018yhy,Bhardwaj:2018vuu} in figure \ref{f:toric-flop} (b).
Here each circle denotes a compact surface. For each compact surface $S_i$, the label $i_n^k$ in the center means that it is a Hirzebruch $\mb{F}_n$, blown up at $k$ points. The Picard group elements of such a surface are denoted by $h$, $e$, $f$ and $x_i$ $(i=1,\dots,k)$. They have the following intersection numbers:
\be
h^2=n\ ,\ e^2=-n\ ,\ f^2=0\ ,\ x_i\cdot x_j=-\delta_{ij}\ ,\ h\cdot e=0\ ,\ h\cdot f=e\cdot f=1\ ,\ x_i\cdot h=x_i\cdot e=x_i\cdot f=0\,.
\ee
Note that the choice of $n$ and the Picard group generators are not unique. Then two circles are connected by a line if the two surfaces $S_i$ and $S_j$ intersect at a curve $C_{ij}$. At the ends of the line, the number in the square box inside the circle $i$ denotes the intersection number $S_i\cdot S_j^2$. We also pressent the Picard group element of the curve $C_{ij}$ over the line. Finally, if three surfaces intersect at a point, we label the triple intersection number at the center of the triangle.

Note that the diagrams in figure \ref{f:toric-flop} (b) of the type used in \cite{Jefferson:2018irk,Bhardwaj:2018yhy,Bhardwaj:2018vuu} do not contain information about the non-compact divisors $D_\alpha$. Nonetheless, $D_\alpha$s are crucial for the determination of the flavor symmetry $G_F$ and the CFD, which will be discussed later. Contrary to that, the description in terms of the surfaces as in figure \ref{f:surface-topology} do contain this information -- and will therefore be key in encoding the flavor symmetries (and thereby the CFDs).

\begin{figure}
\begin{center}

\includegraphics[height =5cm]{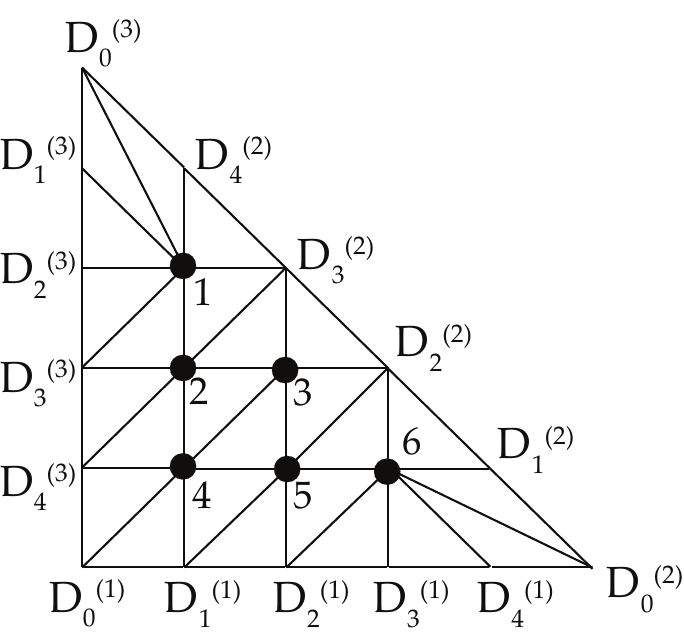}
\caption{The labeling of $T_N$ singularity, with an example of $N=5$.}
 \label{f:TN-labels}
\end{center}
\end{figure}

More generally, the orbifold singularity $\mb{C}^3/(\mb{Z}_N\times\mb{Z}_N)$ associated to $T_N$ theory can be described by a toric fan with rays $(0,0,1)$, $(N,0,1)$ and $(0,N,1)$. The fully resolved toric fan has rays $(x,y,1)$, $(0\leq x\leq N,0\leq y\leq N-x,1)$, and the corresponding divisors are labeled as figure~\ref{f:TN-labels}. The non-compact divisors on the boundary are labeled counterclockwise as $D_0^{(i)},D_1^{(i)},\dots,D_{N-1}^{(i)}$ $(i=1,2,3)$. The compact divisors in the interior are labeled as $S_1,S_2,\dots,S_{(N-1)(N-2)/2}$.

Now we briefly discuss the physics of the 5d $T_N$ theory from M-theory compactified on this toric Calabi-Yau threefold. The \textit{rank} of the 5d theory is equal to the total number of compact divisors and is 
\be
r=\frac{(N-1)(N-2)}{2}\,.
\ee 
The Cartan subalgebra of the superconformal flavor symmetry $G_F$ is given by the non-compact divisors $D_\alpha^{(i)}$ $(i=1,2,3)$ $(\alpha=1,\dots,N-1)$. As there exists three linear equivalence relations~\eqref{linear-eq}, the non-compact divisors $D_0^{(i)}$ are modded out, and the total rank of $G_F$ is
\be
\mathrm{rank}(G_F)=3N-3\,.
\ee
The W-bosons of $G_F$ are given by M2/anti-M2 branes wrapping certain 2-cycles, which become massless in the singular limit. In the case of figure~\ref{f:TN-labels}, these 2-cycles are (also see section~\ref{sec:CFD} for more information):
\be
\ba
&D_1^{(1)}\cdot (S_4+S_5)\ ,\ D_2^{(1)}\cdot (S_5+S_6)\ ,\ D_3^{(1)}\cdot S_6\ ,\ D_4^{(1)}\cdot S_6\cr
&D_1^{(2)}\cdot S_6\ ,\ D_2^{(2)}\cdot (S_3+2S_5+S_6)\ ,\ D_3^{(2)}\cdot (S_1+2S_2+S_3)\ ,\ D_4^{(2)}\cdot S_1\cr
&D_1^{(3)}\cdot S_1\ ,\ D_2^{(3)}\cdot S_1\ ,\  D_3^{(3)}\cdot (S_1+S_2)\ ,\ D_4^{(3)}\cdot (S_2+S_4)\,.
\ea
\ee
As a result, the points in the interior of the boundary lines form the Dynkin diagram of $G_F$. In the $T_N$ case $N>3$, we get exactly
\be
G_F=SU(N)^3\,.
\ee

Finally, we make a comment that the methodology introduced in this section can be generalized to any other toric configuration as well. The only subtlety is that the flavor symmetry read off from the toric diagram may only form a subalgebra of the full $G_F$. For example, in the case of the $T_3$ theory, the flavor symmetry will be enhanced to $E_6$ from $SU(3)^3$. This phenomenon will be explained from the BPS states counting in section~\ref{sec:BPS}.



\subsection{CFDs for $T_N$ and other Toric CY}
\label{sec:CFD}

Combined fiber diagrams (CFDs) were introduced in \cite{Apruzzi:2019vpe,  Apruzzi:2019opn, Apruzzi:2019enx, Apruzzi:2019kgb} to provide an efficient graphical tool to characterize for a given 5d SCFT the following characteristics: 
\begin{itemize}
\item $G_F$ -- UV flavor symmetry: subgraph of $(-2)$-vertices (marked in green)
\item Mass deformations: $(-1)$-vertices (marked in white)
\item weakly coupled descriptions and dualities
\end{itemize}
As we discussed, in the M-theory-SCFT dictionary, $(-2)$-curves inside the compact surfaces $\cup_i S_i$ correspond to flavor symmetries. These are encoded in the marked subgraph of the CFD. Likewise matter multiplets are encoded in $(-1)$-curves, which are precisely the vertices with label $(-1)$. Each vertex of a CFD is labeled not only by an integer (the self-intersection number of the associated curve), but also by the genus $g$. 
Mass deformations correspond to flopping $(-1)$-curves outside of the collection of compact surfaces $\cup_i S_i$. These will translate into operations on the CFDs which map a CFD for one SCFT to a descendent's CFD. Such flops are realized in terms of {\it CFD transitions}.

The initial framework of CFDs was not dependent on a toric description, and simply relied on the resolution of Calabi-Yau singularities (in particular elliptic models). 
Here we will apply this approach to toric Calabi-Yau three-folds, where it turns out the CFDs are also very natural objects to define, which encode all the above data.

We will start by discussing CFDs for a general toric geometry, which is given by a triangulation of a convex  2d polyhedron. In the polyhedron, the internal points are the compact divisors $S_i$ $(i=1,\dots,r)$ and the boundary points are non-compact divisors $D_\alpha$. If an internal point corresponding to $S_i$ is connected to a boundary point corresponding to $D_\alpha$, then the line segment corresponds to the intersection curve $S_i\cdot D_\alpha$. The general procedure of reading off CFDs from geometry was developed in \cite{Apruzzi:2019kgb}, which will be applied here.

In the CFD associated to this geometry, the labels $(n,g)$ for each node $D_\alpha$ are given by:
\be
\label{n-CFDnode}
n(D_\alpha)= (D_\alpha)^2  \cdot \left(\sum_{i=1}^r \xi_{i,\alpha} S_i\right)\,,
\ee
\be
\label{g-CFDnode}
g(D_\alpha)=1+\frac{1}{2}\left[n(D_\alpha)+D_\alpha  \cdot \left(\sum_{i=1}^r \xi_{i,\alpha} S_i\right)^2\right]\,.
\ee
Geometrically, $n$ is the self-intersection number of the intersection curve and $g$ its genus.
Here, $\xi_{i,\alpha}$ is an integral multiplicity factor associated to the curve $S_i\cdot D_\alpha$, which is read off in the following way. 

Given a fixed $D_\alpha$, we plot all the curves $S_i\cdot D_\alpha$ with their self-intersection numbers $S_i^2\cdot D_\alpha$ in $D_\alpha$, which form a loop in the toric case:
\be
 \qquad (S_{i_1}^2\cdot D_\alpha)-(S_{i_2}^2\cdot D_\alpha)-\dots-(S_{i_p}^2\cdot D_\alpha)\,.\label{stringcurve}
\ee
Then we blow down the $(-1)$-curves recursively, until the procedure terminates. Geometrically, this process corresponds to the flop among different $S_i$s, which does not change the CFD and the corresponding SCFT. In the end, the remaining curves are assigned with multiplicity factor one. Then we reverse the process to go back to (\ref{stringcurve}) via a sequence of blow ups, and the multiplicity factor of the new exceptional curve in each step equals to the sum of the multiplicity factors of its neighboring curves. For example, if we have the following sequence of (\ref{stringcurve}) in the beginning -- we list here only the self-intersection numbers of the curves, i.e. $(-2)$ and $(-1)$:
\be
(-2)-(-1)-(-2)\,,
\ee
then after the blow down process, we get:
\be
\begin{array}{c}
\overset{1}{(-2)}-\overset{2}{(-1)}-\overset{1}{(-2)}\\
\downarrow\\
\overset{1}{(-1)}-\overset{1}{(-1)}\\
\downarrow\\
\overset{1}{(0)}
\end{array}
\ee
We already labeled the multiplicity factor for each curve above them. In the toric picture, this process exactly corresponds to the following flops:
\be
\label{212-Flops}
\includegraphics[width=12cm]{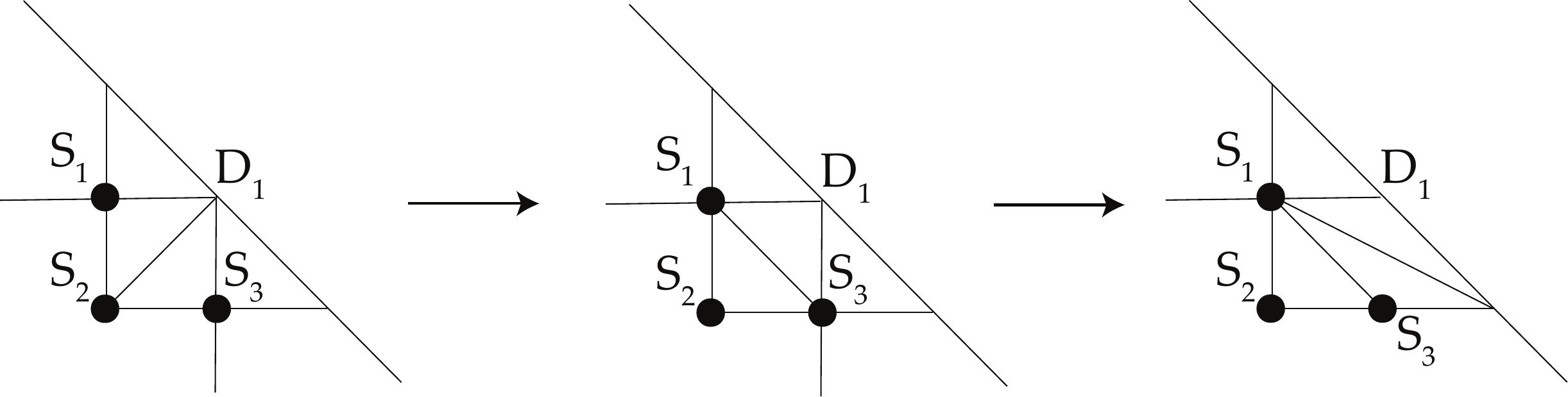}
\ee
One can check that the $(n,g)=(-2,0)$ for $D_1$ is indeed invariant during the process, using the formula (\ref{n-CFDnode},\ref{g-CFDnode}). Hence the complete intersection curve $D_1\cdot (\sum_i\xi_{i,\alpha} S_i)$ always has normal bundle $\mc{O}(0)+\mc{O}(-2)$, and the M2 brane wrapping mode over it exactly gives rise to a W-boson of the non-Abelian flavor symmetry. 

From this definition of CFD vertices, it is straight forward to see that any interior point in a boundary line gives rise to a flavor (marked) vertex with $(n,g)=(-2,0)$ in the CFD. Because in the terminated geometry as in (\ref{212-Flops}), the curve $S_1\cdot D_1$ always has normal bundle $\mc{O}(0)+\mc{O}(-2)$. 
For the vertices of the polyhedron, it corresponds to a CFD vertex with $n\geq -1$ and $g=0$. It always has genus zero in the toric case, since all the toric curves are topologically $\mb{P}^1$.
Finally, the number of edges between two CFD vertices corresponding to $D_\alpha$ and $D_\beta$ is given by:
\be
m_{\alpha\beta}=\sum_{i=1}^r\xi_{i,\alpha}S_i\cdot D_\alpha\cdot D_\beta\,,\label{m-CFDedge}
\ee
which assumed that $\xi_{i,\alpha}=\xi_{i,\beta}$ for all the terms that contribute. In the toric picture, all the relevant terms always have $\xi_{i,\alpha}=\xi_{i,\beta}=1$, because the relevant curves are at the left and right end of the string (\ref{stringcurve}). Hence (\ref{m-CFDedge}) is always well-defined.

\begin{figure}
\centering\includegraphics[width=6cm]{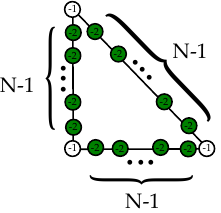} \,.
\caption{CFD for the $T_N$ theory. The $(-2)$-curves are the marked vertices from which the $SU(N)^3$ flavor symmetry can be read off. The $(-1)$ curves correspond to unmarked vertices in the CFD, which are the starting points for CFD-transitions. \label{fig:CFDTN}}
\end{figure}

For the particular $T_N$ theory, we denote by $D_\alpha^{(k)}$, $k=1, 2,3$ and $\alpha = 0, \cdots, N-1$ the non-compact divisors that correspond to the $SU(N)^3$ flavor symmetry. The label $k$ is assigned in a counter-clockwise order. The $\alpha=0$ node can be identified with the affine node. The compact divisors are $S_i$, $i=1,\dots,\frac{(N-1)(N-2)}{2}$. Then the $(n,g)$ for each non-compact divisors are 
\be
\ba
\label{TN-CFDnode}
(n(D_\alpha^{(k)}),g(D_\alpha^{(k)}))= \left\{ 
\ba
(-2,0) & \qquad \alpha \not=0\cr 
(-1,0) & \qquad \alpha =0 
\ea
\right.
\ea
\ee
which will be identified with the marked/unmarked vertices of the CFD. 
The relative intersection numbers read off from $D_\alpha^{(k)}\cdot D_\beta^{(j)}\cdot \sum_i S_i$ yield the CFD in figure \ref{fig:CFDTN}. 
To obtain the complete CFD for the $T_N$ theory that has also the manifest $SU(N)^3$  flavor symmetry, it is key that at least one compact divisor connects to each external vertex.

We consider an example, the $T_5$ theory, with the following toric triangulations and CFD:
\be\label{T5Ex}
\includegraphics[width=11cm]{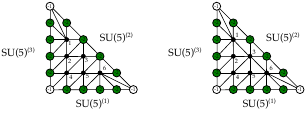}
\ee
This is a rank 6 theory, and we label the 6 compact divisors by $S_i$, from top to bottom, left to right. 
The reduced intersection matrix $S_i (D_\alpha^{(k)})^2$ with the Cartans of the flavor symmetry $SU(5)^3$ with Cartan divisors $D_{\alpha}^{(k)}$, $k=1, 2, 3$ and $\alpha= 0, \cdots, 4$ is
{\scriptsize \be
\begin{array}{c|ccccccccccccccc} 
&
D_0^{(1)} &D_1^{(1)} &D_2^{(1)} &D_3^{(1)} &D_4^{(1)} & 
D_0^{(2)} &D_1^{(2)} &D_2^{(2)} &D_3^{(2)} &D_4^{(2)} &
D_0^{(3)} &D_1^{(4)} &D_2^{(3)} &D_3^{(3)} &D_4^{(3)}  \cr 
\hline
S_1 & 0&0&0&0& 0&        0&0&0&-a& -2 &    -1 & -2 & -2 & -1 & 0  \cr \hline 
S_2 & 0&0&0&0& 0&        0&0&0&-b& 0 &    0 & 0 & 0 & -1 & -1  \cr \hline 
S_3 & 0&0&0&0& 0&        0&0&-a& -c & 0 &    0 & 0 & 0 & 0 & 0  \cr \hline 
S_4 & -1 & -1&0&0& 0&    0&0&0&0& 0 &    0 & 0 & 0 & 0 & -1 \cr \hline 
S_5 &  0&-1&-1&0& 0&        0&0&-b&0& 0 &    0 & 0 & 0 & 0 & 0  \cr \hline 
S_6 & 0&0&-1&-2& -2&        -1&-2&-c&0& 0 &    0 & 0 & 0 & 0 & 0  \cr \hline \hline
 \xi_{i,\alpha^{(k)}} S_i & -1&-2&-2&-2& -2&        -1&-2&-2&-2& -2 &    -1 & -2 & -2 & -2 & -2\cr\hline
\end{array}
\ee
}
In the left triangulation on the LHS of (\ref{T5Ex}), the assignment is $a=c=0$, $b=1$. In the triangulation on the RHS of (\ref{T5Ex}) the assignment is $a=1=c$, $b=0$. Although the two triangulations are inequivalent, they both correspond to the same SCFT fixed point, and thus same CFD, and thus SCFT.

Note that for the left triangulation, there are non-trivial multiplicity factors
\be
\xi_{2,3^{(2)}}=\xi_{5,2^{(2)}}=2\,,
\ee
while all the other multiplicity factors are trivially one.

The approach presented here is not limited to the $T_N$ theories, but is quite generally applicable to toric Calabi-Yau threefolds. 
In particular there is a class of geometries that are closely related to $T_N$, which have the toric diagram given by a square with side lengths $(n+1) \times (N+1)$. For $n=N$ this is simply two copies of $T_N$ glued along the diagonal. The associated CFD and SCFT was discussed in \cite{Apruzzi:2019kgb}, which has $SU(n)^2 \times SU(N)^2$ flavor symmetry and arises as the 5d avatar of the $SU(n)-SU(n)$ conformal matter theory of rank $N$.

\begin{figure}
\centering
\subfloat{(a)}\includegraphics[width=6cm]{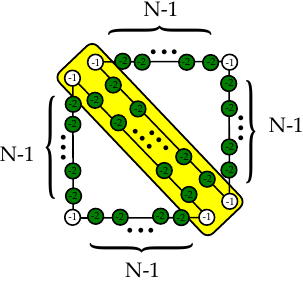}
\subfloat{(b)}\includegraphics[width=5cm]{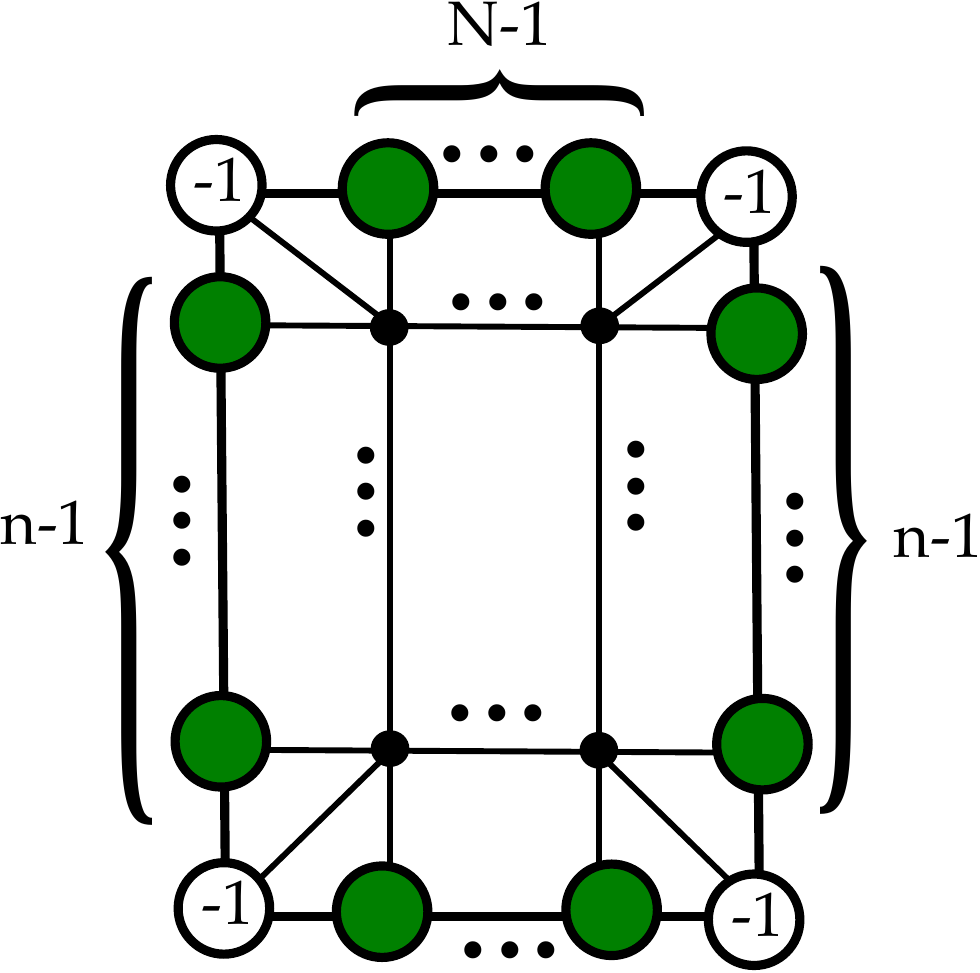}
\caption{(a) Gluing of two $T_N$ along the diagonal. The nodes in the yellow box corresponding to $SU(N)$ flavor are gauged and identified.  (b) An example of a toric model, with flavor $SU(N)^2\times SU(n)^2$, which for $N=n$ corresponds to the gluing in (a).}
\label{fig:TwoT2-Toric&CFD}
\end{figure}

\section{Parents, Descendants, and BPS States of $T_N$}
\label{sec:DescendantsFlops}

CFDs not only encode the flavor symmetries  of the UV fixed point, but also all mass-deformations, i.e. decouplings of hypermultiplets. 
The resulting theory, the so-called descendants, have a description in terms of CFDs again, and for $T_N$ we relate the decoupling to specific toric flops.
In addition to these descendants, obtained by mass deformation, we also identify the parent theories of $T_N$, which are SCFTs in themselves, in 5d. Following up the `genealogy tree' of $T_N$ to the KK-theory, we also identify the 6d uplift. 
To keep with the generic analogy, we also determine the key characteric of these theories, by computing their BPS states using the CFDs and toric resolutions.

\subsection{CFD-Transitions and Flops}

\begin{figure}
\centering
\includegraphics[width=11cm]{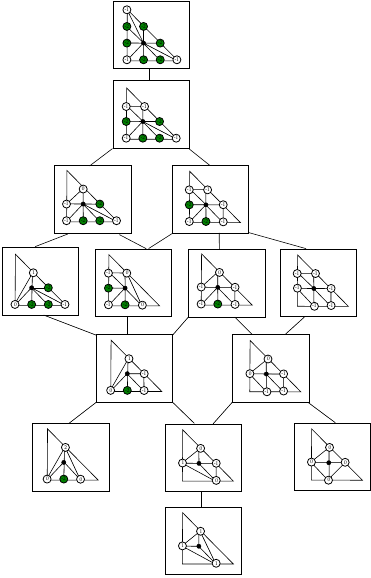}
\caption{All CFD-transitions for $T_3$: each diagram corresponds to an SCFT that is obtained by mass deformations from the initial $T_3$ theory. 
The flavor symmetries are read off from the marked (green) subgraphs. For $T_N$ with $N>3$ this is the non-abelian part of the flavor symmetry. For $T_3$ and its descendants the symmetry is further enhanced as follows from considerations of BPS states.  \label{fig:T3AllFlops}}
\end{figure}

For weakly coupled gauge theories, the descendant theories are obtained after decoupling hypermultiplets and RG-flow. 
For an SCFT we refer to its descendants as the UV completions of the descendants of the gauge theory description. 
In the CFD language, this corresponds to applying CFD-transitions to the $(-1)$ vertices. In the toric case, the CFD-transition rules are particularly simple. Suppose that the $(-1)$-vertex has two neighbors
\be
(m)-(-1)-(n)\,,
\ee
then after the middle $(-1)$-vertex is removed (or flopped out of the compact surfaces), this local configuration is replaced by
\be
(m-1)-(n-1)\,.
\ee
The other parts of the CFD remain unchanged. As one can see, this transition is exactly the same as the blow down of a $(-1)$-curve on a single toric surface. For this reason, the vertex can also be refered to as a ``$(-1)$-curve''. For the more general set of rules, one can see section 2.4 of~\cite{Apruzzi:2019kgb}, but we will not use them in this paper.

An example of a CFD-transition tree is shown for $T_3$ in figure \ref{fig:T3AllFlops}. Each diagram corresponds to a descendant SCFT obtained by mass deformation from $T_3$. These are precisely the descendants of the rank one $E_6$ Seiberg theory. We will discuss the IR-description in the next section. 

In the toric description the CFD-transitions correspond to flops. However, not each flop corresponds to a CFD-transition. CFD-transitions correspond to flops, which change at least one of the intersection numbers $(D_\alpha^{(k)})^2\cdot  (\sum S_i)$. In particular, if curves are flopped between surfaces the associated geometries do not correspond to different SCFTs -- these are merely different weakly coupled Coulomb branch phases, which have however the same UV fixed point. The precise correspondence between 
Coulomb branch equivalence classes and resolutions was discussed in \cite{Apruzzi:2019enx}, and more importantly for the present context of toric triangulations in  
\cite{Braun:2014kla, Braun:2015hkv}.

\begin{figure} 
\centering
\includegraphics[height=0.8\textheight]{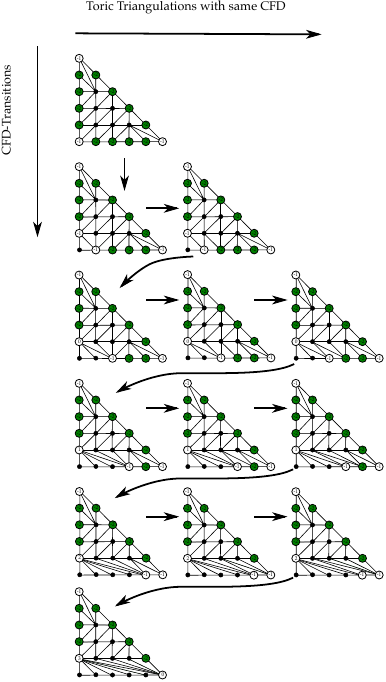}
\caption{CFD-transitions versus Flop Transitions for $T_5$: the vertical axis corresponds to CFD-transitions, which act on the CFD-vertices (out vertices in the toric diagram) only. Not every toric flop corresponds to a CFD-transition: each horizontal line corresponds to the same CFD, but a different toric triangulation. The arrows show how the flops connect various flop phases. \label{fig:CFDFlop}}
\end{figure}

Again let us exemplify this in the $T_5$ theory. Figure \ref{fig:CFDFlop} shows the first few flop transitions starting with the geometry (\ref{T5Ex}) \footnote{We should note that there are several more equivalent descriptions of this geometry which are related by flop transitions that do not change $(D_\alpha^{(k)})^2\cdot  (\sum S_i)$.}. 
We denote the curves by their location in the $(x-y)$-axis. 
By changing the triangulation of the bottom left face, the $(-1)$-curve located at the origin, which was contained in $S_4$, gets flopped out. After the flop it is in fact no longer contained in $\cup_i S_i$. 
This corresponds to a CFD-transition on the $(-1)$-curve. The resulting theory has $SU(5)\times SU(4)^2$ flavor symmetry, and there are two new $(-1)$-curves located at $(1,0)$ and $(0,1)$. To continue the flop transitions, we can in fact not immediately flop the curve at $(1,0)$. This requires first another toric flop, shown in the figure in line two, after which the flop that results in a new CFD is possible. Following this chain of toric flops results in the full CFD-transition tree (of which only a part is shown in the figure -- there are in total 78 descendants of the $T_5$ theory -- i.e. distinct SCFTs and thus CFDs that arise by mass deformation). However, each CFD will correspond to an equivalence class of toric triangulations, and in order to perform a CFD transition, one has to be in the right toric triangulation.

\subsection{Descendant Trees and Non-Lagrangian Theories}
\label{sec:NonLag}

\begin{figure} 
\centering
\subfloat{(a)}\includegraphics[height=4.5cm]{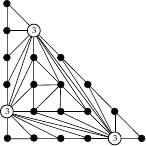} 
\subfloat{(b)}\includegraphics[height=6cm]{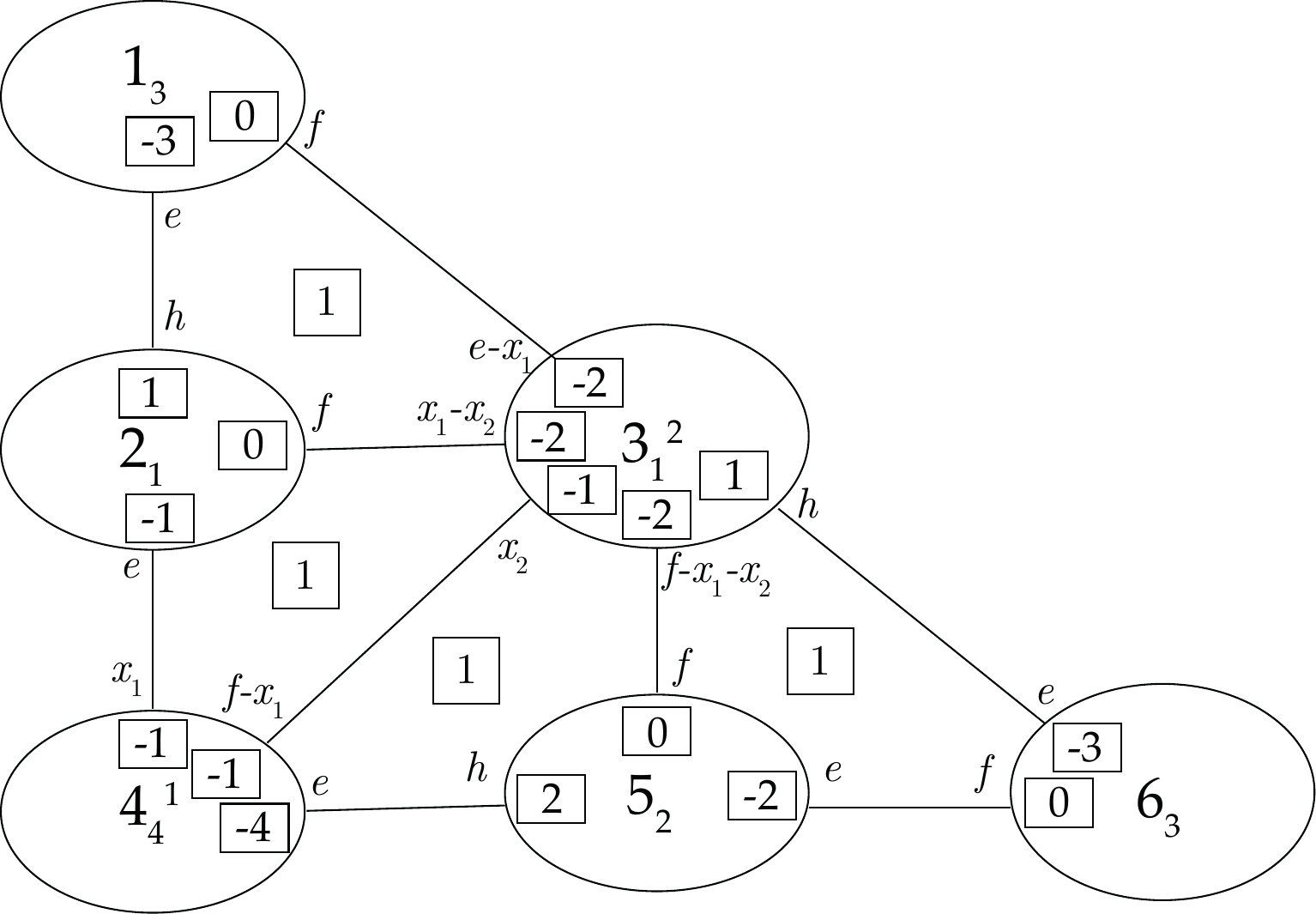}
\caption{A descendant SCFT of the $T_5$ theory, with a triangle-shaped CFD with three $(+3)$ curves. We refer to this theory as $B_5$. 
The left hand side shows the toric triangulation and the right hand side the surface configurations in terms of glued blowups of $\mathbb{F}_m$. \label{f:T5-bottom}}
\end{figure}

\begin{figure}
\centering
\subfloat{(a)}\includegraphics[height=3cm]{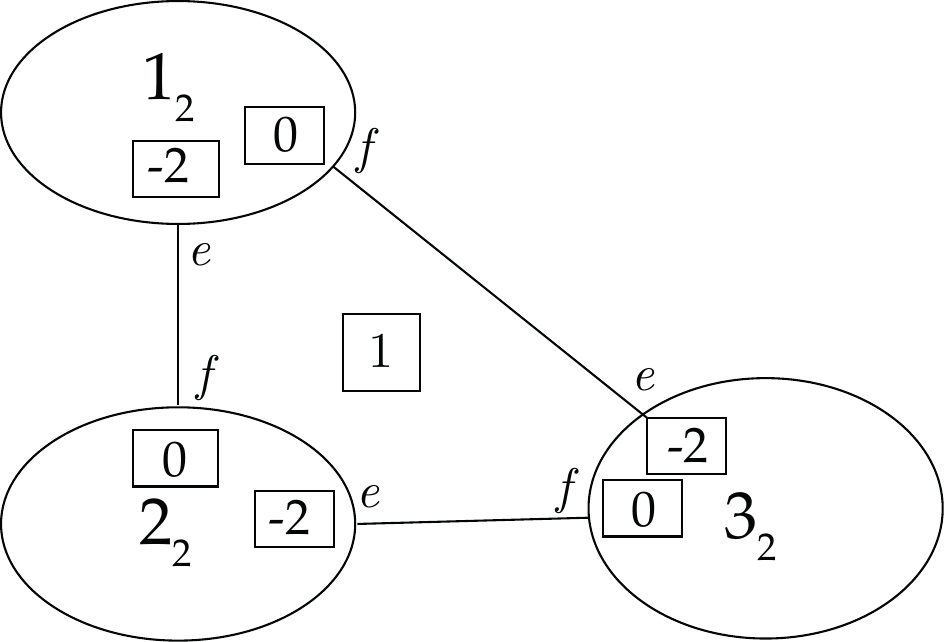}\subfloat{(b)}\includegraphics[height=7cm]{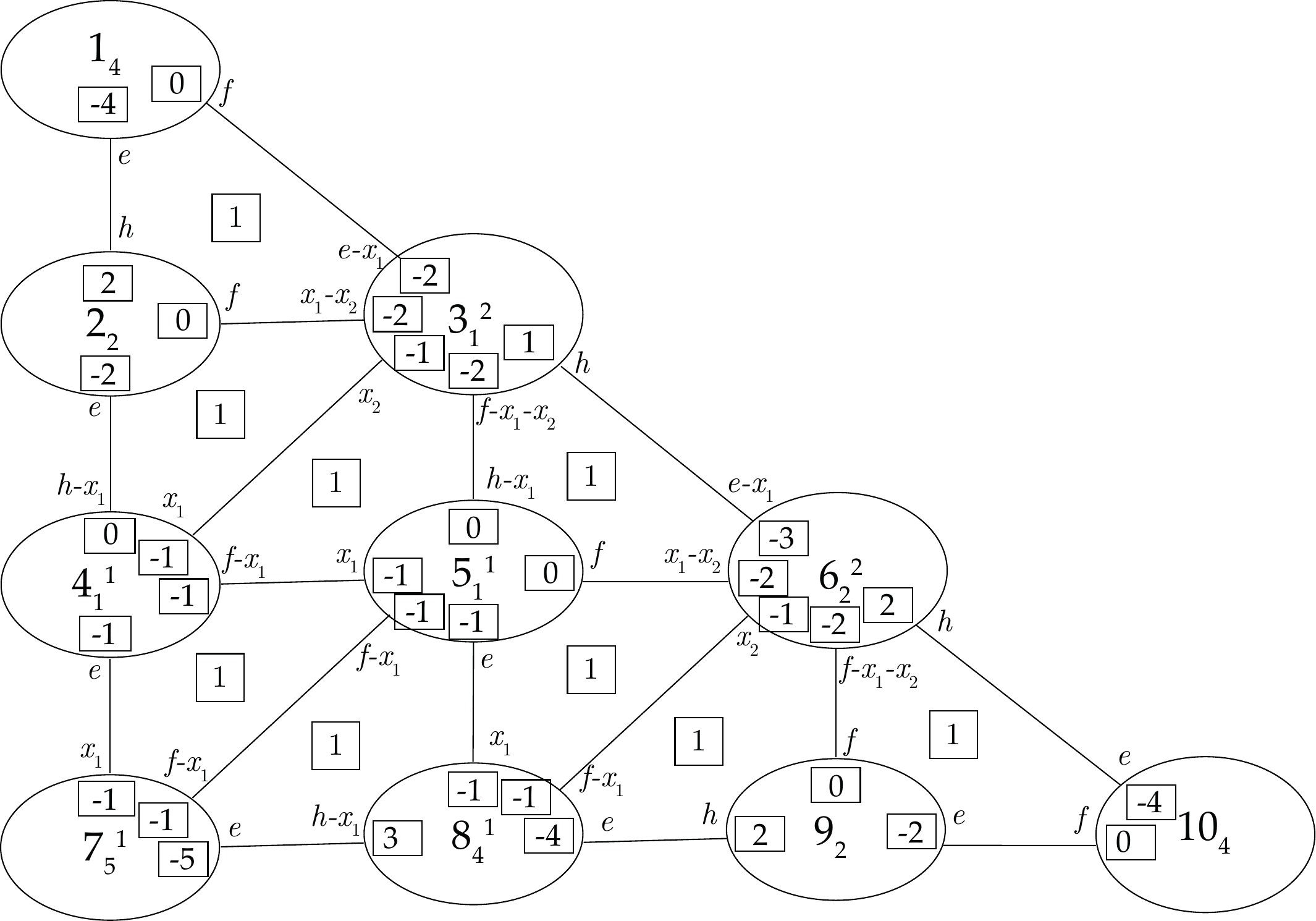}
\caption{The surface components of the theories: (a)  $B_4$, and  (b) $B_6$. \label{fig:B456}}
\end{figure}

\begin{figure}
\centering
\includegraphics[height=11cm]{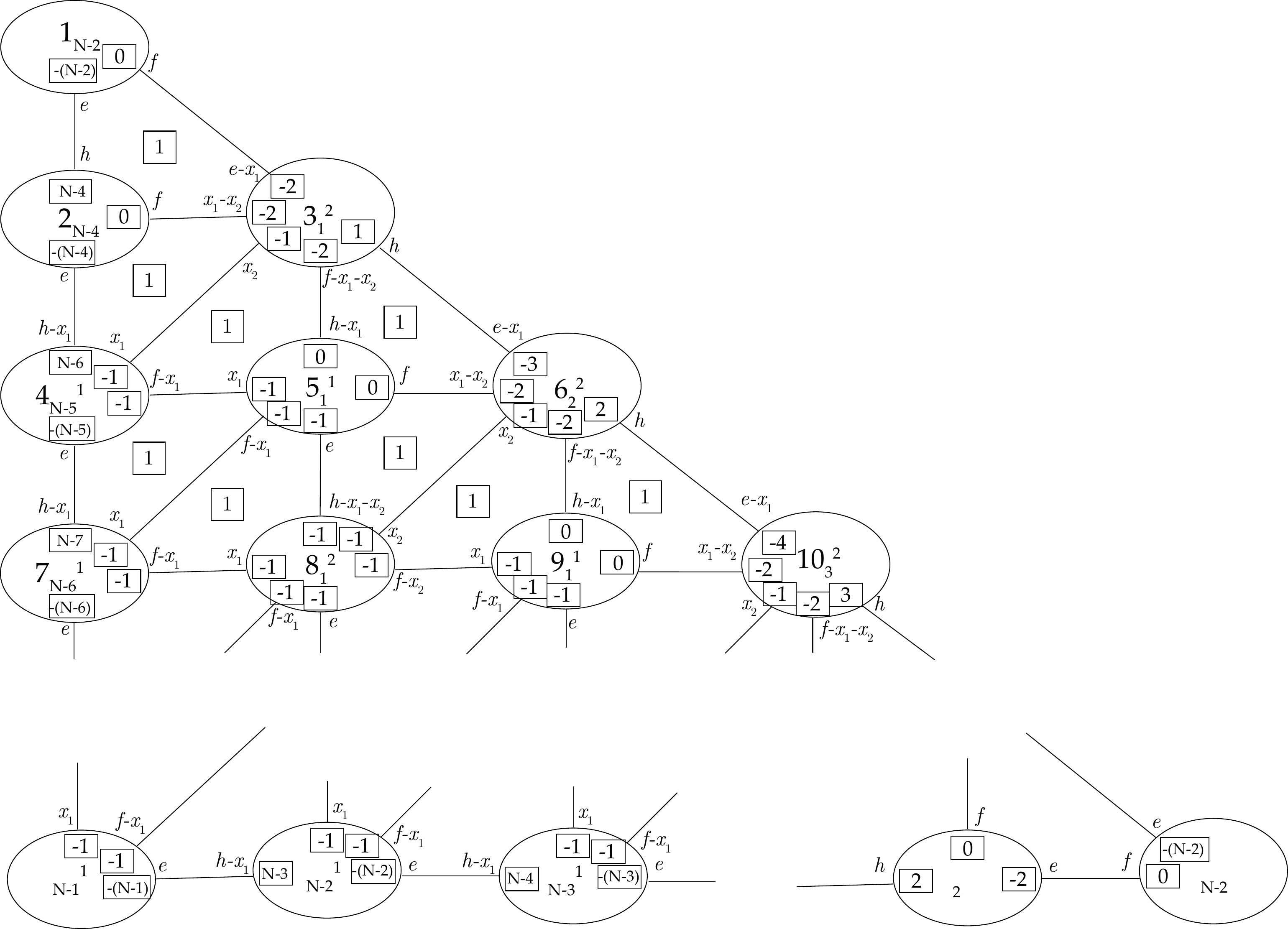}
\caption{The surface component for $B_N$ for general $N$. The last row is always retained for any $N$ (i.e. to obtain the diagram for $B_6$, remove fourth row).\label{fig:BN}}
\end{figure}

At the very bottom of the $T_3$-descendant tree in figure \ref{fig:T3AllFlops} is the theory with CFD given by three $(+1)$ vertices connected in a triangle. This corresponds to the rank one theory, which does not admit an $SU(2)$ weakly coupled description, and whose geometric realization is in terms of a $\mathbb{P}^2$. We will find that all $T_N$ have such a {descendant} with triangular CFD  and all three vertices  labeled by $(N-2)$. We will refer to these as the $B_N$ theory, and they are non-Lagrangian theories with CFD
\be\ba
B_N: & \cr
& \qquad  \includegraphics[width=4.5cm]{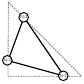} \,.
\ea\ee
The rays of the vertices of the CFD in the toric diagram are
\be
v_1= (N - 1, 0, 1) \,,\qquad 
v_2= (1, N - 1, 1) \,,\qquad 
v_3= (0, 1, 1) \,.
\ee
Such an SCFT always has trivial flavor symmetry $G_F=\emptyset$. We plot the example of the  $N=5$ case in figure~\ref{f:T5-bottom}. From the toric picture, there is no consistent ruling structure of the surfaces. Hence the SCFT is absent of weakly coupled gauge theory description. We also plot the surface intersection diagram for reference in figure \ref{fig:B456}.

For the case of $N=4$, the three compact surfaces that comprise the theory $B_4$ are all Hirzebruch surface $\mb{F}_2$. The section $(-2)$-curve of one surface is always the fiber curve of another surface, as shown in figure \ref{fig:B456} (a). The case of $N=5$ is in figure \ref{f:T5-bottom} and $N=6$ in figure \ref{fig:B456} (b). The general surface components for $B_N$ are shown in figure \ref{fig:BN}.
Note that we did not write out the label $i$ for each surface in the bottom row, but we still keep the superscripts and the subscripts. For example, the bottom right surface should be a Hirzebruch surface $\mb{F}_{N-2}$.
For the case of $N=3$, the single compact surface is just $\mb{P}^2$ and the 5d SCFT is a rank one theory with no gauge theory description. 

\begin{figure}
\centering
\includegraphics*[width=\textwidth]{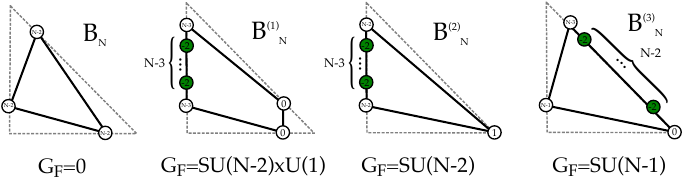}
\caption{CFDs (and toric diagrams) for all RG-flow tree endpoints for $T_N$. $G_F$ is the flavor symmetry and each of these models has no further descendant, and also no IR-gauge theory description. \label{fig:Bottoms}}
\end{figure}

There are several endpoints of the RG-flow tree for $T_N$, which all do not admit an IR gauge theory description. We list them in figure \ref{fig:Bottoms} including their strongly-coupled flavor symmetry $G_F$. The vertices of the CFD are drawn such that the toric diagram can be easily read off from the figure. 

\subsection{5d Parents of $T_N$ and their 6d Origin}

There is a close connection between 5d and 6d SCFTs. In particular, there is ample evidence that all 5d SCFTs arise as dimensional reductions with holonomies in the flavor symmetry/mass deformations, of 6d SCFTs -- at least to this date there are no counter examples to this claim. 
It is thus natural to ask for the 6d avatar of the $T_N$ theories. 

The 6d $(1,0)$ SCFT uplift for 5d $T_N$ theories are given in~\cite{Zafrir:2015rga}, with the following tensor branch:
\be
\ba
&N=3l&&:\quad \overset{\phantom{\mathfrak{su}(3)}}{1}-\overset{\mathfrak{su}(9)}{2}-\overset{\mathfrak{su}(18)}{2}-\dots-\overset{\mathfrak{su}(9l-9)}{2}-[9l]\cr
&N=3l+1&&:\quad \overset{\mathfrak{su}(3)}{1}-\overset{\mathfrak{su}(12)}{2}-\overset{\mathfrak{su}(21)}{2}-\dots-\overset{\mathfrak{su}(9l-6)}{2}-[9l+3]\cr
&N=3l+2&&:\quad \overset{\mathfrak{su}(6)}{1}-\overset{\mathfrak{su}(15)}{2}-\overset{\mathfrak{su}(24)}{2}-\dots-\overset{\mathfrak{su}(9l-3)}{2}-[9l+6]\,.\label{6d-TN-marginal}
\ea
\ee
The corresponding 5d marginal theory, i.e. the 5d theory, which UV completes in the above 6d theories, has the following 5d quiver description
\be\label{TN-marginal-quiver}
[N+2]-SU(N-1)_0-SU(N-2)-\dots-SU(2)-[3]\,,
\ee
which has three more fundamental flavors than the 5d $T_N$ quiver.

\begin{figure}
\centering
\subfloat{(a)}\includegraphics[width=8cm]{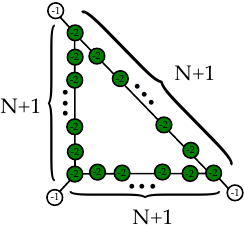}\qquad 
\subfloat{(b)}\includegraphics[width=6cm]{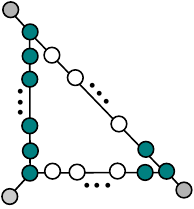}
\caption{(a) CFD for the marginal theory, which after three mass deformations results in the $T_N$ theory. It has flavor symmetry $\widehat{SU(3N)}$. 
(b) The weakly coupled description can be determined from the embedding of the IR-CFDs or BG-CFDs, which is shown here: the CFD  allows embedding of the BG-CFDs for $U(N+2)\times SO(6)$, {where the $U(N+2)$ arises from the left hand side of the diagram, and the $SO(6)$ from the right hand corner.  }
\label{fig:6dTN}}
\end{figure}

\begin{figure}
\centering
\includegraphics[width=10cm]{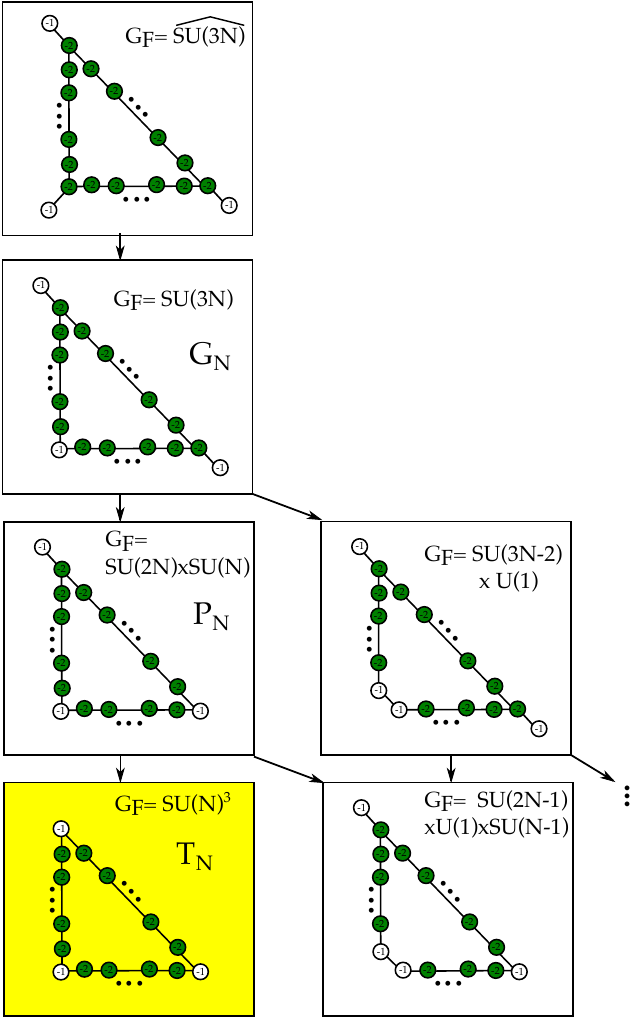}
\caption{Geneaological tree of $T_N$: the theory at the top is the KK-theory, obtained from 6d by circle reduction. The first mass deformation (up to permutation) is the 5d SCFT $G_N$, with flavor symmetry $G_F= SU(3N)$. It has two descendants, with one of them the theory $P_N$ with $G_F= SU(2N)\times SU(N)$. Further mass deformation yields $T_N$. 
\label{fig:ParentsTN}}
\end{figure}

The CFD of the marginal theory (\ref{6d-TN-marginal}) is shown in figure \ref{fig:6dTN} and has exactly affine flavor symmetry $\widehat{SU(3N)}$. As we apply the transition rules on the CFDs, the first descendant $G_N$ has $SU(3N)$ flavor symmetry, and after two more flops we obtain the $T_N$ CFD. These intermediate parent theories are shown in figure \ref{fig:ParentsTN}. {Note that the parent theory of $T_N$ is the $P_N$ theory with flavor symmetry $G_F=SU(2N)\times SU(N)$.}

At this point we should comment on some gauge theoretic aspects that we will expand upon later. From the CFDs we can read off possible IR gauge theory descriptions by embedding IR-analogs of the CFDs, so-called BG-CFDs (Box Graph CFDs) -- we will explain these in detail in section \ref{sec:BGCFD}. These encode the classical flavor symmetries in terms of a collection of $(-2)$-vertices and the matter in terms of $(-1)$-vertices attached to the corresponding Dynkin weights. For $SU(N_c) + N_F\bm{F}$ this is a linear chain of $N_F-1$ $(-2)$-vertices with one $(-1)$-vertex attached at each end. If such a BG-CFD can be embedded into the CFD, by retaining the labels on the vertices, then the corresponding IR description is a potential weakly-coupled description of the SCFT that is associated to the CFD. 

In figure \ref{fig:6dTN}(b) we have shown the embedding in terms of the BG-CFD (grey nodes corresponding to the $(-1)$-vertices and blue nodes to $(-2)$). 
From the embedding of BG-CFDs, it has a weakly-coupled description with flavor symmetry $U(N+2)\times SO(6)$. This is precisely the BG-CFD associated to the IR description in terms of the quiver for the marginal theory in (\ref{TN-marginal-quiver}).

Note that from the symmetry of the CFD, we can confirm that the Chern-Simons level of the $SU(N-1)$ gauge group vanishes. The reason is that in the $U(N)$ BG-CFD, the two gray vertices exactly correspond to fundamental and anti-fundamental matter fields. After the removal of either vertex, the fundamental (or anti-fundamental) matter field is decoupled, and the Chern-Simons level of $SU(N-1)$ is shifted by $\frac{1}{2}$ (or $-\frac{1}{2}$). However, since the positions of the three $(-1)$-vertices are symmetric, there is only a single descendant theory, which is the $G_N$ theory in figure~\ref{fig:ParentsTN}, from the marginal CFD. Thus the Chern-Simons level of $SU(N-1)$ has to be $k=0$, as $k\rightarrow -k$ leads to equivalent theories. This phenomenon was observed in the $(D_k,D_k)$ theory as well~\cite{Apruzzi:2019vpe,Apruzzi:2019opn}.
Furthermore, there exists the following IR duality between two different quiver theories:
\be \label{DualityGN}
\ba
&[N+1]-SU(N-1)_{\frac{1}{2}}-\dots-SU(2)-[3]\cr 
\longleftrightarrow\quad  & [N+2]-SU(N-1)_0-\dots-SU(2)-[2]\,,
\ea
\ee
which both have $G_N$ theory as their UV completion.
Note that the theory on the right hand side is generated by decoupling a fundamental flavor under the $SU(2)$ gauge group.

Similarly, for the $P_N$ theory in figure~\ref{fig:ParentsTN}, there are the following dual IR descriptions 
\be \label{DualityPN}
\ba
&[N]-SU(N-1)_{0}-\dots-SU(2)-[3]\cr 
\longleftrightarrow\quad  & [N+1]-SU(N-1)_{\frac{1}{2}}-\dots-SU(2)-[2]\,.
\ea
\ee

\begin{figure}
\centering
\subfloat{(a)}\includegraphics[width=6.5cm]{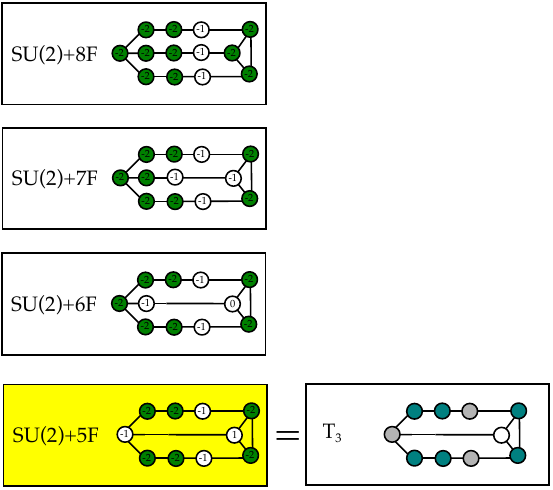}\quad 
\subfloat{(b)}\includegraphics[width=7.5cm]{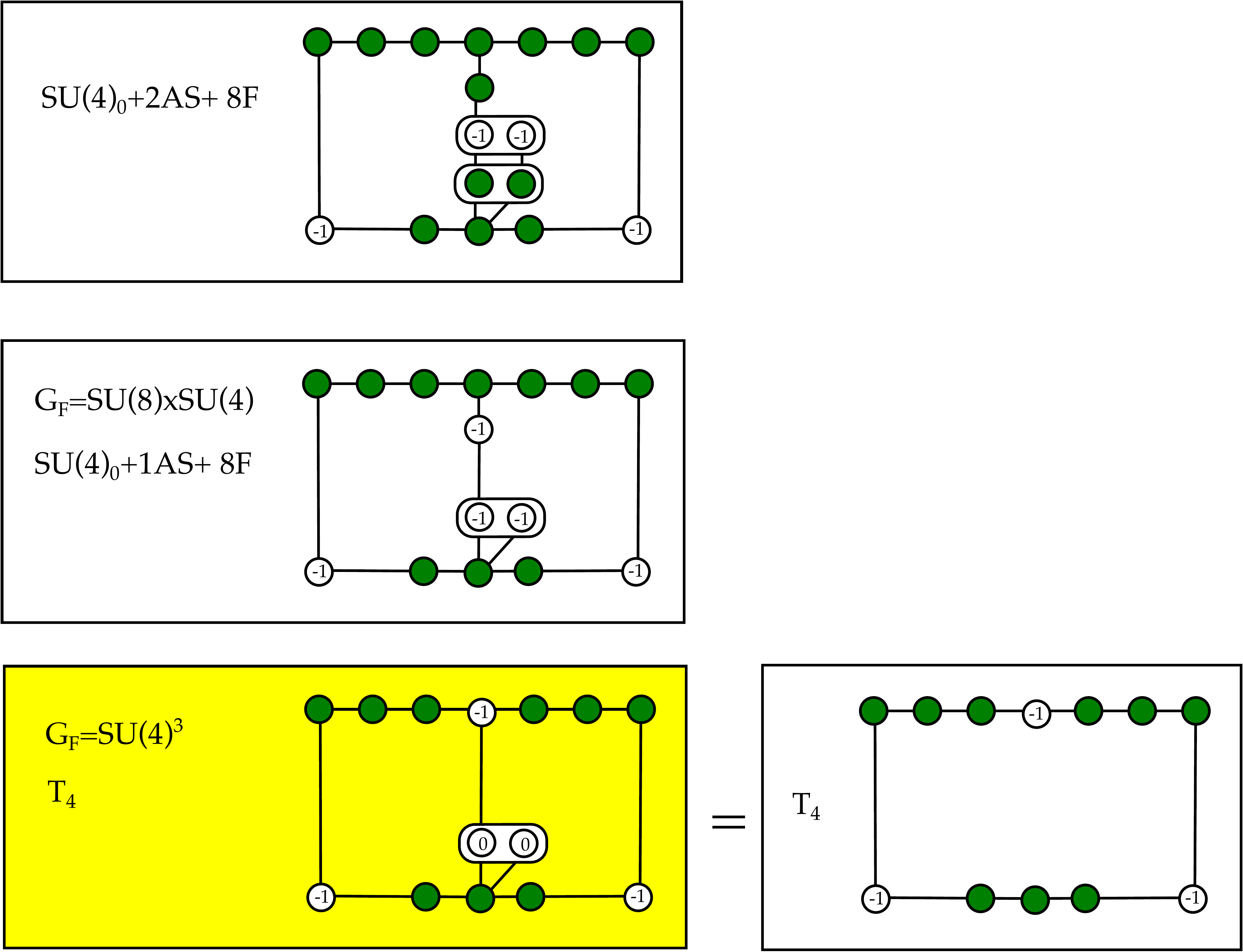}
\caption{(a) $T_3$ as a descendant of the $(E_6, SU(3))$ Rank one conformal matter theory. 
(b) $T_4$ is a rank three theory which is realized as a descendant of the $(E_7, SO(7))$ conformal matter theory. \label{fig:T3fromCM}}
\end{figure}

For the lower rank theories, there exist other marginal descriptions. The rank one $T_3$ theory is a descendant of the rank one E-string theory. Namely, the quiver $1-[SU(9)]$ in (\ref{6d-TN-marginal}) and the CFD in figure~\ref{fig:6dTN} are an equivalent description of the rank one E-string theory with explicit flavor symmetry $SU(9)\subset E_8$.
On the other hand, the $T_3$ theory is also realized as a descendant of the $(E_6,SU(3))$ conformal matter theory, which is another realization of the rank one E-string \cite{Apruzzi:2019opn,Apruzzi:2019kgb}. 
The mass deformations are depicted in figure \ref{fig:T3fromCM}(a). 

For the rank three $T_4$ theory, besides the 6d uplift $\overset{\mathfrak{su}(3)}{1}-[SU(12)]$ in (\ref{6d-TN-marginal}), it is also a descendant of $(E_7, SO(7))$ conformal matter. The marginal CFD of $(E_7, SO(7))$ conformal matter was determined in \cite{Apruzzi:2019enx}. The CFD-transitions to $T_4$ are shown in figure \ref{fig:T3fromCM}(b).

\subsection{BPS States}
\label{sec:BPS}

The BPS states of the 5d SCFTs have contributions from two sources: a subset is determined by the CFD, i.e. curves that fall into representations of the flavor symmetry, and curves, which are intersections between compact surface components. For $T_N$ we will now determine both types of contributions.

\subsubsection{BPS states from the CFD}

Here we briefly review the procedure of reading off BPS states from the geometry and CFD~\cite{Apruzzi:2019vpe,Apruzzi:2019opn}.
In the Calabi-Yau threefold geometry, M2-branes wrapping complex curves give rise to particle BPS states on the Coulomb branch. The spin of such BPS states is given by the moduli space of the curve $C$. In this paper, we only consider the cases where $C$ is a genus 0  complete intersection curve $C=D_1\cdot S_1$, and where $C$ is a linear system on both $D_1$ and $S_1$. Then, if the normal bundle of $C$ is
\be
N_{C|X}=\mc{O}(m)\oplus\mc{O}(-2-m)\,,
\ee
then the moduli space of $C$ is
\be
\mc{M}_C=\mb{P}^{\mathrm{max}(m+1,0)}\times\mb{P}^{\mathrm{max}(-1-m,0)}\,.
\ee
We assume $m\geq -1$, so we simply have $\mc{M}_C=\mb{P}^{m+1}$. The M2 brane wrapping $C$ gives rise to a 5d $\mc{N}=1$ supermultiplet in the representation $(\frac{1}{2},\frac{m+1}{2}) \oplus 2(0,\frac{m+1}{2})$ 
of the 5d massive little group~\cite{Witten:1996qb,Kachru:2018nck,Tian:2018icz}.

Hence the spin-0 hypermultiplet is given by a $\mb{P}^1$ curve with normal bundle $N=\mc{O}(-1)\oplus\mc{O}(-1)$, and a spin-1 vector multiplet is given by a $\mb{P}^1$ curve with normal bundle $N=\mc{O}(0)\oplus\mc{O}(-2)$.

In the CFD, each vertex is associated to a curve
\be
C_\alpha=\sum_{i=1}^r \xi_{i,\alpha}D_\alpha\cdot S_i\,.
\ee
Then a linear combination of the vertices can be defined to be
\be
\sum_\alpha a_\alpha C_\alpha=\sum_\alpha \sum_{i=1}^r a_\alpha\xi_{i,\alpha}D_\alpha\cdot S_i\,.\label{C-linearcomb}
\ee
This curve is obviously a complete intersection curve if all $\xi_{i,\alpha}=1$. If some of the $\xi_{i,\alpha}$ factors are larger than one, then we need to check more carefully. For the $T_N$ theory, we can always choose the triangulation analogous to the right one in (\ref{T5Ex}), such that all $\xi_{i,\alpha}=1$. Hence the linear combination of vertices in the CFD is always well-defined.

From the CFD of $T_N$ in figure~\ref{fig:CFDTN}, we can read off linear combinations of vertices that correspond to the highest weight states of $SU(N)^3$ representations. For the spin-0 BPS states, they are exactly given by the three white vertices $D_0^{(i)}$, $i=1,2,3$. They give rise to the following representations of $SU(N)^3$:
\be
R_{0,\text{CFD}}=(\mbf{N},\mbf{1},\overline{\mbf{N}})\,,(\overline{\mbf{N}},\mbf{N},\mbf{1})\,,(\mbf{1},\overline{\mbf{N}},\mbf{N})\,.\label{R0CFD}
\ee

For the spin-1 BPS states, they are given by the following linear combinations of $D_\alpha^{(i)}$:
\be
\ba
2D_0^{(1)}+D_1^{(1)}+D_{N-1}^{(3)}:\,&\left(\mbf{\frac{N(N-1)}{2}},\mbf{1},\overline{\mbf{\frac{N(N-1)}{2}}}\right)\cr
2D_0^{(2)}+D_1^{(2)}+D_{N-1}^{(1)}:\,&\left(\overline{\mbf{\frac{N(N-1)}{2}}},\mbf{\frac{N(N-1)}{2}},\mbf{1}\right)\cr
2D_0^{(3)}+D_1^{(3)}+D_{N-1}^{(2)}:\,&\left(\mbf{1},\overline{\mbf{\frac{N(N-1)}{2}}},\mbf{\frac{N(N-1)}{2}}\right)\cr
\sum_{\alpha=0}^{N-1}D_\alpha^{(1)}:\,&(\mbf{1},\mbf{N},\overline{\mbf{N}})\cr
\sum_{\alpha=0}^{N-1}D_\alpha^{(2)}:\,&(\overline{\mbf{N}},\mbf{1},\mbf{N})\cr
\sum_{\alpha=0}^{N-1}D_\alpha^{(3)}:\,&(\mbf{N},\overline{\mbf{N}},\mbf{1})\cr
\label{R1CFD}
\ea
\ee

Of course, the green vertices in the CFD correspond to the generators of the flavor symmetry $SU(N)^3$, and they give rise to spin-1 BPS states in the adjoint representation of $SU(N)^3$.

In particular, if we have $N=3$, then the states in (\ref{R0CFD}) are
\be
R_{0,CFD}=(\mbf{3},\mbf{1},\overline{\mbf{3}})\,,(\overline{\mbf{3}},\mbf{3},\mbf{1})\,,(\mbf{1},\overline{\mbf{3}},\mbf{3})\,.
\ee
They can be exactly combined into a single irreducible representation $\mbf{27}$ of $E_6$.
Similarly, the states in (\ref{R1CFD}) can be combined into $\overline{\mbf{27}}$ of $E_6$. This is consistent with the flavor symmetry enhancement $G_F=E_6\supset SU(3)^3$.

\subsubsection{Additional BPS States from Geometry}

From the CFD vertices, we have enumerated the BPS states from a curve in the form of (\ref{C-linearcomb}). Nonetheless, for $N>3$, there are also other curves with normal bundle $\mc{O}(-1)\oplus\mc{O}(-1)$ in the geometry. They are not included in the CFD data, but they also contribute to the BPS spectrum of the SCFT~\cite{Hayashi:2019jvx}. Similarly, there are additional spin-1 BPS states from curves with normal bundle $\mc{O}(0)\oplus\mc{O}(-2)$ as well. In principle, one needs to generate all the possible resolutions, as some representations of BPS states may not appear in a particular resolution, see  \cite{Esole:2018mqb,Esole:2019asj} as an example. Here we will only show a representative of each non-trivial representation under $G_F=SU(N)^3$, in one particular resolution. 

For the $T_5$ theory, we show the additional spin-0 BPS states from $\mc{O}(-1)\oplus\mc{O}(-1)$ curves labeled in red:
\be
\includegraphics[width=0.9\textwidth]{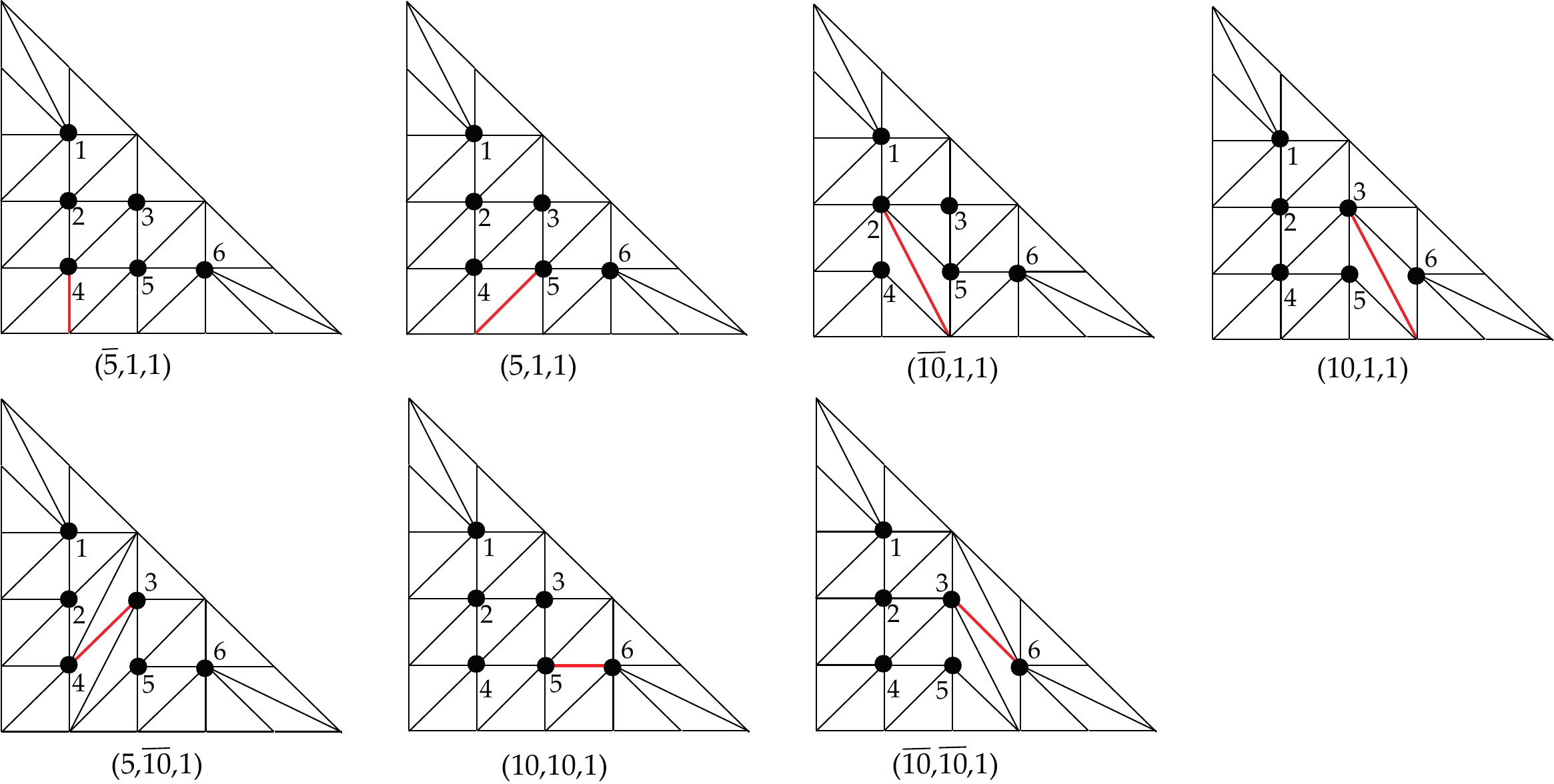}
\ee
Because of the cyclic symmetry among the three $SU(N)$s on the three edges, for any representation $(R_1,R_2,R_3)$ in the above figure, the representations $(R_2,R_3,R_1)$ and $(R_3,R_1,R_2)$ exist as well. 

For the additional spin-1 BPS states from $\mc{O}(0)\oplus\mc{O}(-2)$ curves, they are
\be
\includegraphics[width=0.9\textwidth]{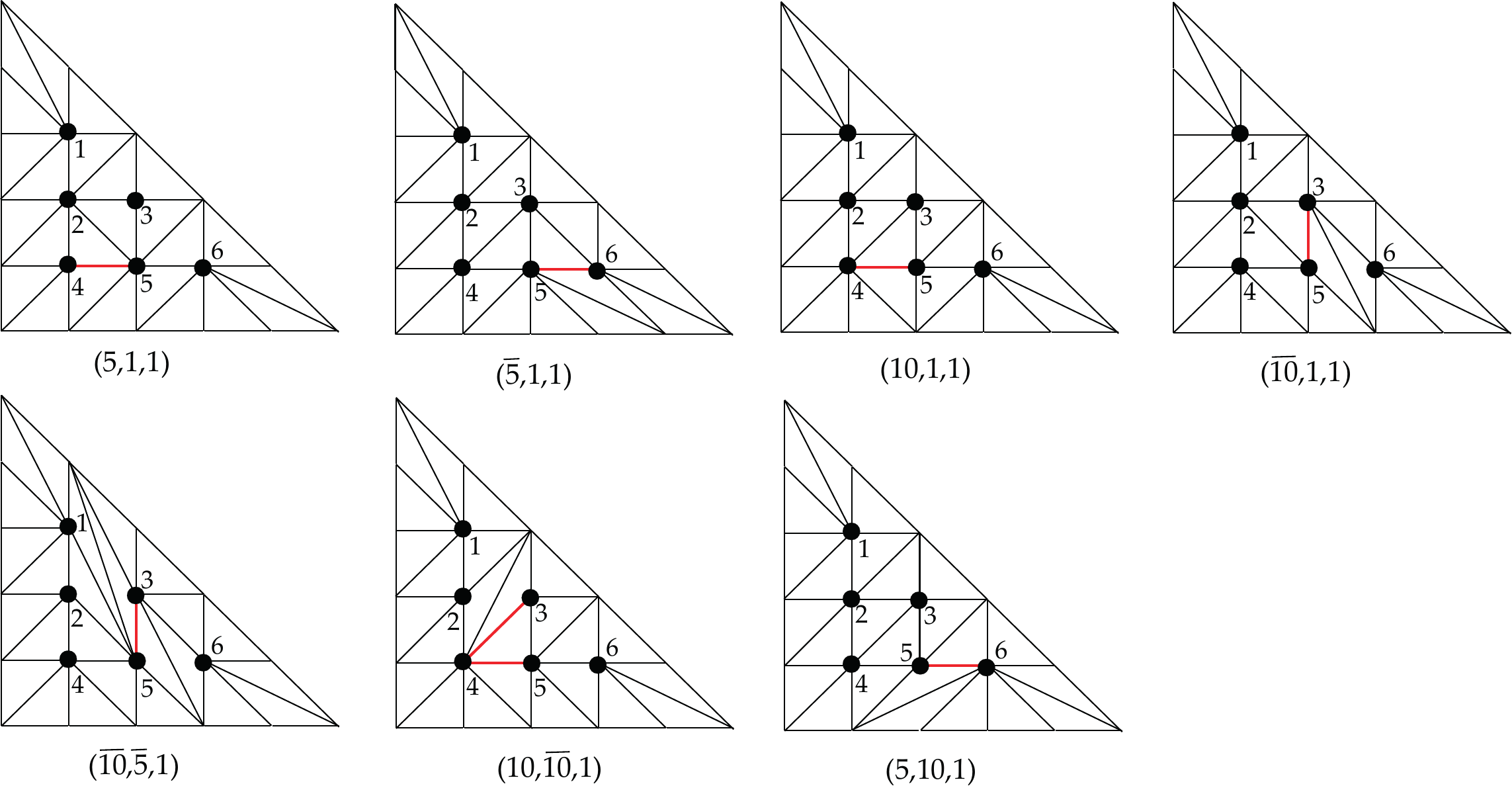}
\ee
Similarly, the representations generated by cyclic permutation also exist.

For general $N$, the additional spin-0 BPS states charged under a single $SU(N)$ are:
\be
R_{0,\text{Add}}^{(1)}=
\left\{ 
\ba(
\Lambda^k,\mbf{1},\mbf{1})\quad&  (k=1,\dots,N-1)\cr 
(\mbf{1},\Lambda^k,\mbf{1})\quad & (k=1,\dots,N-1)  \cr 
(\mbf{1},\mbf{1},\Lambda^k)\quad &(k=1,\dots,N-1) \cr
\ea
\right.
\ee
Here $\Lambda^k$ is the rank-$k$ anti-symmetric representation of $SU(N)$. We do not list all the singlets under $SU(N)^3$.

For the spin-0 states charged under two $SU(N)$s, we have the following general criterion: for a representation $(\Lambda^k,\Lambda^l,\mbf{1})$, we draw a line $L_{k,l}$ in the toric triangle with two end points $(k,0)$ and $(N-l,l)$ (the origin is the left-bottom corner). Then such a representation exists if and only if $L_{k,l}$ does not pass through any integral interior points, and the interior integral points do not entirely locate on one side of $L_{k,l}$. This criterion does not apply to the BPS states from the CFD in (\ref{R0CFD}). 

After this criterion is applied, the other BPS states are generated by the cyclic permutation of $(\Lambda^k,\Lambda^l,\mbf{1})$.
The additional spin-1 BPS states charged under a single $SU(N)$ are 
\be
R_{1,\text{Add}}^{(1)}=
\left\{ 
\ba(
\Lambda^k,\mbf{1},\mbf{1})\quad&  (k=1,\dots,N-1)\cr 
(\mbf{1},\Lambda^k,\mbf{1})\quad & (k=1,\dots,N-1)  \cr 
(\mbf{1},\mbf{1},\Lambda^k)\quad &(k=1,\dots,N-1) \cr
\ea
\right.
\ee

For the spin-1 states charged under two $SU(N)$s, we have the following criterion: for a representation $(\Lambda^k,\Lambda^l,\mbf{1})$, we draw the same line $L_{k,l}$ passing through $(k,0)$ and $(N-l,l)$. Then such a representation exists if and only if $L_{k,l}$ pass through at least one integral interior point. Similarly, the other BPS states are generated by the cyclic permutation of $(\Lambda^k,\Lambda^l,\mbf{1})$. Note that this criterion covers the BPS states from the CFD in (\ref{R1CFD}) as well.

Combining these with the BPS states from the CFD, the total spin-0 BPS states from $\mc{O}(-1)\oplus\mc{O}(-1)$ curves are
\be
R_0=
\left\{
\ba
&(\mbf{N},\mbf{1},\overline{\mbf{N}})\,,(\overline{\mbf{N}},\mbf{N},\mbf{1})\,,(\mbf{1},\overline{\mbf{N}},\mbf{N})\cr 
&(\Lambda^k,\mbf{1},\mbf{1})\quad  (k=1,\dots,N-1)\cr 
&(\mbf{1},\Lambda^k,\mbf{1})\quad  (k=1,\dots,N-1)  \cr 
&(\mbf{1},\mbf{1},\Lambda^k)\quad (k=1,\dots,N-1)\cr
&(\Lambda^k,\Lambda^l,\mbf{1})\cr 
& (\mbf{1},\Lambda^k,\Lambda^l)\cr
&(\Lambda^l,\mbf{1},\Lambda^k)\,.
\ea\right.
\ee
The states charged under two $SU(N)$s are determined by the spin-0 $L_{k,l}$ criterion.

The total spin-1 BPS states from $\mc{O}(0)\oplus\mc{O}(-2)$ curves are
\be
R_1=
\left\{\ba
& (\Lambda^k,\mbf{1},\mbf{1})\quad  (k=1,\dots,N-1)\cr 
&(\mbf{1},\Lambda^k,\mbf{1})\quad  (k=1,\dots,N-1)  \cr 
&(\mbf{1},\mbf{1},\Lambda^k)\quad (k=1,\dots,N-1)\cr
&(\Lambda^k,\Lambda^l,\mbf{1})\cr 
& (\mbf{1},\Lambda^k,\Lambda^l)\cr
&(\Lambda^l,\mbf{1},\Lambda^k)\cr
&(\mbf{N^2-1},\mbf{1},\mbf{1})\,,(\mbf{1},\mbf{N^2-1},\mbf{1})\,,(\mbf{1},\mbf{1},\mbf{N^2-1})\cr 
\ea\right.
\ee
The states charged under two $SU(N)$s are determined by the spin-1 $L_{k,l}$ criterion, and we have included the adjoint of $SU(N)^3$ as well.


\section{Coulomb Branch, IR-descriptions, and RG-flows}

\subsection{Geometry and Rulings}
\label{sec:IR-ruling}

In this section, we present the geometric construction of the IR non-Abelian gauge theory description of $T_N$ theory, following the methodology of~\cite{Intriligator:1997pq}. For each compact surface $S_i$, we find a ruling by a $\mb{P}^1 = f_i$ over curves (the sections). The ruling curve is an irreducible rational curve with self-intersection number $f_i\cdot_{S_i}f_i=0$ if $S_i$ is $\mb{F}_n$. If $S_i$ is a blow up of $\mb{F}_n$, then $f_i$ can be written as a linear combination of rational curves with negative self-intersection number on $S_i$ . The component curves in this linear combination are then also labeled as ``ruling curves'' of $S_i$. 

The fully resolved geometry $X_\Sigma$ has an IR non-Abelian gauge theory description if and only if: for any curve $S_i\cdot S_j$, it is either assigned as a ruling curve on both $S_i$ and $S_j$, or assigned as a section curve on both $S_i$ and $S_j$. If this condition can be satisfied, then after the volume of all the ruling curves Vol$(f_i)\rightarrow 0$, the M2-brane wrapping modes over $f_i$ become massless. They exactly correspond to the W-bosons of the enhanced non-Abelian gauge group $G_{\rm gauge}$. From the triple intersection numbers $S_i^2\cdot S_j$, one can read off the gauge group $G_{\rm gauge}$ as follows:
\begin{enumerate}
\item {Take each $S_i$ as a Cartan node in the Dynkin diagram of $G_{\rm gauge}$, which has the Cartan matrix $\mc{C}_{ij}$.}
\item {For each intersection curve $S_i\cdot S_j$ assigned as a section curve, draw a connection in the Dynkin diagram. More precisely, the off-diagonal entries of $\mc{C}$ are exactly given by \be
\mc{C}_{ij}=f_i\cdot_{S_i} (S_i\cdot S_j)\ , \quad\ \mc{C}_{ji}=f_j\cdot_{S_j} (S_i\cdot S_j)\,.
\ee
}
\end{enumerate}

If the Dynkin diagram is reducible, then it describes a quiver gauge theory description. The connection of gauge nodes in the quiver is determined by the bifundamental hypermultiplets between each gauge group, which are given by the M2 brane wrapping modes over the ruling curve components with normal bundle $\mc{O}(-1)\oplus\mc{O}(-1)$. Other charged matter hypermultiplets arise in a similar way as well.

\begin{figure}
\begin{center}

\includegraphics[height =5cm]{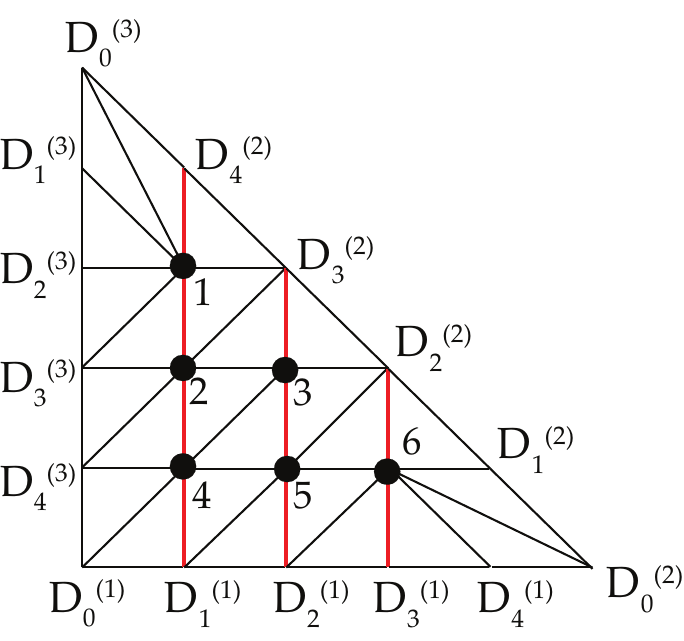}
\caption{The ruling structure of compact surfaces in the example of $T_5$ theory. Each red vertical line denotes the section curve of the ruling. The quiver gauge theory description is $[5]-SU(4)_0-SU(3)_0-SU(2)-[2]$.}
 \label{f:TN-ruling}
\end{center}
\end{figure}

In the toric picture, the section curve can be chosen as straight lines, while the other curves correspond to the ruling curves. For the $T_5$ example, they can be chosen as in figure~\ref{f:TN-ruling}. 
Then one can check that the gauge group $G_{\rm gauge} =SU(4)\times SU(3)\times SU(2)$. The Cartan nodes of $SU(4)$ correspond to the compact surfaces $S_1$, $S_2$, $S_4$. The Cartan nodes of $SU(3)$ correspond to $S_3$, $S_5$ and the Cartan node of $SU(2)$ corresponds to $S_6$. From the information of $\mc{O}(-1)\oplus\mc{O}(-1)$ curves, we see that the quiver gauge theory description is
\be
[5]-SU(4)_0-SU(3)_0-SU(2)-[2]\,.
\ee
Here the two fundamentals attached to the $SU(2)$ gauge node are given by
\be
D_0^{(2)}\cdot S_6\ ,\ (D_0^{(2)}+D_1^{(2)})\cdot S_6\,.
\ee
The five fundamentals attached to the $SU(4)$ gauge node are given by
\be
\ba
&D_0^{(3)}\cdot S_1\ ,\ (D_0^{(3)}+D_1^{(3)})\cdot S_1\ ,\ (D_0^{(3)}+D_1^{(3)}+D_2^{(3)})\cdot S_1\cr
&(D_0^{(3)}+D_1^{(3)}+D_2^{(3)}+D_3^{(3)})\cdot (S_1+S_2)\ ,\ (D_0^{(3)}+D_1^{(3)}+D_2^{(3)}+D_3^{(3)}+D_4^{(3)})\cdot (S_1+S_2+S_4)\,.
\ea
\ee
This analysis can be generalized to arbitary $T_N$ theories, and the quiver gauge theory description is always
\be \label{LinQuiver}
T_N:\qquad [N]-SU(N-1)_0- SU(N-2)_0- \dots-SU(2)-[2]\,.
\ee
The total flavor rank is
\be
\ba
r_F&=N+2+(N-3)+(N-2)\cr
&=3N-3\,,
\ea
\ee
since each bifundamental matter and each gauge group factor has contribution one. 

\begin{figure}
\begin{center}

\includegraphics[height =5cm]{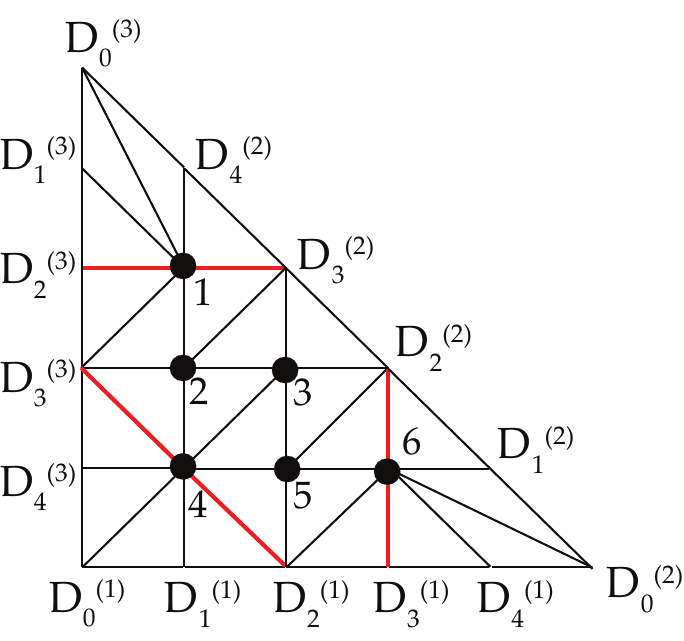}
\caption{The ruling structure of compact surfaces in the example of $T_5$ theory that realizes the $SO(4)^3$ IR flavor symmetry. Each red line denotes the section curve of the ruling. The quiver description is not weakly coupled.}
 \label{f:TN-ruling-2}
\end{center}
\end{figure}

Besides this weakly coupled description, one can also choose the ruling structure in figure~\ref{f:TN-ruling-2}. One can see that the three compact surfaces $S_1$, $S_4$ and $S_6$ in the corner give rise to three $SU(2)$ gauge groups. Each of them has two fundamental flavors:
\be
\ba
&S_1:\quad D_0^{(3)}\cdot S_1\ ,\ (D_0^{(3)}+D_1^{(3)})\cdot S_1\cr
&S_4:\quad D_0^{(1)}\cdot S_4\ ,\ (D_0^{(1)}+D_1^{(1)})\cdot S_4\cr
&S_6:\quad D_0^{(2)}\cdot S_6\ ,\ (D_0^{(2)}+D_1^{(2)})\cdot S_6\,.
\ea
\ee
This gauge theory description exactly has $SO(4)^3$ IR flavor symmetry. However, for $N>4$, the other compact surfaces do not have consistent section/ruling assignments. Hence this ``quiver gauge theory'' description is
\be \label{StrongQuiv}
\begin{array}{c}
 [2]\\
\vert  \\
 SU(2) \\
 \vert  \\
\,[2]-SU(2)-\boxed{\text{Strongly\ coupled\ matter}}-SU(2)-[2]\\
\end{array}
\ee
For the general $T_N$ theory $(N>4)$, the strongly coupled matter in the middle has Coulomb branch rank $\frac{(N-1)(N-2)}{2}-3$ and flavor rank $3N-12$. 

{On the other hand, when $N=4$, the matter field among the three $SU(2)$ gauge groups is a trifundamental half-hypermultiplet in the representation $\frac{1}{2}(\mathbf{2},\mathbf{2},\mathbf{2})$ of $G_{\rm gauge}=SU(2)\times SU(2)\times SU(2)$\footnote{We thank Fabio Apruzzi for pointing this out.}. Hence the theory has a weakly coupled gauge theory description.}

\subsection{IR Descriptions and BG-CFDs} 
\label{sec:BGCFD}

\begin{figure}
\centering
\includegraphics[width=14cm]{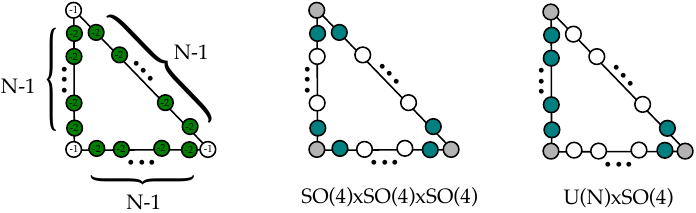}
\caption{BG-CFDs for $T_N$: On the left hand side we show the full CFD for $T_N$. The two diagrams on the right are the possible embeddings of BG-CFDs. The IR flavor symmetries are shown below each diagram. \label{fig:TNBGCFDs}}
\end{figure}

Weakly coupled descriptions can be obtained either, as in the last section, from the ruling of the surfaces in the geometry. Alternatively, as developed in \cite{Apruzzi:2019enx} they are constrained by the embedding of IR-versions of CFDs -- called {\it box graph CFDs (BG-CFDs)}.  
The BG-CFDs are subgraphs of the CFD, which encode the classical flavor symmetries $G_F^{\rm IR}$ for gauge theories with matter (or more generally quivers). Here we will briefly review some BG-CFDs used in this paper. Note that the $(-2)$-vertices are torquoise colored and the $(-1)$-vertices are gray.
For example, the BG-CFD for $SU(N_c)+ N_F \bm{F}$ $(N_c\geq 3)$ is
\be
\includegraphics[width=4cm]{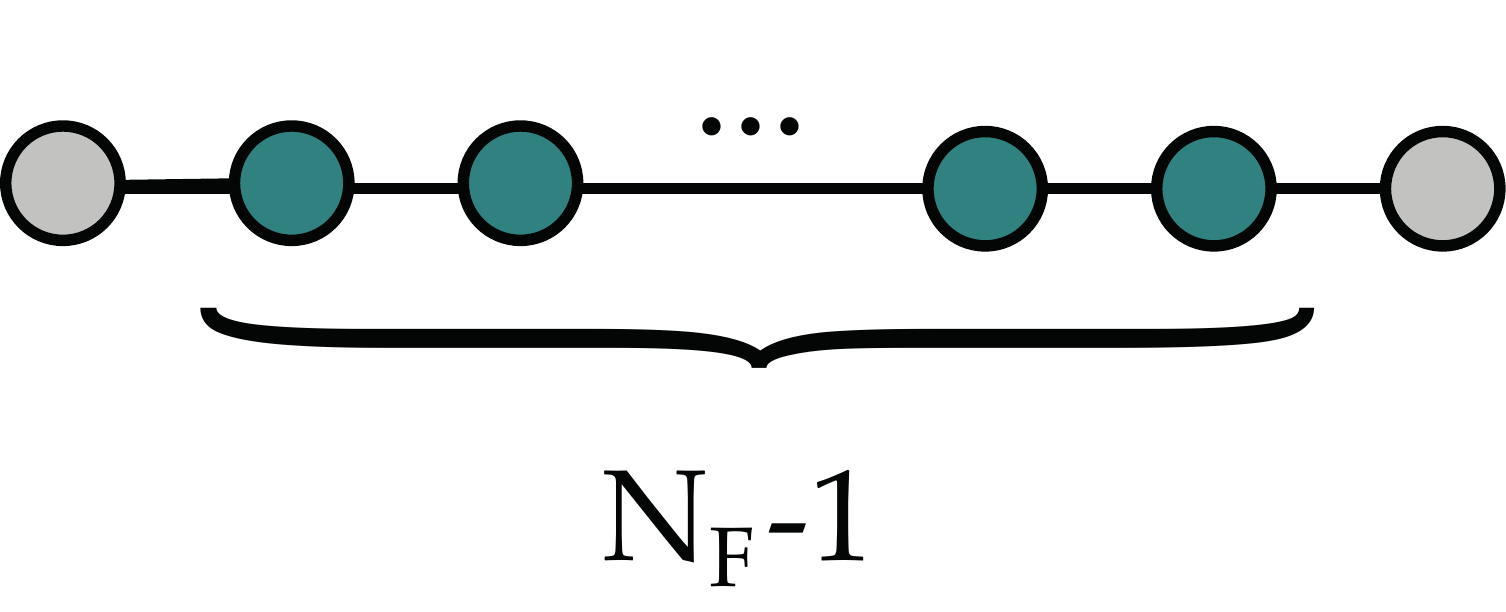} \,,
\ee
with $N_F-1$ $(-2)$-vertices, which corresponds to the $U(N_F)$ classical flavor symmetry. The $(-1)$-vertices at the ends are precisely the hypermultiplets in the fundamental representation of $SU(N_c)$.
For $N_c=2$ and $N_F=2$, the $SU(2)+ 2 \bm{F}$ has classical flavor symmetry $SO(4)$ with the BG-CFD
\be
\includegraphics[height=0.4cm]{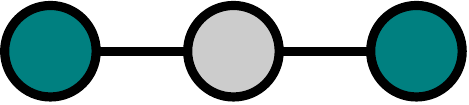} \,.
\ee
In \cite{Apruzzi:2019enx} we showed that embedding of a BG-CFD into the CFD for a 5d SCFT is a necessary requirement for the existence of the associated IR-description. 

The BG-CFDs for the $T_N$ theories can be read off in the usual way from  the $T_N$ CFD and are shown in figure \ref{fig:TNBGCFDs}: 
\begin{itemize}
\item The linear quiver (\ref{LinQuiver}) corresponds to the embedding of the $G_{F}^{\text{IR}}=U(N)\times SO(4)$ classical flavor symmetry. 
\item Finally, there is for $N>2$ an embedding of $SO(4)^3$, which corresponds to gluing the $T_N$ theory out of $T_2$ theories.  
This corresponds to the quiver with a strongly-coupled sector (\ref{StrongQuiv}), as derived from the geometry. 
\end{itemize}

\subsection{Extended Coulomb Branch and Box Graphs}

\label{sec:BoxGraphs}

Associated to the linear quiver 
\be
[N] - SU(N-1)_0   - SU(N-2)_0 - \cdots  SU(2) - [2]\,,
\ee
we can consider the extended Coulomb branch phases for the gauge groups and IR flavor symmetries. Extended here refers to the fact that we include mass deformations into the Coulomb branch description, which will allow us to also consider the decouplings of hypermultiplets within the framework of the Coulomb branch. 

The most efficient way to characterize Coulomb branch phases is in terms of the box graphs \cite{Hayashi:2013lra,Hayashi:2014kca,Braun:2014kla,Braun:2015hkv}, which comprise the data of the matter representations $(\bm{R}_{\text{gauge}}, \bm{R}_{F}^{\text{IR}})$.
For quiver gauge theories, we need to also include bi-fundamental matter representations between the gauge nodes. 
The box graphs encode the data of the Coulomb branch in terms of a sign-decorated representation graph. 
For the instances of $SU(N_c) + N_F\bm{F}$, we consider the representation graph for $(\bm{N_c}, \bm{N_F})$ 
\be\label{SUnN}
\includegraphics*[width=6cm]{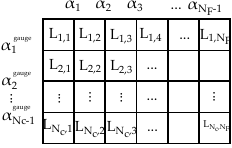}
\ee
To each box we associate a weight $L_{i, j}$, which is the sums of fundamental weights of $SU(N_c)$ and $SU(N_F)$ respectively and the roots $\alpha$ act between the boxes. 
A consistent extended Coulomb branch phase is given by a sign assignment that satisfies the following basic rules (for an in depth analysis and generalizations see \cite{Hayashi:2014kca,Apruzzi:2019enx}) 
\begin{enumerate}
\item A weight $\lambda$ with the sign assignment $\epsilon = \pm1 $ (blue/yellow) corresponds to the associated weights $\epsilon \lambda$ being part of the Coulomb branch. 
\item For $SU(N)$ representation graphs: sign assignments have to satisfy the {\it flow rules}:
\be
\includegraphics[width=4cm]{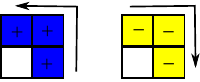}
\ee
These follow straight-forwardly from the fact that the simple roots are part of the Coulomb branch and adding these to weights that are in the Coulomb branch has to result in consistent sign assignments. 
\end{enumerate}
For the current problem, we require only the box graphs for $SU(N-1) - [N]$ as well as $SU(2)-[2]$. 

For $SU(N-1)$ with $N$ fundamentals, the box graphs take precisely the form shown in (\ref{SUnN}). 
For $SU(2)-[2]$ we need to consider the box graphs for $Sp$ gauge groups with fundamental matter. In this specific case, the situation is quite simple: the theory has classical flavor symmetry $SO(4)$, and we consider representation graphs of $(\bm{2}, \bm{4})$ under $SU(2) \times SO(4)$. The representation graph is 
\be
\includegraphics[width=2cm]{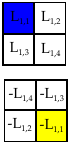}
\ee
We already colored/sign-assigned two of the boxes as these are fixed from the get-go. 
In addition there are the bifundamentals, which are simply box graphs for $(\bm{N-m}, \bm{N-m-1})$ for $SU(N-m) \times SU(N-m-1)$. 

\begin{figure}
\centering
\includegraphics[width=\textwidth]{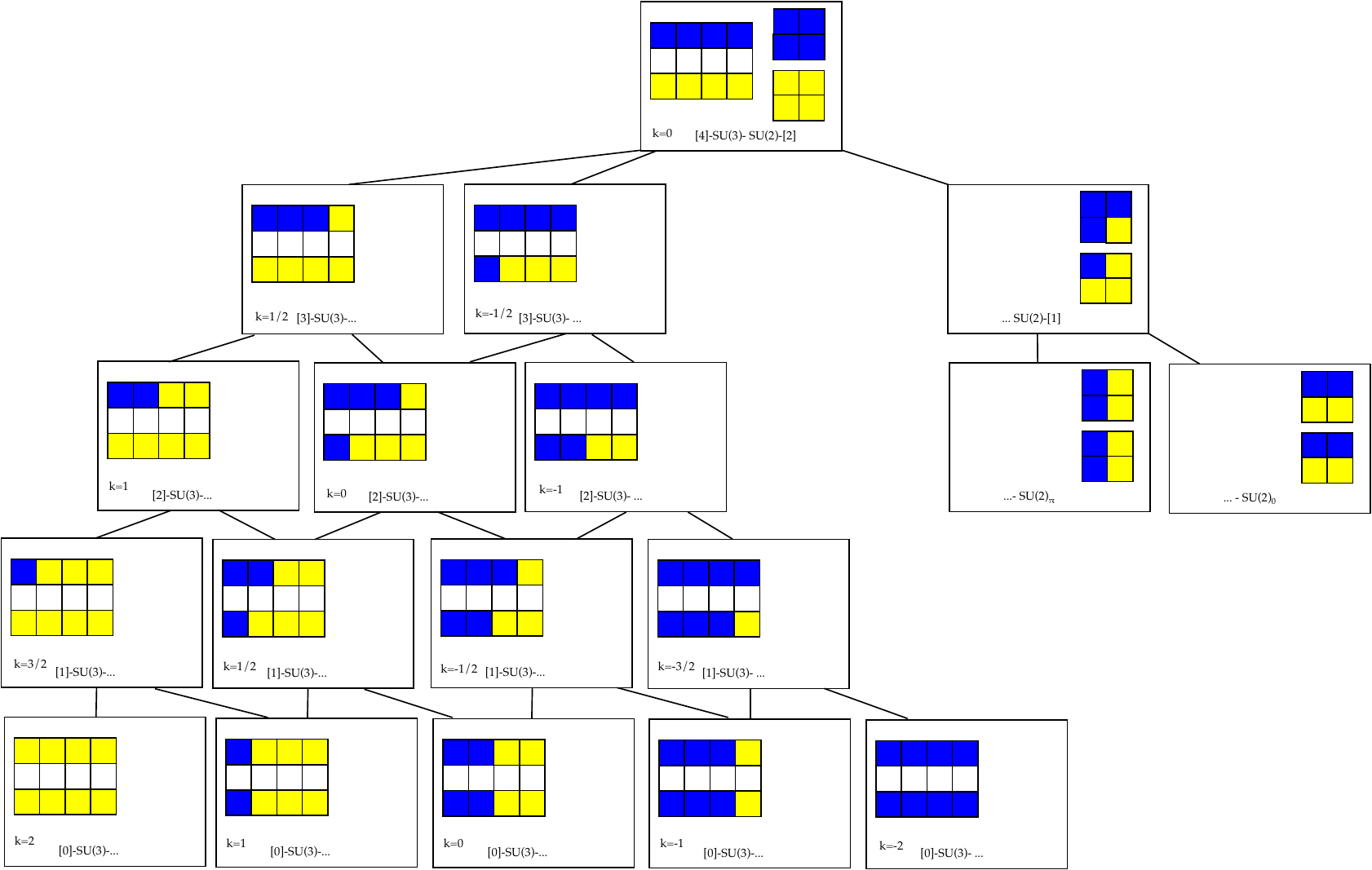}
\caption{Box graphs for $T_4$ in the weakly coupled description as a quiver $[4]-SU(3)-SU(2)-[2]$. Each extended Coulomb branch phase is characterized by a pair of box graphs for $SU(3)$ (left hand branch) and for $SU(2)$ (right hand branch). The CS-level of the $SU(3)$ is denoted by $k$.  \label{fig:QuiverBGPhases}}
\end{figure}

What we are interested in are those gauge group phases, that have the same flavor symmetries in the UV -- in \cite{Apruzzi:2019enx} these were called {\it flavor equivalence classes} of Coulomb branch phases. The bifundamental matter does not influence the flavor symmetry and we thus only need to consider the gauge theory phases for $SU(N-1) - [N]$ and  $SU(2)-[2]$. For the $SU(2)-[2]$ these are shown on the right hand branch of figure \ref{fig:QuiverBGPhases}. For the $SU(N-1)-[N]$, the only relevant information is encoded in the reduced box graphs, which correspond to the first and last rows of (\ref{SUnN}), as these determine the flavor symmetry. 

These are summarized in figure \ref{fig:QuiverBGPhases} for  $T_4$, i.e. $[4]-SU(3)_0-SU(2)-[2]$. 
The box graphs shown there are the different phases for the $SU(3)-[4]$ as well as the $SU(2)-[2]$, which can be combined in all possible ways. 
In the next section we will describe the corresponding toric triangulations.

\begin{figure}
\centering
\includegraphics[width=11cm]{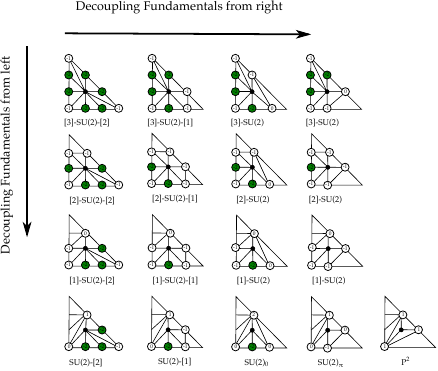}
\caption{CFD-transitions and decoupling for $T_3$, which have a gauge-theoretic interpretation. The final flop is of course not realized in the gauge theory. Note also that there are further flops, see figure \ref{fig:T3AllFlops}, which do not have an interpretation in terms of the quiver. \label{fig:T3Flop}}
\end{figure}


\section{Higgs Branch, Magnetic Quivers, and Hasse Diagrams}

In this section we study the Higgs branch of $T_N$ and its descendants using magnetic quivers and associated Hasse diagrams.
For 5d gauge theories, the magnetic quivers (MQs) can be motivated from various points of view. They are the 3d mirror dual quivers to the 3d $\mathcal{N}=4$ gauge theories obtained by $T^2$-compactification \cite{Hori:1997zj}. 
It was found in \cite{Cabrera:2018jxt} that the MQs can also be read off from $(p,q)$ 5-brane webs, which we briefly review. This approach allows us to study the Higgs branch of both the strongly coupled theory and its quiver description.

\subsection{$(p,q)$ 5-Brane Webs}

$(p,q)$ 5-brane webs \cite{Aharony:1997ju,Aharony:1997bh,DeWolfe:1999hj} provide an effective description of 5d $\mathcal{N}=1$ gauge theories inside type IIB string theory. Each 5-brane is a $\half$-BPS object and a general configuration breaks supersymmetry completely. However, the following setup, up to Lorentz- and $SL(2,\Z)$-invariance, preserves 8 supercharges. We denote the ten space-time dimensions by $(x_0,x_1,\dots,x_9)$, where $x_0$ is the time direction.
\begin{enumerate}
\item
All $(p,q)$ 5-branes, defined as a bound state of $p$ D5- and $q$ NS5-branes, share five spacetime dimensions $(x_0,\dots,x_4)$ and lie at an angle
\be
\tan\theta=\frac{p}{q}
\ee
inside the $(x_5$-$x_6)$-plane.\footnote{With this we fix $\tau_{\text{IIB}}=\tau_1+i\tau_2=i$. In general we have $\tan\theta=\frac{p+q\tau_1}{q\tau_2}$.}
\item
At each junction, consisting of three incoming $(p,q)$ 5-branes, there is charge conservation
\be
\sum_{i=1}^3 \left(p_i,q_i\right)=(0,0)\,.
\ee
\item
Infinite $(p,q)$ 5-branes can be made finite by letting them end on $(p,q)$ 7-branes, extended in the $(x_0,\dots,x_4,x_7,x_8,x_9)$-directions.
\end{enumerate}
Then, there is an effective 5d $\mathcal{N}=1$ gauge theory in the $(x_0,\dots,x_4)$ space-time directions.\footnote{We can further generalize this setup by letting multiple 5-branes end on the same 7-brane, usually as a consequence of Hanany-Witten brane creations. For a summary see e. g. \cite{Benini:2009gi,Bergman:1998ej}.}

\begin{figure}
\centering
\resizebox{.8\textwidth}{!}{%
{\large
\begin{tikzpicture}[
roundnode/.style={circle, draw=black, thick, fill=white, minimum size=5mm},
]

\node at ($(0,0)$) (1) {};
\node at ($(-1,1)$) (2) {};
\node at ($(-1,2)$) (3) {};
\node at ($(-2,3)$) (4) {};
\node at ($(-2,4)$) (5) {};
\node at ($(-3,5)$) (6) {};
\node at ($(-3,6)$) (7) {};
\node at ($(-2,7)$) (8) {};
\node at ($(0,8)$) (9) {};
\node at ($(1,8)$) (10) {};
\node at ($(1,4)$) (11) {};
\node at ($(2,3)$) (12) {};
\node at ($(2,2)$) (13) {};
\node at ($(3,1)$) (14) {};
\node at ($(3,0)$) (15) {};
\node at ($(4,-1)$) (16) {};
\node at ($(5,-1)$) (17) {};
\node at ($(8,-4)$) (18) {};
\node at ($(9,-4)$) (19) {};
\node at ($(10,-3)$) (20) {};
\node at ($(11,-1)$) (21) {};
\node at ($(11,0)$) (22) {};
\node at ($(5,0)$) (23) {};
\node at ($(4,1)$) (24) {};
\node at ($(4,3)$) (25) {};

\node[roundnode] at ($(2)+(-2,0)$) (26) {};
\node[roundnode] at ($(4)+(-2,0)$) (27) {};
\node[roundnode] at ($(6)+(-2,0)$) (28) {};
\node[roundnode] at ($(7)+(-2,0)$) (29) {};
\node[roundnode] at ($(8)+(-2,0)$) (30) {};

\node[roundnode] at ($(9)+(2,2)$) (31) {};
\node[roundnode] at ($(10)+(2,2)$) (32) {};
\node[roundnode] at ($(25)+(2,2)$) (33) {};
\node[roundnode] at ($(22)+(2,2)$) (34) {};
\node[roundnode] at ($(21)+(2,2)$) (35) {};

\node[roundnode] at ($(1)+(0,-2)$) (36) {};
\node[roundnode] at ($(16)+(0,-2)$) (37) {};
\node[roundnode] at ($(18)+(0,-2)$) (38) {};
\node[roundnode] at ($(19)+(0,-2)$) (39) {};
\node[roundnode] at ($(20)+(0,-2)$) (40) {};

\node at ($(-2,10)$) (41) {\Huge (a)};

\draw[thick] (0,0) -- (-1,1);
\draw[thick] (-1,2) -- (-1,1);
\draw[thick] (-1,2) -- (-2,3);
\draw[thick] (-2,4) -- (-2,3);
\draw[thick] (-2,4) -- (-3,5);
\draw[thick] (-3,6) -- (-3,5);
\draw[thick] (-3,6) -- (-2,7);
\draw[thick] (0,8) -- (-2,7);
\draw[thick] (0,8) -- (1,8);
\draw[thick] (1,4) -- (1,8);
\draw[thick] (1,4) -- (2,3);
\draw[thick] (2,2) -- (2,3);
\draw[thick] (2,2) -- (3,1);
\draw[thick] (3,0) -- (3,1);
\draw[thick] (3,0) -- (4,-1);
\draw[thick] (5,-1) -- (4,-1);
\draw[thick] (5,-1) -- (8,-4);
\draw[thick] (9,-4) -- (8,-4);
\draw[thick] (9,-4) -- (10,-3);
\draw[thick] (11,-1) -- (10,-3);
\draw[thick] (11,-1) -- (11,0);
\draw[thick] (5,0) -- (11,0);
\draw[thick] (5,0) -- (4,1);
\draw[thick] (4,3) -- (4,1);
\draw[thick] (-2,4) -- (1,4);
\draw[thick] (-1,2) -- (2,2);
\draw[thick] (0,0) -- (3,0);
\draw[thick] (2,3) -- (4,3);
\draw[thick] (3,1) -- (4,1);
\draw[thick] (5,-1) -- (5,0);

\draw[thick] (-1,1) -- (26);
\draw[thick] (-2,3) -- (27);
\draw[thick] (-3,5) -- (28);
\draw[thick] (-3,6) -- (29);
\draw[thick] (-2,7) -- (30);
\draw[thick] (0,8) -- (31);
\draw[thick] (1,8) -- (32);
\draw[thick] (4,3) -- (33);
\draw[thick] (11,0) -- (34);
\draw[thick] (11,-1) -- (35);
\draw[thick] (0,0) -- (36);
\draw[thick] (4,-1) -- (37);
\draw[thick] (8,-4) -- (38);
\draw[thick] (9,-4) -- (39);
\draw[thick] (10,-3) -- (40);
%
%

\node at ($(32)+(10,4)$) (1) {};
\node[roundnode] at ($(1)+(-2,0)$) (2) {};
\node[roundnode] at ($(1)+(-4,0)$) (3) {};
\node[roundnode] at ($(1)+(-6,0)$) (4) {};
\node[roundnode] at ($(1)+(-8,0)$) (5) {};
\node[roundnode] at ($(1)+(-10,0)$) (6) {};
\node[roundnode] at ($(1)+(0,-2)$) (7) {};
\node[roundnode] at ($(1)+(1.5,1.5)$) (8) {};
\node at ($(1)+(2,0)$) (9) {};
\node[roundnode] at ($(9)+(0,-2)$) (10) {};
\node[roundnode] at ($(9)+(1.5,1.5)$) (11) {};
\node at ($(9)+(2,0)$) (12) {};
\node[roundnode] at ($(12)+(0,-2)$) (13) {};
\node[roundnode] at ($(12)+(1.5,1.5)$) (14) {};
\node at ($(12)+(2,0)$) (15) {};
\node[roundnode] at ($(15)+(0,-2)$) (16) {};
\node[roundnode] at ($(16)+(0,-2)$) (17) {};
\node[roundnode] at ($(15)+(1.5,1.5)$) (18) {};
\node[roundnode] at ($(18)+(1.5,1.5)$) (19) {};

\node at ($(4)+(0,2)$) (20) {\Huge (b)};

\draw[thick] ($(1)$) -- (2) node[midway,above] {5};
\draw[thick] (3) -- (2) node[midway,above] {4};
\draw[thick] (3) -- (4) node[midway,above] {3};
\draw[thick] (5) -- (4) node[midway,above] {2};
\draw[thick] (5) -- (6) node[midway,above] {1};

\draw[thick] ($(1)$) -- ($(9)$)  node[midway,above] {4};
\draw[thick] ($(9)$) -- ($(12)$)  node[midway,above] {3};
\draw[thick] ($(12)$) -- ($(15)$)  node[midway,above] {2};

\draw[thick] ($(1)$) -- (7);
\draw[thick] ($(1)$) -- (8);
\draw[thick] ($(9)$) -- (10);
\draw[thick] ($(9)$) -- (11);
\draw[thick] ($(12)$) -- (13);
\draw[thick] ($(12)$) -- (14);
\draw[thick] ($(15)$) -- (16) node[midway,right] {2};
\draw[thick] (17) -- (16) node[midway,right] {1};;
\draw[thick] ($(15)$) -- (18) node[midway,above left] {2};
\draw[thick] (19) -- (18) node[midway,above left] {1};

%
%

\node at ($(19)+(-2,-12)$) (1) {};
\node[roundnode] at ($(1)+(-2,0)$) (2) {};
\node[roundnode] at ($(1)+(-4,0)$) (3) {};
\node[roundnode] at ($(1)+(-6,0)$) (4) {};
\node[roundnode] at ($(1)+(-8,0)$) (5) {};
\node[roundnode] at ($(1)+(-10,0)$) (6) {};
\node[roundnode] at ($(1)+(0,-2)$) (7) {};
\node[roundnode] at ($(1)+(0,-4)$) (8) {};
\node[roundnode] at ($(1)+(0,-6)$) (9) {};
\node[roundnode] at ($(1)+(0,-8)$) (10) {};
\node[roundnode] at ($(1)+(0,-10)$) (11) {};
\node[roundnode] at ($(1)+(1,1)$) (12) {};
\node[roundnode] at ($(1)+(2,2)$) (13) {};
\node[roundnode] at ($(1)+(3,3)$) (14) {};
\node[roundnode] at ($(1)+(4,4)$) (15) {};
\node[roundnode] at ($(1)+(5,5)$) (16) {};

\node at ($(4)+(0,2)$) (17) {\Huge (c)};

\draw[thick] ($(1)$) -- (2) node[midway,above] {5};
\draw[thick] (3) -- (2) node[midway,above] {4};
\draw[thick] (3) -- (4) node[midway,above] {3};
\draw[thick] (5) -- (4) node[midway,above] {2};
\draw[thick] (5) -- (6) node[midway,above] {1};
\draw[thick] ($(1)$) -- (12) node[midway,below right=-.1] {5};
\draw[thick] (13) -- (12) node[midway,below right=-.1] {4};
\draw[thick] (13) -- (14) node[midway,below right=-.1] {3};
\draw[thick] (14) -- (15) node[midway,below right=-.1] {2};
\draw[thick] (16) -- (15) node[midway,below right=-.1] {1};
\draw[thick] ($(1)$) -- (7) node[midway,left] {5};
\draw[thick] (7) -- (8) node[midway,left] {4};
\draw[thick] (8) -- (9) node[midway,left] {3};
\draw[thick] (9) -- (10) node[midway,left] {2};
\draw[thick] (10) -- (11) node[midway,left] {1};

\end{tikzpicture}

}
}
\caption{Brane web for $T_5$ at a) a general point on the extended Coulomb branch, b) in the weakly coupled quiver description and c) at the superconformal fixed point.
 \label{fig:T5BraneWeb}}
\end{figure}
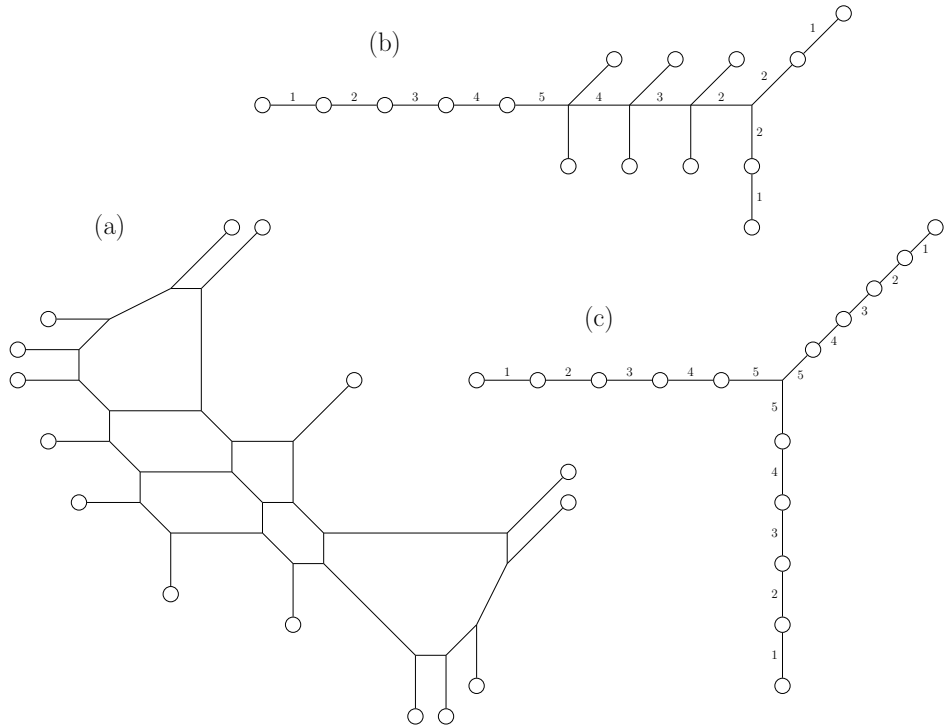

If the 5d theory has a description in terms of toric geometry, see section~\ref{sec:toric}, then the brane web is the dual  graph of the toric fan.
Only the configuration of the external 7-branes is fixed by the geometry - the internal web is determined by the chosen triangulation. These correspond to different gauge theory phases. For example, the brane web for $T_5$, with the triangulation as in figure \ref{T5ToricIntersections}, is shown in figure \ref{fig:T5BraneWeb} (a). This generalizes naturally to any $N$  \cite{Benini:2009gi}.
Clearly, the web has much more information than the toric diagram as the latter does not show all the length scales explicitly. Nonetheless, these length scales parametrize the extended Coulomb branch of the gauge theory. In the brane web picture, the Coulomb branch parameters are proportional to the heights of the compact faces, which correspond to the internal points of the dual toric diagram. The mass parameters and gauge couplings are given by the positions of the external 7-branes, where three of them are fixed by Lorentz-invariance and supersymmetry.
In this language, we confirm that
\be
\dim \mathcal{C}={{N}\choose{2}}\,,\qquad \dim \mathcal{C}_{\text{ext}}=\dim \mathcal{C}+3(N-1)\,.
\ee
From the alignment of compact faces in the brane web, we can also find the weakly coupled quiver description
\be \label{QuiverDescription}
[N]- SU(N-1)_0  - SU(N-2)_0 - \cdots - SU(2) - [2] \,.
\ee
This was shown in detail in \cite{Bergman:2014kza}.
We see that at a certain point on the extended Coulomb branch, corresponding to the ruling in figure \ref{f:TN-ruling}, the brane web of $T_5$ is given in figure \ref{fig:T5BraneWeb} (b).
In this description we can clearly see the weakly coupled $U(N)\times SO(4)$ flavor symmetry. In the left part of the graph, there is a stack of five 5-branes corresponding to the $U(5)$ flavor symmetry. In the right part of the graph, there are two stacks of two 5-branes corresponding to the $SO(4)=SU(2)\times SU(2)$. The parameters of the extended Coulomb branch are the $(N-2)$ gauge coupling parameters, the masses of the $(N+2)$ fundamental and that of the $(N-3)$ bi-fundamental matter fields.

The SCFT sits at the origin of the extended Coulomb branch, i.e. all the parameters are set to zero. In terms of toric geometry, this corresponds to the singular limit, where the toric fan is untriangulated. The corresponding web is depicted in figure \ref{fig:T5BraneWeb} (c). We can immediately see the classical $SU(N)^3$ flavor symmetry with matter coming from the string modes between the $N$ coincident 5-branes.

\subsection{Magnetic Quiver and Hasse Diagram of $T_N$}

From the brane web it is possible to read off the magnetic quiver \cite{Cabrera:2018jxt}. The $SU(2)_R$ R-symmetry of the 5d $\mathcal{N}=1$ theories can be interpreted in the brane web as a rotation in $\R^3_R$ spanned by $(x_7,x_8,x_9)$. The R-symmetry is broken along the Higgs branch. In the brane web picture, this is done by displacing individual 5-branes away from the origin in $\R^3_R$. However, we can only do this as long as the web at each point in $\R^3_R$ is consistent. To map out the full Higgs branch, we thus need to divide the brane web into its possible subwebs.
Each subweb corresponds to a node in the magnetic quiver, representing a unitary gauge group whose rank equals the multiplicity of the subweb.
The edges between nodes, corresponding to bifundamental matter, have contributions from the intersection between the subwebs as well as 7-branes:
\begin{enumerate}
\item The intersection between a $(p_1,q_1)$-  and a $(p_2,q_2)$ 5-brane contributes as $+\left|\det\begin{pmatrix} p_1 & p_2 \\ q_1 & q_2 \end{pmatrix}\right|$.
\item The contribution from 7-branes is $+1$ if two 5-branes end on opposite sides of a 7-brane and $-1$ if they end on the same side.
\end{enumerate}

Each node in the magnetic quiver has a balance, which can be computed from the node multiplicities $n_i$ and the edge multiplicities $k_{ij}$
\be
\beta_i =  - 2 n_i+\sum_{j} k_{ij}\ n_j \,.
\ee
Usually, the balanced nodes form the Dynkin diagram of the flavor group of the theory.\footnote{Counter-examples to this rules exist but we will not encounter them here. The full flavor group can still be reliably computed from the MQ using the Hilbert series \cite{Cremonesi:2013lqa}.} We will denote unbalanced nodes with a cross.

The Hasse diagrams describes the foliation structure of the Higgs branch into symplectic singularities. It is derived from the magnetic quiver by an application of simple combinatorial rules, namely, a repeated quiver subtraction \cite{Cabrera:2018ann} of ADE affine Dynkin diagrams, corresponding to nilpotent orbits, and subsequent rebalancing of the nodes \cite{Bourget:2019aer,Bourget:2019rtl}.\footnote{Another allowed subtraction are Kleinian singularities $A_{N-1}$, represented by two $U(1)$ nodes connected by $N$ bifundamentals.}

\begin{figure}
\centering
\resizebox{.5\textwidth}{!}{%
{\large

\begin{tikzpicture}[
roundnode/.style={circle, draw=black, thick, fill=white, minimum size=5mm},
]

\node[roundnode, label=below:\large $N$] (1) {};
\node at (1) {\huge$\times$};

\node[roundnode, label=below:\large $N$-1] (2) [right=of 1] {};
\node (3) [right=.5of 2] {};
\node (4) [right=.5of 3] {};
\node[roundnode, label=below:\large 1] (5) [right=.5of 4] {};
\node  at ($(3)!0.5!(4)$) {$\hdots$} (6);

\node[roundnode, label=below:\large $N$-1] (7) [left=of 1] {};
\node (8) [left=.5of 7] {};
\node (9) [left=.5of 8] {};
\node[roundnode, label=below:\large 1] (10) [left=.5of 9] {};
\node  at ($(8)!0.5!(9)$) {$\hdots$} (11);

\node[roundnode, label=left:\large $N$-1] (12) [above=.7of 1] {};
\node (13) [above=.5of 12] {};
\node (14) [above=.5of 13] {};
\node[roundnode, label=left:\large 1] (15) [above=.35of 14] {};
\node  at ($(13)!0.5!(14)$) {$\vdots$} (16);

\draw[thick] (1) -- (2);
\draw[thick] (2) -- (3);
\draw[thick] (4) -- (5);

\draw[thick] (1) -- (7);
\draw[thick] (7) -- (8);
\draw[thick] (9) -- (10);

\draw[thick] (1) -- (12);
\draw[thick] (12) -- (13);
\draw[thick] (14) -- (15);

\end{tikzpicture}
}
}
\caption{The magnetic quiver for the $T_N$ theory. The unbalanced nodes will be marked with a cross. For $N=3$ the central node is also balanced and the flavor symmetry is $G_{F, T_3}=E_6$.  \label{fig:TNMQ}}
\end{figure}
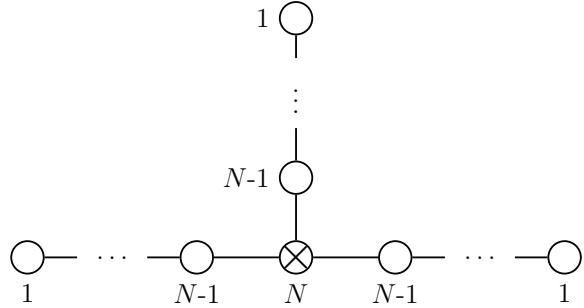

\begin{figure}

\subfloat{(a)}
\resizebox{.5\textwidth}{!}{%
{\large
\begin{tikzpicture}[roundnode/.style={circle, draw=black, thick, fill=white, minimum size=5mm},]

\node[roundnode, label=below:\large 4] (1) {};
\node at (1) {\huge$\times$};

\node[roundnode, label=below:\large 3] (2) [right=of 1] {};
\node[roundnode, label=below:\large 2] (3) [right=of 2] {};
\node[roundnode, label=below:\large 1] (4) [right=of 3] {};

\node[roundnode, label=below:\large 3] (5) [left=of 1] {};
\node[roundnode, label=below:\large 2] (6) [left=of 5] {};
\node[roundnode, label=below:\large 1] (7) [left=of 6] {};

\node[roundnode, label=left:\large 3] (8) [above=.7of 1] {};
\node[roundnode, label=left:\large 2] (9) [above=.7of 8] {};
\node[roundnode, label=left:\large 1] (10) [above=.7of 9] {};

\draw[thick] (1) -- (2);
\draw[thick] (2) -- (3);
\draw[thick] (3) -- (4);

\draw[thick] (1) -- (5);
\draw[thick] (5) -- (6);
\draw[thick] (6) -- (7);

\draw[thick] (1) -- (8);
\draw[thick] (8) -- (9);
\draw[thick] (9) -- (10);


\node (11) at ($(1)+(-1.5,-1.5)$) {};
\node (12) at ($(11)+(-2,-2)$) {};

\node (13) at ($(1)+(1.5,-1.5)$) {};
\node (14) at ($(13)+(2,-2)$) {};

\draw[thick,->] (11) -- (12) node[midway,left] {$e_6$};
\draw[thick,->] (13) -- (14) node[midway,left] {$e_7$}  node[midway,right] {three options};



\node[roundnode, label=left:\large 1] (15) at ($(12)+(-1.5,0)$) {};
\node at (15) {\huge$\times$};

\node[roundnode, label=left:\large 1] (16) [below=.7of 15] {};
\node[roundnode, label=left:\large 1] (17) [below=.7of 16] {};
\node[roundnode, label=left:\large 1] (18) [below=.7of 17] {};
\node[roundnode, label=below:\large 1] (19) [below=.7of 18] {};
\node at (19) {\huge$\times$};

\node[roundnode, label=below:\large 1] (20) [right=of 19] {};
\node[roundnode, label=below:\large 1] (21) [right=of 20] {};
\node[roundnode, label=below:\large 1] (22) [right=of 21] {};

\node[roundnode, label=below:\large 1] (23) [left=of 19] {};
\node[roundnode, label=below:\large 1] (24) [left=of 23] {};
\node[roundnode, label=below:\large 1] (25) [left=of 24] {};

\draw[thick] (15) -- (16);
\draw[thick] (16) -- (17);
\draw[thick] (17) -- (18);
\draw[thick] (18) -- (19);

\draw[thick] (19) -- (20);
\draw[thick] (20) -- (21);
\draw[thick] (21) -- (22);

\draw[thick] (19) -- (23);
\draw[thick] (23) -- (24);
\draw[thick] (24) -- (25);

\draw[thick] (15) -- (22);
\draw[thick] (15) -- (25);


\node[roundnode, label=above:\large 1] (26) at ($(14)+(1.5,-2)$) {};
\node at (26) {\huge$\times$};

\node[roundnode, label=below:\large 2] (27) [below=.7 of 26] {};
\node[roundnode, label=below:\large 1] (28) [right=of 27] {};
\node[roundnode, label=below:\large 1] (29) [left=of 27] {};

\draw[thick,double distance=3pt] (26) -- (27);
\draw[thick] (27) -- (28);
\draw[thick] (27) -- (29);


\node (30) at ($(19)+(0,-1.5)$) {};
\node (31) at ($(30)+(2,-2)$) {};

\node (32) at ($(30)+(10,0)$) {};
\node (33) at ($(32)+(-2,-2)$) {};

\draw[thick,->] (30) -- (31) node[midway,left] {$a_7$} node[midway,right] {three options};
\draw[thick,->] (32) -- (33) node[midway,right] {$a_1$};


\node[roundnode, label=above:\large 1] (34) at ($(31)+(3,-1.5)$) {};
\node[roundnode, label=below:\large 1] (35) [below= of 34] {};
\node[roundnode, label=below:\large 1] (36) [right= of 35] {};
\node[roundnode, label=below:\large 1] (37) [left= of 35] {};

\draw[thick] (34) -- (36);
\draw[thick] (34) -- (37);
\draw[thick] (35) -- (36);
\draw[thick] (35) -- (37);

\end{tikzpicture}
}}\qquad 
\subfloat{(b)} \includegraphics[width=5cm]{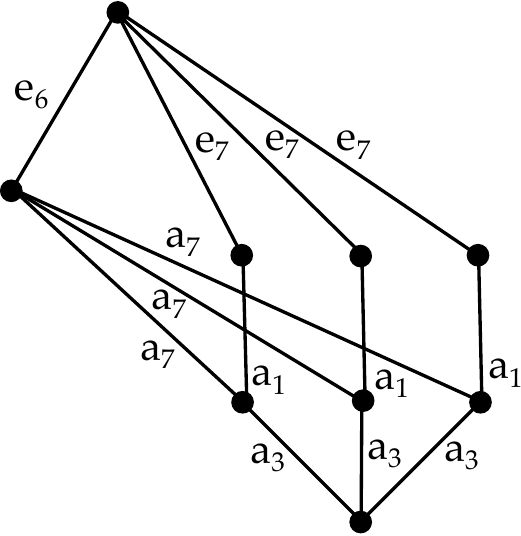}

\caption{(a) Subtractions leading to the Hasse diagram of $T_4$. We only show one of the three equivalent pathways. (b) The Hasse diagram for $T_4$.
\label{fig:T4HasseExplicit}}
\end{figure}
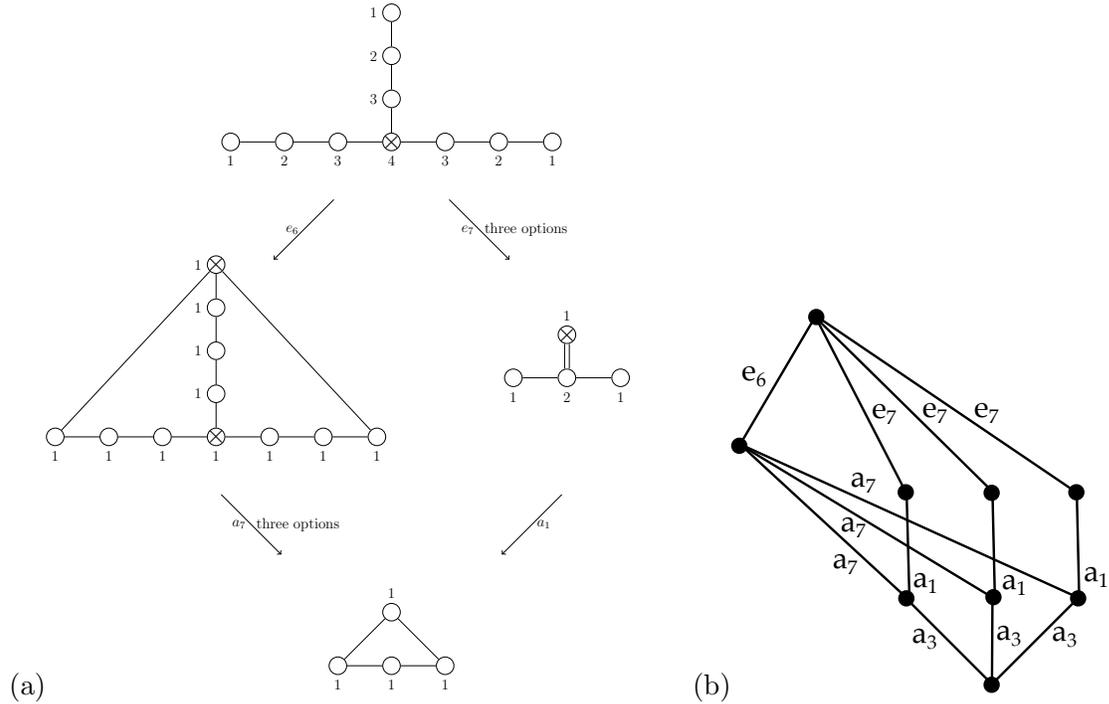

The magnetic quivers for the $T_N$ theories were determined in \cite{Benini:2009gi} and are shown in figure \ref{fig:TNMQ}.
The subgraph of balanced nodes reproduces the flavor symmetry $G_{F, T_N} = SU(N)^3$. For $N=3$ the central node with label $N$ is also balanced  (the magnetic quiver is in fact given by the affine $E_6$ Dynkin diagram), and therefore the flavor symmetry is correctly reproduced as $G_{F, T_3}= E_6$.  In this case the Hasse diagram consists of only one leaf given by $\mathfrak{e}_6$.
For $T_4$ the possible subtractions and the final Hasse diagram are illustrated in figure \ref{fig:T4HasseExplicit}. An arrow with $\mathfrak{g}$ indicates the subtraction of the affine $\mathfrak{g}$ Dynkin diagram and rebalancing. 
%

\begin{figure}
\centering
\includegraphics[width=15cm]{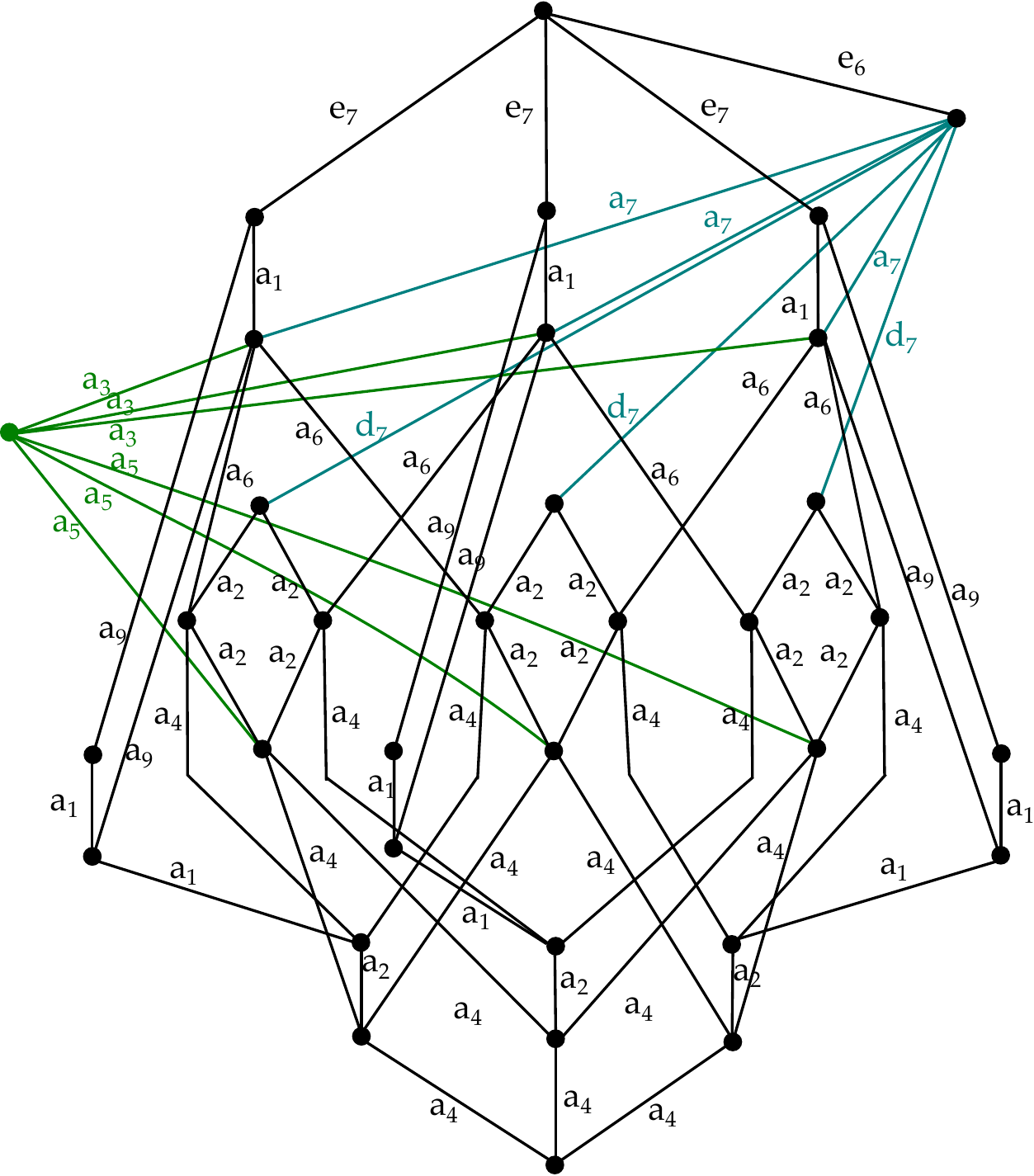}
\caption{The Hasse diagram for $T_5$. The colors in this figure are simply there for ease of visualization. 
\label{fig:T5Hasse}}
\end{figure}

For $T_5$ the Hasse diagram has somewhat increased complexity, shown in figure \ref{fig:T5Hasse}.

\begin{figure}
\centering
\subfloat{(a)} \resizebox{ .7\textwidth}{!}{%
{\large

\begin{tikzpicture}[
roundnode/.style={circle, draw=black, thick, fill=white, minimum size=5mm},
]

\node[roundnode, label=below:\large $N$-1] (1) {};

\node[roundnode, label=below:\large $N$-2] (2) [left=of 1] {};
\node (3) [left=.5of 2] {};
\node (4) [left=.5of 3] {};
\node[roundnode, label=below:\large 1] (5) [left=.5of 4] {};
\node  at ($(3)!0.5!(4)$) {$\hdots$} (6);

\node[roundnode, label=below:\large 2] (7) [right=2of 1] {};
\node at (7) {\huge$\times$};
\node[roundnode, label=below:\large 1] (8) [right=of 7] {};
\node[roundnode, label=above:\large 1] (9) [above=of 7] {};

\node[roundnode, label=left:\large 1] at ($(1)+(-1,1)$) (10) {};
\node[roundnode, label=right:\large 1] at ($(1)+(1,1)$) (11) {};
\node at (10) {\huge$\times$};
\node at (11) {\huge$\times$};
\node at ($(10)!0.5!(11)$) {$\hdots$} (12);

\draw[thick] (1) -- (2);
\draw[thick] (2) -- (3);
\draw[thick] (4) -- (5);

\draw[thick] (1) -- (7);
\draw[thick] (7) -- (8);
\draw[thick] (9) -- (7);

\draw[thick] (10) -- (1);
\draw[thick] (11) -- (1);

\draw[decoration={calligraphic brace,amplitude=5pt}, decorate, line width=1.25pt] ($(10)+(-.3,.5)$) -- ($(11)+(.3,.5)$);
\node at ($(10)!0.5!(11) + (0,1)$) {$N$-2} (13);

\end{tikzpicture}
}
\qquad\qquad \subfloat{\Large(b)}\includegraphics*[width=4cm]{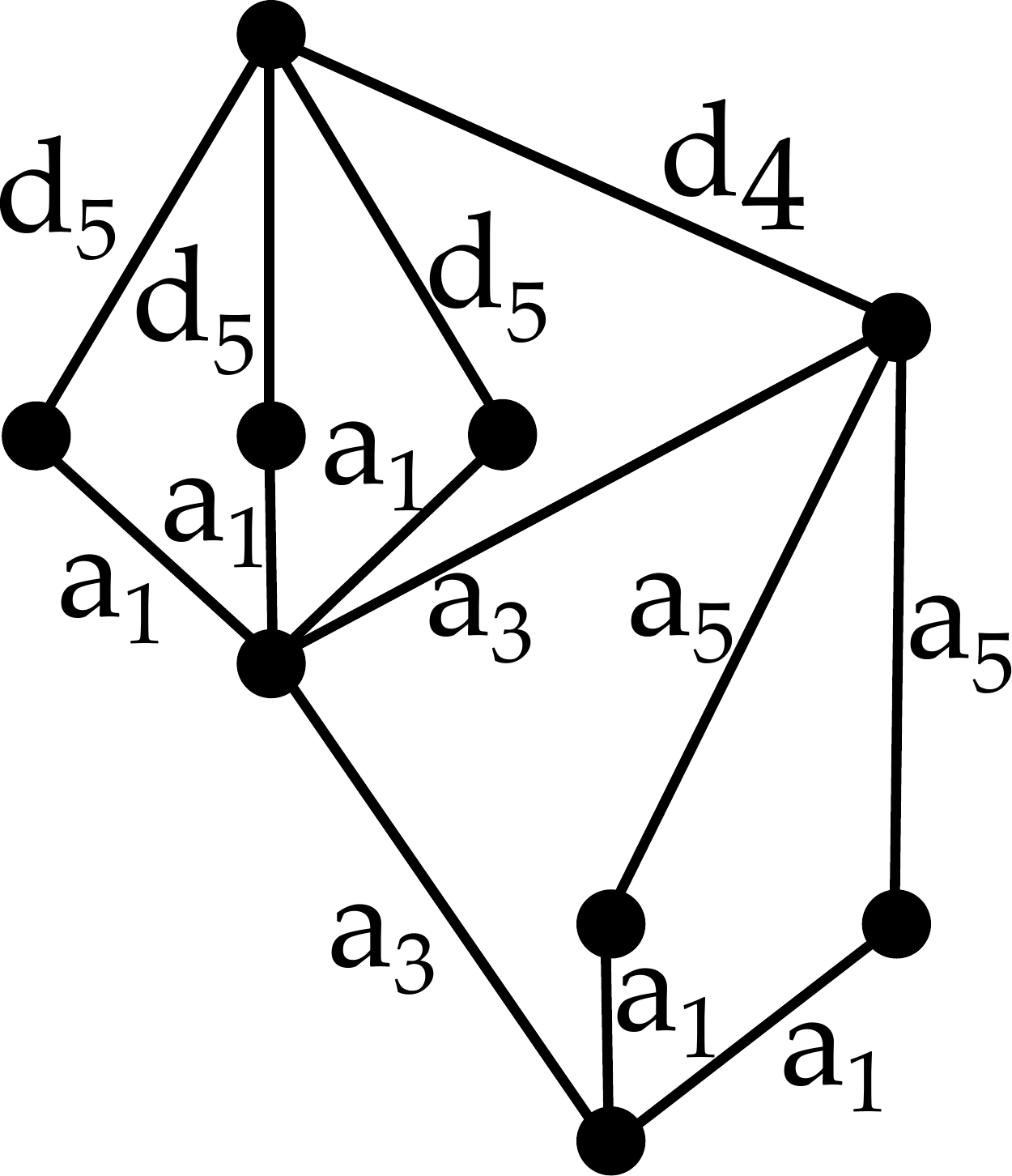}
}
\caption{(a) Magnetic quiver for the weakly coupled quiver description of $T_N$ and (b) Hasse diagram for $[4]-SU(3)_0-SU(2)-[2]$.
\label{fig:MQQuiver}}
\end{figure}
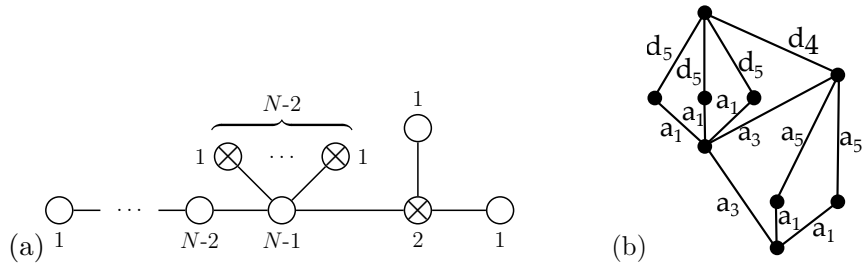

Of course we can also determine the magnetic quiver for the weakly coupled quiver description with brane web shown in figure \ref{fig:T5BraneWeb} (b). For general $N$ it is shown in figure \ref{fig:MQQuiver}, and we also give the Hasse diagram for $N=4$.

\subsection{Magnetic Quivers for Ancestors and Descendants}

So far we considered the magnetic quivers, i.e. the Higgs branch, for the $T_N$ theory. In earlier sections we saw that the extended Coulomb branch 
gives an intricate network of theories, that are related by decoupling and RG-flow. We identified the ancestors of $T_N$, the theories $G_N$ and $P_N$, as well as their 6d origin. On the other hand we also determined the descendants of $T_N$, which are obtained by decoupling hypermultiplets. The structure of the CFDs and toric diagrams allowed us to relate the Coulomb branches of these theories in a simple fashion. Extending this descendant structure to the Higgs branch is therefore a natural question, i.e. to uncover a relationship among the magnetic quivers in the $T_N$ genealogical tree. 

The brane webs for the weakly couple theories admit a simple way to decouple hypermultiplets, and therefore allow us to compute the magnetic quivers for the descendants as well. Although we do not present a general principle how to decouple in the magnetic quiver directly, the data provided here may be useful in identifying such a general principle and finding a description of the moduli space of these theories, that combine both Higgs and Coulomb branch. 
Motivated by this, we now determine the magnetic quivers for the ancestors and descendants of $T_N$.

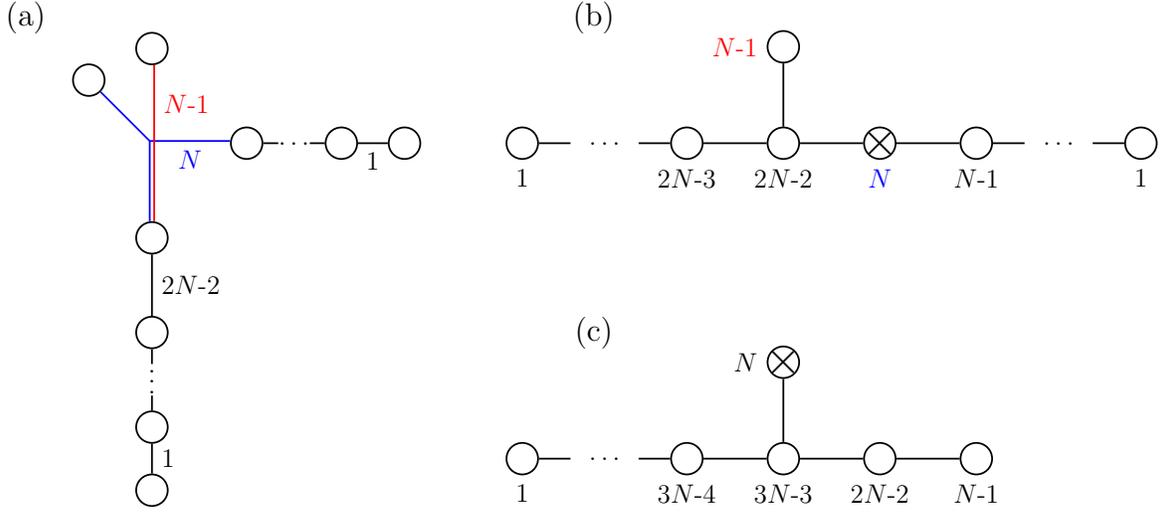
\begin{figure}
\centering
\resizebox{\textwidth}{!}{%
{\large
\begin{tikzpicture}[
roundnode/.style={circle, draw=black, thick, fill=white, minimum size=5mm},
]

\node (1) {};
\node[roundnode] at ($(1)+(1.5,0)$) (2) {};
\node[roundnode] at ($(1)+(3,0)$) (3) {};
\node[roundnode] at ($(1)+(4,0)$) (4) {};

\node[roundnode] at ($(1)+(0,-1.5)$) (5) {};
\node[roundnode] at ($(1)+(0,-3)$) (6) {};
\node[roundnode] at ($(1)+(0,-4.5)$) (7) {};
\node[roundnode] at ($(1)+(0,-5.5)$) (8) {};

\node[roundnode] at ($(1)+(0,1.5)$) (9) {};
\node[roundnode] at ($(1)+(-1,1)$) (10) {};

\node at (2.25,0) {$\hdots$};
\node at (0,-3.65) {$\vdots$};

\node at ($(1)+(-2,2)$) {\Large (a)};

\draw[thick,blue] ($(1)+(-1pt,1pt)$) -- ($(2.west)+(0,1pt)$) node[midway,below] {$N$};
\draw[thick] (2,0) -- (2);
\draw[thick] (2.5,0) -- (3);
\draw[thick] (4) -- (3) node[midway,below] {1};

\draw[thick,blue] ($(1)+(-1pt,1pt)$) -- ($(5.north)+(-1pt,0)$) {};
\draw[thick] (5) -- (6) node[midway,right] {$2N$-2};
\draw[thick] (0,-3.5) -- (6);
\draw[thick] (0,-4) -- (7);
\draw[thick] (7) -- (8) node[midway,right] {1};

\draw[thick,blue] ($(1)+(-1pt,1pt)$) -- (10) {};

\draw[thick,red] ($(9.south)+(1pt,0)$) -- ($(5.north)+(1pt,0)$) node[near start,right] {$N$-1};

\node[roundnode, label=below:\large $2N$-2] at ($(10,0)$) (1) {};

\node[roundnode, label=below:\textcolor{blue}{\large $N$}] (2) [right=of 1] {};
\node at (2) {\huge$\times$};
\node[roundnode, label=below:\large $N$-1] (3) [right=of 2] {};
\node (4) [right=.5of 3] {};
\node (5) [right=.5of 4] {};
\node[roundnode, label=below:\large 1] (6) [right=.5of 5] {};
\node  at ($(4)!0.5!(5)$) {$\hdots$};

\node[roundnode, label=below:\large $2N$-3] (7) [left=of 1] {};
\node (8) [left=.5of 7] {};
\node (9) [left=.5of 8] {};
\node[roundnode, label=below:\large 1] (10) [left=.5of 9] {};
\node  at ($(8)!0.5!(9)$) {$\hdots$} (11);

\node[roundnode, label=left:\textcolor{red}{\large $N$-1}] (12) [above=of 1] {};

\node at ($(1)+(-3,2)$) {\Large (b)};

\draw[thick] (1) -- (2);
\draw[thick] (2) -- (3);
\draw[thick] (3) -- (4);
\draw[thick] (5) -- (6);

\draw[thick] (1) -- (7);
\draw[thick] (7) -- (8);
\draw[thick] (9) -- (10);

\draw[thick] (1) -- (12);

\node[roundnode, label=below:\large $3N$-3] at ($(10,-5)$) (1) {};

\node[roundnode, label=below:\large $2N$-2] (2) [right=of 1] {};
\node[roundnode, label=below:\large $N$-1] (3) [right=of 2] {};

\node[roundnode, label=below:\large $3N$-4] (4) [left=of 1] {};
\node (5) [left=.5of 4] {};
\node (6) [left=.5of 5] {};
\node[roundnode, label=below:\large 1] (7) [left=.5of 6] {};
\node  at ($(5)!0.5!(6)$) {$\hdots$} (8);

\node[roundnode, label=left:\large $N$] (9) [above=of 1] {};
\node at (9) {\huge$\times$};

\node at ($(1)+(-3,2)$) {\Large (c)};

\draw[thick] (1) -- (2);
\draw[thick] (2) -- (3);

\draw[thick] (1) -- (4);
\draw[thick] (4) -- (5);
\draw[thick] (6) -- (7);

\draw[thick] (1) -- (9);

\end{tikzpicture}
}
}
\caption{a) Strongly coupled brane web for $P_N$ with subwebs and magnetic quivers for b) $P_N$ and c) $G_N$.
 \label{fig:MQsGNPN}}
\end{figure}

First, let us start with the (grand)parent theories $G_N$ and $P_N$, whose CFD is shown in figure \ref{fig:ParentsTN}. The starting point is the quiver description \eqref{TN-marginal-quiver} for the marginal theory, with one or two decoupled flavors,  which results in $G_N$ and $P_N$, respectively. Associated to these we can construct the brane web from the toric diagram and perform some Hanany-Witten moves that allow us to take the strong-coupling limit. The magnetic quiver is then obtained by the same methods as for $T_N$, i.e. by successively subtracting off sub-diagrams in the web. In figure \ref{fig:MQsGNPN} a) and b) we present the strongly coupled brane web for $P_N$ with the subwebs represented in colours, leading to the corresponding magnetic quiver. For $G_N$ the brane web is somewhat more intricate but can equivalently be obtained from the weakly coupled quiver description and we state the magnetic quiver in figure \ref{fig:MQsGNPN} c).

Using this method we can also confirm the dualities in \eqref{DualityGN} and \eqref{DualityPN}. 
We clearly see that $G_N$ has flavor symmetry $SU(3N)$ whereas $P_N$ has flavor symmetry $SU(2N) \times SU(N)$. For $N=3$ these are enhanced to $E_8$ and $E_7$ respectively.

\begin{figure}
\centering
\resizebox{.8\textwidth}{!}{%
{\large
\begin{tikzpicture}[
roundnode/.style={circle, draw=black, thick, fill=white, minimum size=5mm},
]

\node (1) {};
\node[roundnode] at ($(1)+(-2,0)$) (2) {};
\node[roundnode] at ($(1)+(-3.5,0)$) (3) {};
\node[roundnode] at ($(1)+(-4.5,0)$) (4) {};

\node[roundnode] at ($(1)+(0,-2)$) (5) {};
\node[roundnode] at ($(1)+(0,-3.5)$) (6) {};
\node[roundnode] at ($(1)+(0,-4.5)$) (7) {};

\node[roundnode] at ($(1)+(1,1)$) (8) {};
\node[roundnode] at ($(1)+(2,2)$) (9) {};
\node[roundnode] at ($(1)+(3,3)$) (10) {};
\node[roundnode] at ($(1)+(4,4)$) (11) {};

\node at ($(1)+(-1,-1)$) (12) {};
\node[roundnode] at ($(1)+(-1,-2)$) (13) {};
\node[roundnode] at ($(1)+(-2,-1)$) (14) {};

\node at (-2.75,0) {$\hdots$};
\node at (0,-2.65) {$\vdots$};
\node at (2.5,2.65) {$\iddots$};

\node at ($(1)+(-2,2)$) {\Large (a)};

\draw[thick,blue] ($(1)+(0,1pt)$) -- ($(2.east)+(0,1pt)$) node[midway,above] {$N$-1};
\draw[thick] (-2.5,0) -- (2);
\draw[thick] (-3,0) -- (3);
\draw[thick] (4) -- (3) node[midway,below] {1};

\draw[thick,blue] ($(1)+(0,1pt)$) -- (5) {};
\draw[thick] (0,-2.5) -- (5);
\draw[thick] (0,-3) -- (6);
\draw[thick] (6) -- (7) node[midway,right] {1};

\draw[thick,blue] ($(1)+(0,1pt)$) -- ($(8.south west)+(-1pt,1pt)$) {};
\draw[thick] (8) -- (9) node[midway,below right=-.2] {$N$-1};
\draw[thick] (9) -- (2.25,2.25);
\draw[thick] (10) -- (2.75,2.75);
\draw[thick] (10) -- (11) node[midway,below right] {1};

\draw[thick,red] ($(8.south west)+(1pt,-1pt)$) -- ($(-1,-1)+(1pt,-1pt)$);
\draw[thick,red] ($(13.north)+(1pt,0)$) -- ($(-1,-1)+(1pt,-1pt)$) node[midway,above right] {1};
\draw[thick,red] ($(14.east)+(0,-1pt)$) -- ($(-1,-1)+(1pt,-1pt)$);

\node[roundnode, label=below: \textcolor{blue}{\large$N$-1}] at ($(1)+(9,-4)$) (1) {};
\node at (1) {\huge$\times$};

\node[roundnode, label=below:\large $N$-2] (2) [right=of 1] {};
\node (3) [right=.5of 2] {};
\node (4) [right=.5of 3] {};
\node[roundnode, label=below:\large 1] (5) [right=.5of 4] {};
\node  at ($(3)!0.5!(4)$) {$\hdots$} (6);

\node[roundnode, label=below:\large $N$-2] (7) [left=of 1] {};
\node (8) [left=.5of 7] {};
\node (9) [left=.5of 8] {};
\node[roundnode, label=below:\large 1] (10) [left=.5of 9] {};
\node  at ($(8)!0.5!(9)$) {$\hdots$} (11);

\node[roundnode, label=left:\large $N$-1] (12) [above=of 1] {};
\node[roundnode, label=left:\large $N$-2] (13) [above=of 12] {};
\node (14) [above=.5of 13] {};
\node (15) [above=.5of 14] {};
\node[roundnode, label=left:\large 1] (16) [above=.5of 15] {};
\node  at ($(14)!0.5!(15)$) {$\vdots$} (17);

\node[roundnode, label=right:\textcolor{red}{\large 1}] (18) [right=of 12] {};
\node at (18) {\huge$\times$};

\node at ($(1)+(-3,2)$) {\Large (b)};

\draw[thick] (1) -- (2);
\draw[thick] (2) -- (3);
\draw[thick] (4) -- (5);

\draw[thick] (1) -- (7);
\draw[thick] (7) -- (8);
\draw[thick] (9) -- (10);

\draw[thick] (1) -- (12);
\draw[thick] (12) -- (13);
\draw[thick] (13) -- (14);
\draw[thick] (15) -- (16);

\draw[thick] (12) -- (18);

\end{tikzpicture}

}}
\caption{a) Brane web and b) magnetic quiver for the first descendant of $T_N$.}
 \label{fig:1stDescTN}
\end{figure}
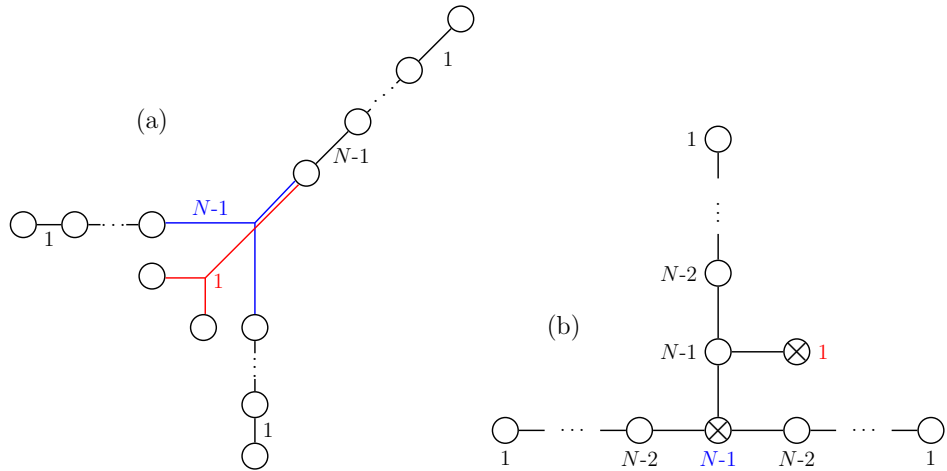

As explained above the first descendant of $T_N$ is obtained by flopping out one curve, i.e. changing a triangulation in the toric diagram. In this description, it corresponds to the first arrow in figure \ref{fig:CFDFlop}. In the brane web picture, we can describe this procedure by turning on the corresponding mass parameter. The resulting brane web and magnetic quiver are shown in figure \ref{fig:1stDescTN}. We explicitly show the decomposition into the subwebs in color, which is reflected in the labels in the MQ.
The Hasse diagram for the first descendant of $T_4$ is
\be
\includegraphics*[width=4cm]{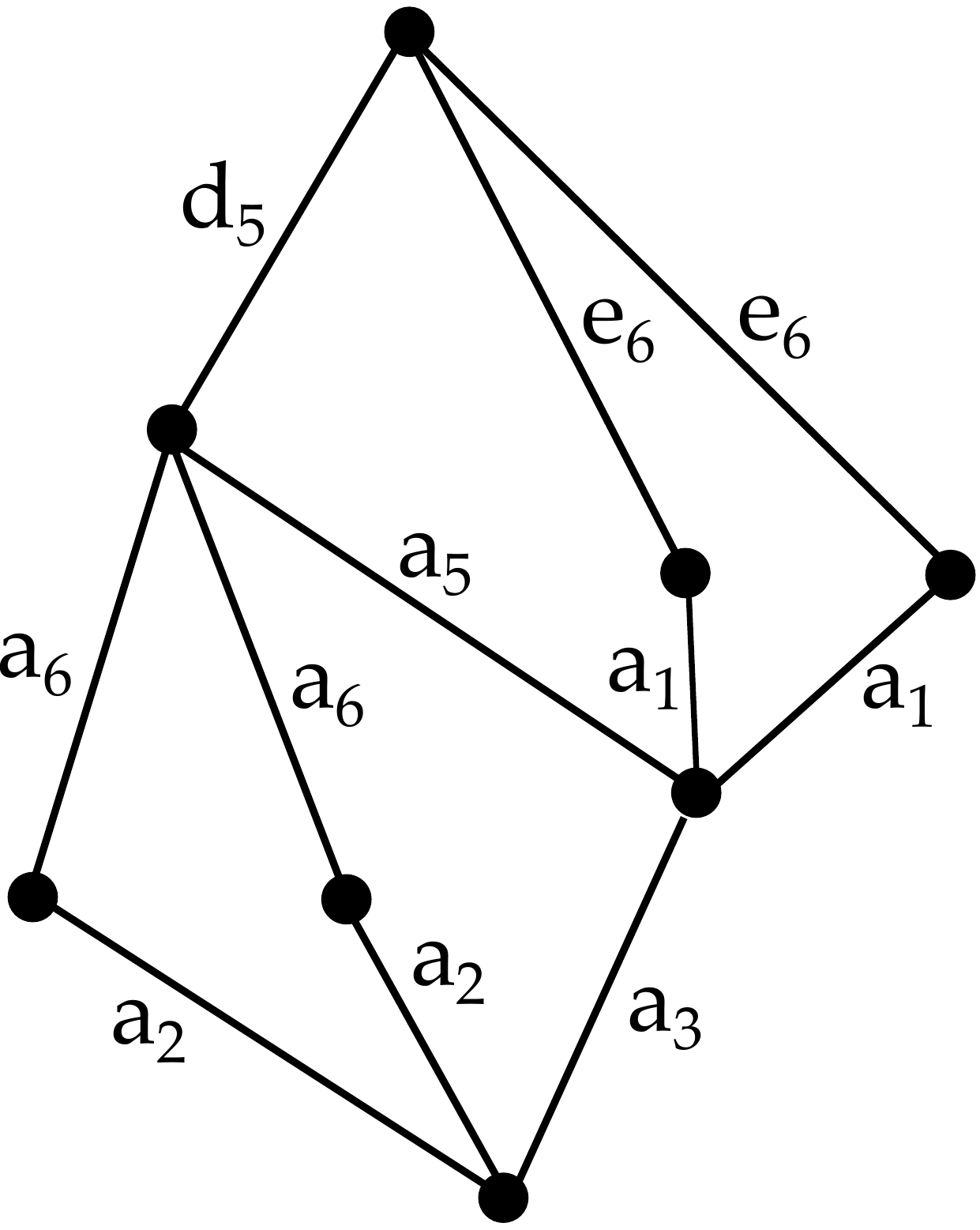} \,.
\ee
As expected the flavor group is given by $SU(N)\times SU(N-1)^2$ from the balanced nodes. For $N=3$ all nodes are balanced and the flavor symmetry is enhanced to $SO(10)=E_5$.

We can continue this logic to find the magnetic quivers for the entire descendant tree of $T_N$. In practice, this is quite arduous and depends heavily on $N$. However, the first few descendants are completely universal, and we now present the first three descendants and their magnetic quivers explicitly. 

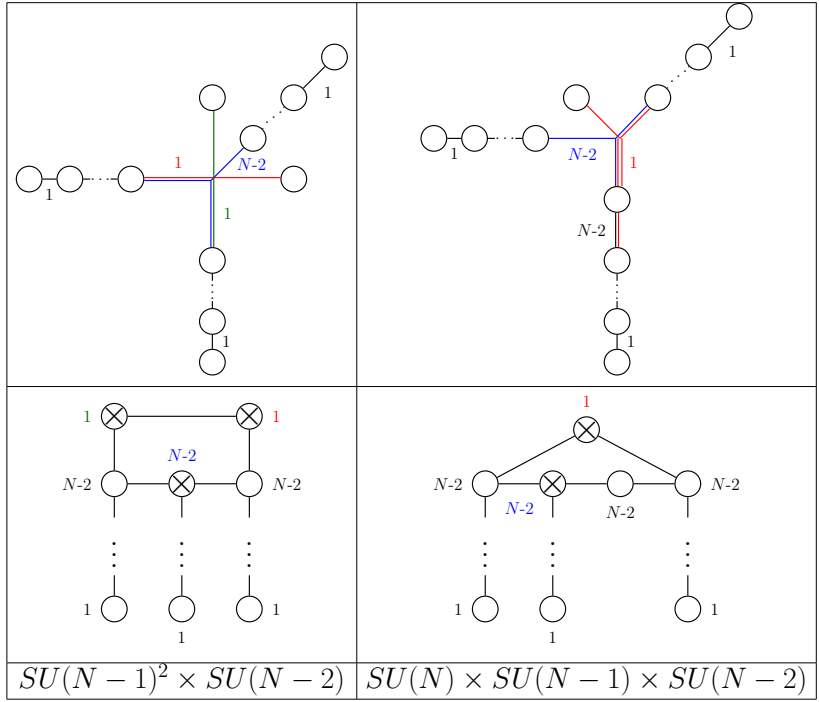
\begin{figure}
\centering

\resizebox{.7\textwidth}{!}{%
{\huge
\begin{tabular}{| c | c | }

\hline

\begin{tikzpicture}[
roundnode/.style={circle, draw=black, thick, fill=white, minimum size=5mm},
]

\node (1) {};
\node[roundnode] at ($(1)+(-2,0)$) (2) {};
\node[roundnode] at ($(1)+(-3.5,0)$) (3) {};
\node[roundnode] at ($(1)+(-4.5,0)$) (4) {};

\node[roundnode] at ($(1)+(0,-2)$) (5) {};
\node[roundnode] at ($(1)+(0,-3.5)$) (6) {};
\node[roundnode] at ($(1)+(0,-4.5)$) (7) {};

\node[roundnode] at ($(1)+(1,1)$) (8) {};
\node[roundnode] at ($(1)+(2,2)$) (9) {};
\node[roundnode] at ($(1)+(3,3)$) (10) {};

\node[roundnode] at ($(1)+(2,0)$) (11) {};
\node[roundnode] at ($(1)+(0,2)$) (12) {};

\node at (-2.75,0) {\large$\hdots$};
\node at (0,-2.65) {\large$\vdots$};
\node at (1.5,1.65) {\large$\iddots$};

\draw[thick,blue] ($(1)+(-1pt,-1pt)$) -- ($(2.east)+(0,-1pt)$) {};
\draw[thick] (-2.5,0) -- (2);
\draw[thick] (-3,0) -- (3);
\draw[thick] (4) -- (3) node[midway,below] {\large 1};

\draw[thick,blue] ($(1)+(-1pt,-1pt)$) -- ($(5.north)+(-1pt,0)$) {};
\draw[thick] (0,-2.5) -- (5);
\draw[thick] (0,-3) -- (6);
\draw[thick] (6) -- (7) node[midway,right] {\large 1};

\draw[thick,blue] ($(1)+(-1pt,-1pt)$) -- (8) node[midway, right] {\large$N$-2};
\draw[thick] (8) -- (1.25,1.25);
\draw[thick] (9) -- (1.75,1.75);
\draw[thick] (9) -- (10) node[midway,below right] {\large 1};

\draw[thick,red] ($(2.east)+(0,1pt)$) -- ($(11.west)+(0,1pt)$) node[near start, above] {\large 1};
\draw[thick, black!60!green] ($(5.north)+(1pt,0)$) -- ($(12.south)+(1pt,0)$) node[near start, right] {\large 1};

\end{tikzpicture}

&

\begin{tikzpicture}[
roundnode/.style={circle, draw=black, thick, fill=white, minimum size=5mm},
]

\node at ($(11)+(8,0)$) (1) {};
\node[roundnode] at ($(1)+(-2,0)$) (2) {};
\node[roundnode] at ($(1)+(-3.5,0)$) (3) {};
\node[roundnode] at ($(1)+(-4.5,0)$) (4) {};

\node[roundnode] at ($(1)+(0,-1.5)$) (5) {};
\node[roundnode] at ($(1)+(0,-3)$) (12) {};
\node[roundnode] at ($(1)+(0,-4.5)$) (6) {};
\node[roundnode] at ($(1)+(0,-5.5)$) (7) {};

\node[roundnode] at ($(1)+(1,1)$) (8) {};
\node[roundnode] at ($(1)+(2,2)$) (9) {};
\node[roundnode] at ($(1)+(3,3)$) (10) {};

\node[roundnode] at ($(1)+(-1,1)$) (11) {};

\node at (7.25,0) {\large$\hdots$};
\node at (10,-3.65) {\large$\vdots$};
\node at (11.5,1.65) {\large$\iddots$};

\draw[thick,blue] ($(1)+(-1pt,0)$) -- (2) node[midway,below] {\large$N$-2};
\draw[thick] (7.5,0) -- (2);
\draw[thick] (7,0) -- (3);
\draw[thick] (4) -- (3) node[midway,below] {\large1};

\draw[thick,blue] ($(1)+(-1pt,0)$) -- ($(5.north)+(-1pt,0)$) {};
\draw[thick] ($(5.south)+(-1pt,0)$) -- ($(12.north)+(-1pt,0)$)  node[midway,left] {\large$N$-2};
\draw[thick] (10,-3.5) -- (12);
\draw[thick] (10,-4) -- (6);
\draw[thick] (6) -- (7) node[midway,right] {\large1};

\draw[thick,blue] ($(1)+(-1pt,0)$) -- ($(8.south west)+(-1pt,1pt)$) {};
\draw[thick] (8) -- (11.25,1.25);
\draw[thick] (9) -- (11.75,1.75);
\draw[thick] (9) -- (10) node[midway,below right] {\large1};

\draw[thick,red] ($(1)+(2pt,0)$) -- ($(11.south east)+(1pt,1pt)$) {};
\draw[thick,red] ($(1)+(2pt,0)$) -- ($(8.south west)+(1pt,-1pt)$) {};
\draw[thick,red, double distance=2pt] ($(1)+(2pt,0)$) -- ($(5.north)+(2pt,0)$) node[midway,right] {\large1};
\draw[thick,red] ($(5.south)+(1pt,0)$) -- ($(12.north)+(1pt,0)$) {};

\end{tikzpicture}

\\

\hline

     \begin{tikzpicture}[
roundnode/.style={circle, draw=black, thick, fill=white, minimum size=5mm},
]

\node[roundnode, label=above:\textcolor{blue}{\large $N$-2}] (1) {};
\node at (1) {\Huge$\times$};
\node[roundnode, label=right:\large $N$-2] (2) [right=of 1] {};
\node[roundnode, label=left:\large $N$-2] (3) [left=of 1] {};

\node (4) [below=.5of 1] {};
\node (5) [below=.5of 4] {};
\node[roundnode, label=below:\large 1] (6) [below=.5of 5] {};
\node  at ($(4)!0.5!(5)$) {$\vdots$} (7);

\node (8) [below=.5of 2] {};
\node (9) [below=.5of 8] {};
\node[roundnode, label=right:\large 1] (10) [below=.5of 9] {};
\node  at ($(8)!0.5!(9)$) {$\vdots$} (11);

\node (12) [below=.5of 3] {};
\node (13) [below=.5of 12] {};
\node[roundnode, label=left:\large 1] (14) [below=.5of 13] {};
\node  at ($(12)!0.5!(13)$) {$\vdots$} (15);

\node[roundnode, label=right:\textcolor{red}{\large 1}] (16) [above=of 2] {};
\node[roundnode, label=left:\textcolor{ black!60!green}{\large 1}] (17) [above=of 3] {};
\node at (16) {\Huge$\times$};
\node at (17) {\Huge$\times$};

\draw[thick] (1) -- (2);
\draw[thick] (1) -- (3);
\draw[thick] (1) -- (4);
\draw[thick] (5) -- (6);
\draw[thick] (2) -- (8);
\draw[thick] (9) -- (10);
\draw[thick] (3) -- (12);
\draw[thick] (13) -- (14);
\draw[thick] (2) -- (16);
\draw[thick] (3) -- (17);
\draw[thick] (16) -- (17);

\end{tikzpicture}

&

\begin{tikzpicture}[
roundnode/.style={circle, draw=black, thick, fill=white, minimum size=5mm},
]

\node[roundnode, label=below left:\textcolor{blue}{\large $N$-2}] (1) {};
\node at (1) {\Huge$\times$};
\node[roundnode, label=below:\large $N$-2] (17) [right=of 1] {};

\node[roundnode, label=right:\large $N$-2] (2) [right=of 17] {};
\node[roundnode, label=left:\large $N$-2] (3) [left=of 1] {};

\node (4) [below=.5of 1] {};
\node (5) [below=.5of 4] {};
\node[roundnode, label=below:\large 1] (6) [below=.5of 5] {};
\node  at ($(4)!0.5!(5)$) {$\vdots$} (7);

\node (8) [below=.5of 2] {};
\node (9) [below=.5of 8] {};
\node[roundnode, label=right:\large 1] (10) [below=.5of 9] {};
\node  at ($(8)!0.5!(9)$) {$\vdots$} (11);

\node (12) [below=.5of 3] {};
\node (13) [below=.5of 12] {};
\node[roundnode, label=left:\large 1] (14) [below=.5of 13] {};
\node  at ($(12)!0.5!(13)$) {$\vdots$} (15);

\node[roundnode, label=above:\textcolor{red}{\large 1}] (16) [above=of $(1)!0.5!(17)$] {};
\node at (16) {\Huge$\times$};

\draw[thick] (1) -- (17);
\draw[thick] (2) -- (17);
\draw[thick] (1) -- (3);
\draw[thick] (1) -- (4);
\draw[thick] (5) -- (6);
\draw[thick] (2) -- (8);
\draw[thick] (9) -- (10);
\draw[thick] (3) -- (12);
\draw[thick] (13) -- (14);
\draw[thick] (2) -- (16);
\draw[thick] (3) -- (16);

\end{tikzpicture}

\\ \hline
$SU(N-1)^2\times SU(N-2)$ & $SU(N)\times SU(N-1) \times SU(N-2)$  \\ \hline
\end{tabular}

}
}

\caption{Brane webs and magnetic quivers for second descendants of $T_{N}$ with corresponding flavor group.
 \label{fig:MQ2nddescTN}}
\end{figure}

There are two different second descendants, which are generated by two mass deformations from the $T_N$ theory. In the toric diagram these are obtained by flopping two curves either on the same or different corners.
Note that because of the symmetry we only need to consider one possible configuration each. From the toric diagram we can infer the brane web and compute the magnetic quiver. This is shown in figure \ref{fig:MQ2nddescTN}. For $N=3$ both descendants are equivalent and the flavor group is enhanced to $SU(5)=E_4$. From the toric diagram this equivalence is not immediately obvious.

\begin{figure}
\centering

\resizebox{.7\textwidth}{!}{%
{\huge
\begin{tabular}{| c | c |  }
\hline

     \begin{tikzpicture}[
roundnode/.style={circle, draw=black, thick, fill=white, minimum size=5mm},
]

\node[roundnode, label=above right:\large $N$-2] (1) {};
\node at (1) {\Huge$\times$};

\node[roundnode, label=below:\large $N$-3] (2) [right=of 1] {};
\node (3) [right=.5of 2] {};
\node (4) [right=.5of 3] {};
\node[roundnode, label=below:\large 1] (5) [right=.5of 4] {};
\node  at ($(3)!0.5!(4)$) {$\hdots$} (6);

\node[roundnode, label=below:\large $N$-3] (7) [left=of 1] {};
\node (8) [left=.5of 7] {};
\node (9) [left=.5of 8] {};
\node[roundnode, label=below:\large 1] (10) [left=.5of 9] {};
\node  at ($(8)!0.5!(9)$) {$\hdots$} (11);

\node[roundnode, label=left:\large $N$-3] (12) [above=of 1] {};
\node (13) [above=.5of 12] {};
\node (14) [above=.5of 13] {};
\node[roundnode, label=left:\large 1] (15) [above=.5of 14] {};
\node  at ($(13)!0.5!(14)$) {$\vdots$} (16);

\node[roundnode, label=left:\large 1] (17) [below=of 1] {};
\node at (17) {\Huge$\times$};

\draw[thick] (1) -- (2);
\draw[thick] (2) -- (3);
\draw[thick] (4) -- (5);

\draw[thick] (1) -- (7);
\draw[thick] (7) -- (8);
\draw[thick] (9) -- (10);

\draw[thick] (1) -- (12);
\draw[thick] (12) -- (13);
\draw[thick] (14) -- (15);

\draw[thick, double distance=3pt] (1) -- (17);

\end{tikzpicture}
\quad

     \begin{tikzpicture}[
roundnode/.style={circle, draw=black, thick, fill=white, minimum size=5mm},
]

\node[roundnode, label=below:\large $N$-3] (1) {};
\node at (1) {\Huge$\times$};

\node[roundnode, label=below:\large $N$-3] (2) [right=of 1] {};
\node (3) [right=.5of 2] {};
\node (4) [right=.5of 3] {};
\node[roundnode, label=below:\large 1] (5) [right=.5of 4] {};
\node  at ($(3)!0.5!(4)$) {$\hdots$} (6);

\node[roundnode, label=below:\large $N$-3] (7) [left=of 1] {};
\node (8) [left=.5of 7] {};
\node (9) [left=.5of 8] {};
\node[roundnode, label=below:\large 1] (10) [left=.5of 9] {};
\node  at ($(8)!0.5!(9)$) {$\hdots$} (11);

\node[roundnode, label=left:\large $N$-3] (12) [above=2of 1] {};
\node (13) [above=.5of 12] {};
\node (14) [above=.5of 13] {};
\node[roundnode, label=left:\large 1] (15) [above=.5of 14] {};
\node  at ($(13)!0.5!(14)$) {$\vdots$} (16);

\node[roundnode, label=left:\large 1] (17) [below=of 1] {};
\node at (17) {\Huge$\times$};
\node[roundnode, label=right:\large 1] (18) [above=of 2] {};
\node at (18) {\Huge$\times$};
\node[roundnode, label=left:\large 1] (19) [above=of 7] {};
\node at (19) {\Huge$\times$};

\draw[thick] (1) -- (2);
\draw[thick] (2) -- (3);
\draw[thick] (4) -- (5);

\draw[thick] (1) -- (7);
\draw[thick] (7) -- (8);
\draw[thick] (9) -- (10);

\draw[thick] (1) -- (12);
\draw[thick] (12) -- (13);
\draw[thick] (14) -- (15);

\draw[thick] (17) -- (18);
\draw[thick] (17) -- (19);
\draw[thick] (19) -- (18);

\draw[thick] (17) -- (7);
\draw[thick] (18) -- (2);
\draw[thick] (19) -- (12);

\end{tikzpicture}

\\
\hline

$SU(N-2)^3$

\\

\hline
\hline

 \begin{tikzpicture}[
roundnode/.style={circle, draw=black, thick, fill=white, minimum size=5mm},
]

\node[roundnode, label=below:\large $N$-3] (1) {};

\node[roundnode, label=below:\large $N$-3] (2) [right=of 1] {};
\node[roundnode, label=below:\large $N$-3] (3) [right=of 2] {};
\node at (3) {\Huge$\times$};
\node (4) [right=.5of 3] {};
\node (5) [right=.5of 4] {};
\node[roundnode, label=below:\large 1] (6) [right=.5of 5] {};
\node  at ($(5)!0.5!(4)$) {$\hdots$} (7);

\node[roundnode, label=below:\large $N$-3] (8) [left=of 1] {};
\node (9) [left=.5of 8] {};
\node (10) [left=.5of 9] {};
\node[roundnode, label=below:\large 1] (11) [left=.5of 10] {};
\node  at ($(10)!0.5!(9)$) {$\hdots$} (12);

\node[roundnode, label=right:\large $N$-3] (13) [above=of 3] {};
\node[roundnode, label=right:\large $N$-3] (14) [above=of 13] {};
\node (15) [above=.5of 14] {};
\node (16) [above=.5of 15] {};
\node[roundnode, label=right:\large 1] (17) [above=.5of 16] {};
\node  at ($(15)!0.5!(16)$) {$\vdots$} (18);

\node[roundnode, label=left:\large 1] at ($(8)+(0,3.3)$) (19)  {};
\node at (19) {\Huge$\times$};

\draw[thick] (1) -- (2);
\draw[thick] (2) -- (3);
\draw[thick] (3) -- (4);
\draw[thick] (5) -- (6);

\draw[thick] (1) -- (8);
\draw[thick] (9) -- (8);
\draw[thick] (11) -- (10);

\draw[thick] (3) -- (13);
\draw[thick] (14) -- (13);
\draw[thick] (14) -- (15);
\draw[thick] (16) -- (17);

\draw[thick] (8) -- (19);
\draw[thick] (14) -- (19);

\end{tikzpicture}
\\

\hline

$SU(N)\times SU(N-1) \times SU(N-3)$\\ 

\hline \hline

 \begin{tikzpicture}[
roundnode/.style={circle, draw=black, thick, fill=white, minimum size=5mm},
]

\node[roundnode, label=above:\large $N$-3] (1) {};

\node[roundnode, label=below:\large $N$-3] (2) [right=of 1] {};
\node at (2) {\Huge$\times$};
\node[roundnode, label=above:\large $N$-2] (3) [right=of 2] {};
\node (4) [right=.5of 3] {};
\node (5) [right=.5of 4] {};
\node[roundnode, label=above:\large 1] (6) [right=.5of 5] {};
\node  at ($(5)!0.5!(4)$) {$\hdots$} (7);

\node[roundnode, label=above:\large $N$-3] (8) [left=of 1] {};
\node (9) [left=.5of 8] {};
\node (10) [left=.5of 9] {};
\node[roundnode, label=above:\large 1] (11) [left=.5of 10] {};
\node  at ($(10)!0.5!(9)$) {$\hdots$} (12);

\node[roundnode, label=right:\large $N$-4] (13) [above=of 2] {};
\node (14) [above=.5of 13] {};
\node (15) [above=.5of 14] {};
\node[roundnode, label=right:\large 1] (16) [above=.5of 15] {};
\node  at ($(14)!0.5!(15)$) {$\vdots$} (17);

\node[roundnode, label=below:\large 1] (18) [below=of 2] {};
\node[roundnode, label=below:\large 1] (19) [below=of 3] {};
\node at (18) {\Huge$\times$};
\node at (19) {\Huge$\times$};

\draw[thick] (1) -- (2);
\draw[thick] (2) -- (3);
\draw[thick] (3) -- (4);
\draw[thick] (5) -- (6);

\draw[thick] (1) -- (8);
\draw[thick] (9) -- (8);
\draw[thick] (11) -- (10);

\draw[thick] (2) -- (13);
\draw[thick] (14) -- (13);
\draw[thick] (16) -- (15);

\draw[thick] (3) -- (19);
\draw[thick] (3) -- (18);
\draw[thick] (19) -- (18);
\draw[thick] (18) -- (8);

\end{tikzpicture}

\\
\hline

$SU(N-1)^2\times SU(N-3)$
\\
\hline
\hline

\begin{tikzpicture}[
roundnode/.style={circle, draw=black, thick, fill=white, minimum size=5mm},
]

\node[roundnode, label=below:\large $N$-3] (1) {};

\node[roundnode, label=below:\large $N$-3] (2) [right=of 1] {};
\node at (2) {\Huge$\times$};
\node[roundnode, label=below:\large $N$-3] (3) [right=of 2] {};
\node (4) [right=.5of 3] {};
\node (5) [right=.5of 4] {};
\node[roundnode, label=below:\large 1] (6) [right=.5of 5] {};
\node  at ($(5)!0.5!(4)$) {$\hdots$} (7);

\node (9) [left=.5of 1] {};
\node (10) [left=.5of 9] {};
\node[roundnode, label=below:\large 1] (11) [left=.5of 10] {};
\node  at ($(10)!0.5!(9)$) {$\hdots$} (12);

\node[roundnode, label=right:\large 1] (13) [above=of 3] {};
\node at (13) {\Huge$\times$};
\node[roundnode, label=right:\large $N$-3] (14) [above=of 13] {};
\node (15) [above=.5of 14] {};
\node (16) [above=.5of 15] {};
\node[roundnode, label=right:\large 1] (17) [above=.5of 16] {};
\node  at ($(15)!0.5!(16)$) {$\vdots$} (18);

\node[roundnode, label=left:\large 1] (19) [above=of 1] {};
\node at (19) {\Huge$\times$};
\node[roundnode, label=left:\large $N$-3] (20) [left=of 14] {};

\draw[thick] (1) -- (2);
\draw[thick] (2) -- (3);
\draw[thick] (3) -- (4);
\draw[thick] (5) -- (6);

\draw[thick] (1) -- (9);
\draw[thick] (11) -- (10);

\draw[thick] (3) -- (13);
\draw[thick] (14) -- (13);
\draw[thick] (14) -- (15);
\draw[thick] (16) -- (17);

\draw[thick] (1) -- (19);
\draw[thick] (14) -- (20);
\draw[thick, double distance=3pt] (19) -- (13);
\draw[thick] (20) -- (2);

\end{tikzpicture}

\quad

\begin{tikzpicture}[
roundnode/.style={circle, draw=black, thick, fill=white, minimum size=5mm},
]

\node[roundnode, label=below:\large $N$-3] (1) {};

\node[roundnode, label=below:\large $N$-2] (2) [right=of 1] {};
\node at (2) {\Huge$\times$};
\node[roundnode, label=below:\large $N$-3] (3) [right=of 2] {};
\node (4) [right=.5of 3] {};
\node (5) [right=.5of 4] {};
\node[roundnode, label=below:\large 1] (6) [right=.5of 5] {};
\node  at ($(5)!0.5!(4)$) {$\hdots$} (7);

\node (9) [left=.5of 1] {};
\node (10) [left=.5of 9] {};
\node[roundnode, label=below:\large 1] (11) [left=.5of 10] {};
\node  at ($(10)!0.5!(9)$) {$\hdots$} (12);

\node[roundnode, label=right:\large $N$-2] (13) [above=of 2] {};
\node (14) [above=.5of 13] {};
\node (15) [above=.5of 14] {};
\node[roundnode, label=right:\large 1] (16) [above=.5of 15] {};
\node  at ($(14)!0.5!(15)$) {$\vdots$} (17);

\node[roundnode, label=left:\large 1] (18) [above=of 1] {};
\node at (18) {\Huge$\times$};

\draw[thick] (1) -- (2);
\draw[thick] (2) -- (3);
\draw[thick] (3) -- (4);
\draw[thick] (5) -- (6);

\draw[thick] (1) -- (9);
\draw[thick] (11) -- (10);

\draw[thick] (2) -- (13);
\draw[thick] (14) -- (13);
\draw[thick] (16) -- (15);

\draw[thick] (13) -- (18);
\draw[thick] (2) -- (18);

\end{tikzpicture}

\\

\hline

 $SU(N-1)\times SU(N-2)^2$  \\

\hline

\end{tabular}

}
}

\caption{MQs for third descendants of $T_{N>3}$ with corresponding flavor group.
 \label{fig:MQ3rddescTN}}
\end{figure}
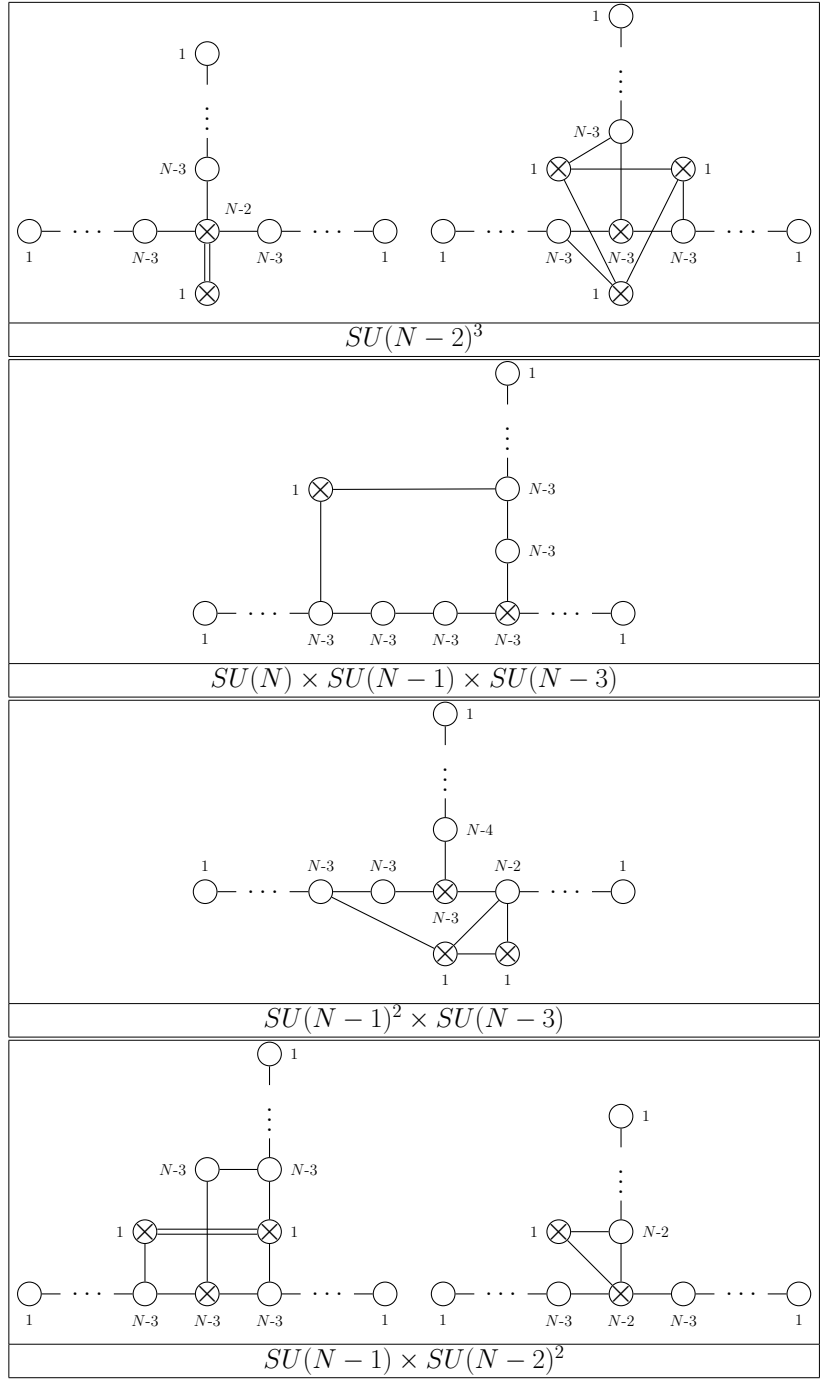

Finally, there are four third descendants. They are shown in figure \ref{fig:MQ3rddescTN} and all feature the expected flavor symmetry. For $N=3$ there is again only one descendant with flavor group $SU(3)\times SU(2)=E_3$. We can most easily see this by setting $N=3$ in the two branches of the first or last quiver in figure \ref{fig:MQ3rddescTN}. This process can be further applied to determine the magnetic quivers of all the descendants of $T_N$. We close by focusing on the theories that arise at the bottom of the descendant tree, $B_N^{(i)}$.

\subsection{Magnetic Quivers for $B_N^{(i)}$ Theories}

In the last subsection we showed how the descendant structure is realized along the Higgs branch, by computing the magnetic quivers and Hasse diagrams. 
In this final section we will determine the magnetic quivers for the infinite series of models, that are non-Lagrangian, and were introduced in section \ref{sec:NonLag}. 
These theories are furthermore of interest as they have higher form symmetries \cite{Morrison:2020ool} and an interesting questions is whether this is realized (unbroken) on the Higgs branch. Note that the theories $B^{(2)}_N$ and $B^{(3)}_{N-1}$ have the same Higgs branches. 

\begin{figure}
\centering
\resizebox{\textwidth}{!}{%
{\large
\begin{tikzpicture}[
roundnode/.style={circle, draw=black, thick, fill=white, minimum size=5mm},
]

\node[roundnode, label=below:\large $N$-3] at ($(0,0)$) (1) {};

\node[roundnode, label=below:\large 1] (2) [right=of 1] {};
\node at (2) {\huge$\times$};

\node[roundnode, label=below:\large $N$-4] (4) [left=of 1] {};
\node (5) [left=.5of 4] {};
\node (6) [left=.5of 5] {};
\node[roundnode, label=below:\large 1] (7) [left=.5of 6] {};
\node  at ($(5)!0.5!(6)$) {$\hdots$} (8);

\node[roundnode, label=left:\large $1$] (9) [above=of 1] {};
\node at (9) {\huge$\times$};

\node at ($(1)+(-3,2)$) {\Large (a)};

\draw[thick] (1) -- (2);

\draw[thick] (1) -- (4);
\draw[thick] (4) -- (5);
\draw[thick] (6) -- (7);

\draw[thick] (1) -- (9) node[midway,right] {\large$N$-3};

\node[roundnode, label=below:\large $N$-3] at ($(8,2.5)$) (1) {};

\node[roundnode, label=below:\large 1] (2) [right=1.5of 1] {};
\node at (2) {\huge$\times$};

\node[roundnode, label=below:\large $N$-4] (4) [left=of 1] {};
\node (5) [left=.5of 4] {};
\node (6) [left=.5of 5] {};
\node[roundnode, label=below:\large 1] (7) [left=.5of 6] {};
\node  at ($(5)!0.5!(6)$) {$\hdots$} (8);

\node at ($(1)+(-3,1)$) {\Large (b)};

\draw[thick] (1) -- (2) node[midway,above] {\large$N$-2};

\draw[thick] (1) -- (4);
\draw[thick] (4) -- (5);
\draw[thick] (6) -- (7);

\node[roundnode, label=below:\large $N$-2] at ($(8,0)$) (1) {};

\node[roundnode, label=below:\large 1] (2) [right=1.5of 1] {};
\node at (2) {\huge$\times$};

\node[roundnode, label=below:\large $N$-3] (4) [left=of 1] {};
\node (5) [left=.5of 4] {};
\node (6) [left=.5of 5] {};
\node[roundnode, label=below:\large 1] (7) [left=.5of 6] {};
\node  at ($(5)!0.5!(6)$) {$\hdots$} (8);

\node at ($(1)+(-3,1)$) {\Large (c)};

\draw[thick] (1) -- (2) node[midway,above] {\large$N$-1};

\draw[thick] (1) -- (4);
\draw[thick] (4) -- (5);
\draw[thick] (6) -- (7);

\end{tikzpicture}
}
}
\caption{MQs for a) $B_N^{(1)}$, b) $B_N^{(2)}$ and c) $B_N^{(3)}$ for $N>3$.
 \label{fig:MQBottoms}}
\end{figure}
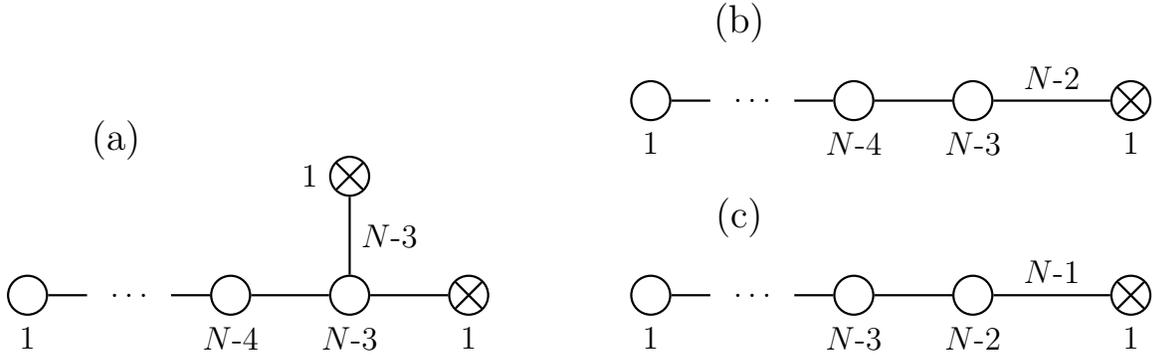

In a similar fashion to the previous section we can compute the magnetic quiver of the $B_N^{(i)}$ theories from the toric descriptions in figure \ref{fig:Bottoms}. For $B_N$ it is trivial, i.e. it consists of a single $U(1)$ node, but for the other theories we obtain the MQs summarized in figure \ref{fig:MQBottoms}, which all feature a connection of edge multiplicity of order $N$. The balanced nodes reproduce the expected flavour symmetry.

\begin{figure}
\centering
\includegraphics*[width= 10cm]{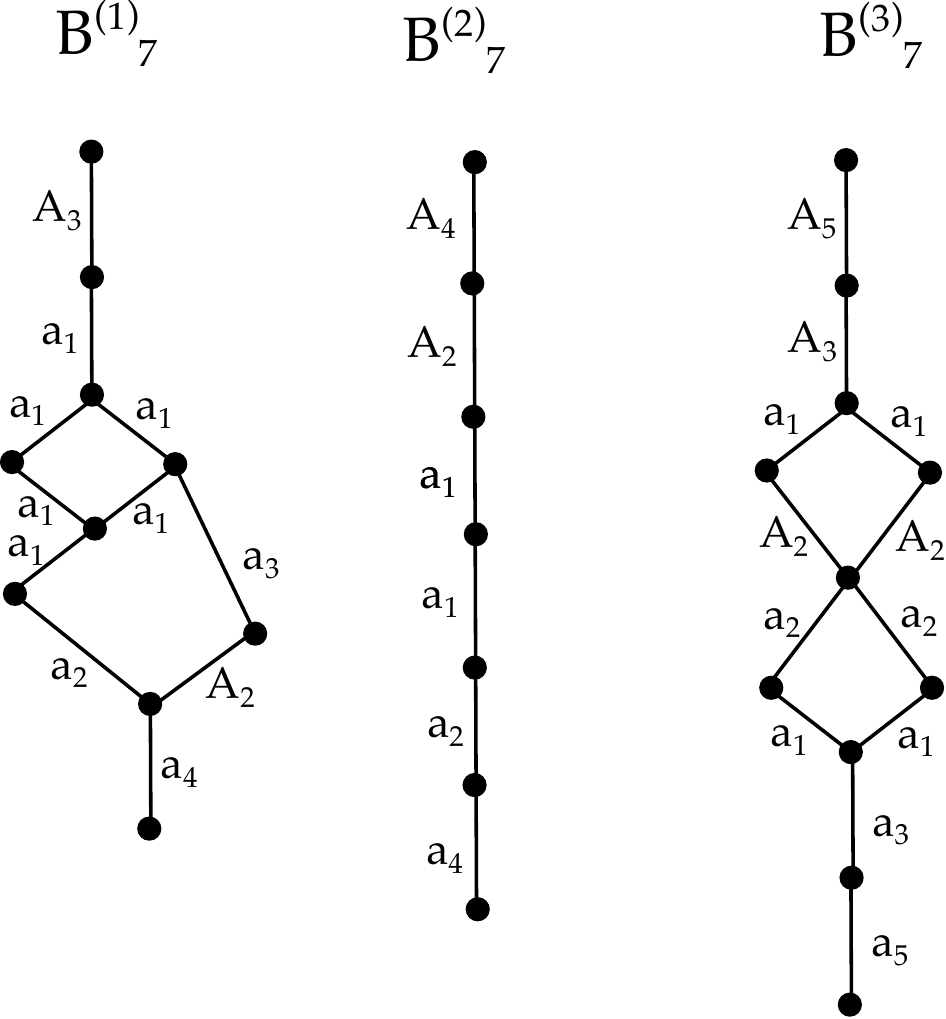}
\caption{The Hasse diagrams for $B_7^{(1)}$, $B_7^{(2)}$ and $B_7^{(3)}$. The $A_N$ are the Kleinian singularities, whereas $a_n$ are the standard $SU(n+1)$ singularities. \label{fig:HasseBN}}
\end{figure}

From the magnetic quivers we can again compute the Hasse diagram, i.e. the foliation structure of the Higgs branch by symplectic singularities, for these non-Lagrangian theories. These are shown in figure \ref{fig:HasseBN}, where in particular the non-abelian flavor symmetry appears as the last leaf. It would very interesting to confirm this structure from a direct geometric computation or otherwise. 

\section{Summary, Conclusions and Outlook}

{The main purpose of this paper was to explore the physics {of} $T_N$ and closely related SCFTs, using all recently developed approaches to  5d gauge theories and SCFTs}. The exposition should give a largely self-contained explanation of the advances in the Coulomb branch and geometry, as well as the application of brane webs to study the Higgs branch, magnetic quivers and Hasse diagrams. 
The $T_N$ and related theories that we discussed are particularly useful for such a comprehensive approach, as they have realizations in 
a multitude of different frameworks: both toric geometry, brane web, CFDs and gauge theories.

We hope that by discussing the CFDs in the context of toric Calabi-Yau singularities, their properties and uses are more accessible to some readers than the approach in the original work, where the geometries arose from non-flat resolutions of elliptic Calabi-Yau threefolds with non-minimal singularities. In comparison, the toric approach is limited to a subclass of theories, but the concepts of CFDs and BG-CFDs have hopefully been put into a more familiar framework to most readers. 

We determined the descendant trees for $T_N$, using both toric flops and CFD-transitions, and showed that the bottom of the CFD tree has in fact several theories that have no weakly-coupled gauge theory descriptions, much like the rank one $\mathbb{P}^2$ theory. By utilizing brane web techniques, we also studied the magnetic quivers of the theories related to $T_N$ by mass deformations.  One of the goals would be to combine these approaches to obtain a global picture of the Higgs and Coulomb branches. 

The focus of this paper has been the theories related to $T_N$, but a similar analysis can be performed for any theory realized in terms of a toric Calabi-Yau singularity. We exemplified this by gluing two $T_N$ theories to get a new toric model. For a general convex toric diagram, the CFDs will be comprised of the  boundary. Likewise the web can be obtained from the toric diagram and thereby the magnetic quiver. It would interesting to develop this complete approach, both in the Coulomb branch and Higgs branch for all toric models.

\subsection*{Acknowledgements}

We thank F.~Apruzzi, M.~van Beest, A.~Bourget, C.~Closset, S.~Giacomelli, D.~Morrison and B.~Willett for discussions.  This work is supported by the ERC Consolidator Grant number 682608 ``Higgs bundles: Supersymmetric Gauge Theories and Geometry (HIGGSBNDL)''.


\providecommand{\href}[2]{#2}\begingroup\raggedright\endgroup

\end{document}